\documentclass[11pt,a4paper]{article}
\pdfoutput=1
\usepackage{jheppub}
\usepackage{amsthm,amsbsy,amsfonts,mathrsfs,enumerate,float,wrapfig}

\title{More on 5d descriptions of 6d SCFTs}

\author[a]{Hirotaka Hayashi,}
\author[b,c]{Sung-Soo Kim,}
\author[c]{Kimyeong Lee,}
\author[d]{Masato Taki,}
\author[c]{and Futoshi Yagi}

\affiliation[a]{Departamento de F\'isica Te\'orica and Instituto de F\'isica Te\'orica UAM/CSIC,\\ Universidad Aut\'onoma de Madrid, Cantoblanco, 28049 Madrid, Spain}
\affiliation[b]{School of Physical Electronics,
University of Electronic Science and Technology of China, Chengdu, Sichuan 611731, China}
\affiliation[c]{Korea Institute for Advanced Study, 
85 Hoegi-ro Dongdaemun-gu, Seoul, 02455, Korea}
\affiliation[d]{
iTHES Research Group and Mathematical Physics Laboratory, RIKEN Nishina Center, Saitama 351-0198, Japan
}

\emailAdd{h.hayashi@csic.es}
\emailAdd{sungsoo.kim@kias.re.kr}
\emailAdd{klee@kias.re.kr}
\emailAdd{taki@riken.jp}
\emailAdd{fyagi@kias.re.kr}

\abstract{We propose new five-dimensional gauge theory descriptions of six-dimensional $\mathcal{N}=(1,0)$ superconformal field theories arising from type IIA brane configurations including an $ON^0$-plane. The new five-dimensional gauge theories may have $SO$, $Sp$, and $SU$ gauge groups and further broaden the landscape of ultraviolet complete five-dimensional $\mathcal{N}=1$ supersymmetric gauge theories. When we include an $O8^-$-plane in addition to an $ON^0$-plane, T-duality yields two $O7^-$-planes at the intersections of an $ON^0$-plane and two $O5^0$-planes. We propose a novel resolution of the $O7^-$-plane with four D7-branes in such a configuration, which enables us to obtain three different types of five-dimensional gauge theories, depending on whether we resolve either none or one or two $O7^-$-planes.
Such different possibilities yield a new five-dimensional duality between a D-type $SU$ quiver and an $SO-Sp$ quiver theories. We also claim that a twisted circle compactification of a six-dimensional superconformal field theory may lead to a five-dimensional gauge theory different from those obtained through a simple circle compactification.  
}
\arxivnumber{1512.08239}

\begin{document}
\preprint{
\begin{flushright}
\tt 
IFT-UAM/CSIC-15-134\\
KIAS-Q15006\\
RIKEN-STAMP-23
\end{flushright}
}
\maketitle


\section{Introduction}\label{sec:intro}

Various aspects of six-dimensional (6d) ${\cal N}=(1,0)$ superconformal field theories (SCFTs) have been investigated recently. 
In particular, a circle compactification of 6d $\mathcal{N}=(1, 0)$ SCFTs lead to new five-dimensional (5d) $\mathcal{N}=1$ supersymmetric gauge theories whose ultraviolet (UV) completion is given by the original 6d SCFT \cite{Hayashi:2015fsa, Yonekura:2015ksa, Gaiotto:2015una, Zafrir:2015rga, Hayashi:2015zka, Ohmori:2015tka}. While the new 5d gauge theories lie beyond the classification in \cite{Intriligator:1997pq}, the manifestation   of these 5d theories can be seen explicitly  through the 5-brane web diagrams 
\cite{Hayashi:2015fsa, Yonekura:2015ksa, Gaiotto:2015una, Bergman:2015dpa,Hayashi:2015zka, Zafrir:2015rga,Ohmori:2015tka}. The   Tao web diagram  extension of these 5-brane webs provides a 
realization of the 6d SCFTs in type IIB string theory  \cite{Kim:2015jba,Hayashi:2015fsa,Hayashi:2015zka}.

A connection between the new 5d gauge theories and 6d SCFTs may be understood by T-duality in string theory. In some cases, it is possible to obtain the 5-brane web diagram realizing a 5d theory with a 6d UV completion by performing T-duality to a type IIA brane setup yielding the 6d SCFT on $S^1$ \cite{Hayashi:2015fsa, Zafrir:2015rga, Hayashi:2015zka, Ohmori:2015tka}. For example, the 6d $Sp(N)$ gauge theory with $2N+8$ hypermultiplets in the fundamental representation coupled to a tensor multiplet may be obtained as a worldvolume theory on $2N$ fractional D6-branes between an $O8^-$-plane and an NS5-brane \cite{Hanany:1997gh, Brunner:1997gf}. T-duality of the type IIA brane system yields a 5-brane web realizing the 5d $SU(N+2)$ gauge theory with $2N+8$  hypermultiplets in the fundamental representation and the vanishing Chern-Simons (CS) level \cite{Hayashi:2015fsa}. The 5-brane web is obtained by resolving two $O7^-$-planes which are T-dual to one $O8^-$-plane. In fact, resolving one $O7^-$-plane also yields another 5d gauge theory, namely the 5d $Sp(N+1)$ theory with $4N+16$ half-hypermultiplets \cite{Hayashi:2015zka}, This shows the 5d $SU$-$Sp$ duality proposed in \cite{Gaiotto:2015una}. The generalization by including more NS5-branes or considering 6d Higgsing lead to further new 5d gauge theories with a 6d UV completion \cite{Zafrir:2015rga, Hayashi:2015zka, Ohmori:2015tka} and also new 5d dualities \cite{Hayashi:2015zka}.

In this work, we continue to expand our analysis by including a so-called $ON^0$-plane in the type IIA brane setup. An $ON^0$-plane in type IIB string theory is an object which is S-dual to a combination of an $O5^-$-plane and a D5-brane \cite{Kutasov:1995te, Sen:1996na}. It also has a perturbative description which enables us to study its property \cite{Sen:1998rg, Sen:1998ii, Kapustin:1998fa, Hanany:1999sj}. T-duality implies that the $ON^0$-plane also exists in type IIA string theory. It has been known that the inclusion of the $ON^0$-plane realizes a further large class of 6d SCFTs \cite{Hanany:1999sj}. The main aim of this paper is to take T-duality to the type IIA brane system with an $ON^0$-plane as well as other orientifolds and obtain further new 5d gauge theories on new 5-brane webs. Indeed, we will find that, in some cases with an $ON^0$-plane, T-duality yields new 5d gauge theories. In the previous cases without an $ON^0$-plane, T-duality gave 5d quiver theories with mainly $SU$ gauge groups. In the current cases, the T-duality yields quiver gauge theories with $Sp$ and $SO$ gauge groups, and widens the landscape of 5d gauge theories with different types of gauge groups. 

It is possible to introduce different types of orientifolds such as an $ON^0$-plane, an $O8^-$-plane and $O6$-planes simultaneously in the type IIA brane setup \cite{Hanany:1999sj}. An $ON^0$-plane is located at the intersection between the $O8^-$-plane and the $O6$-plane. T-duality may give two $O7^-$-planes located at the intersections of an $ON^0$-plane and two $O5^0$-planes each of which we define as a combination of an $O5^-$-plane and a D5-brane. Originally, it has been known that an isolated $O7^-$-plane consists of two 7-branes and can be resolved into a $[1, -1]$ 7-brane and a $[1, 1]$ 7-brane by the quantum effects with respect to the string coupling \cite{Sen:1996vd}. When an NS5-brane is attached on an $O7^-$-plane, another resolution into a $[0, -1]$ 7-brane and a $[2, 1]$ 7-brane was proposed \cite{Zafrir:2015rga, Hayashi:2015zka}. In this paper, we will propose further novel resolution of an $O7^-$-plane with four D7-branes at the intersection between an $ON^0$-plane and an $O5^0$-plane and also between an $ON^-$-plane, which we define as an object S-dual to an $O5^-$-plane, and an $O5^-$-plane. The resolution of the $O7^-$-plane in the presence of the different orientifolds give rise to different five-dimensional gauge theories. Namely, we will have three dual five-dimensional gauge theories depending on either we decompose no, one or two $O7^-$-plane. The difference will yield a new five-dimensional duality between a D-type $SU$ quiver theory and an $SO$-$Sp$ quiver theory.

We also consider yet another generalization by changing the geometry of the compactification. In the analysis so far, we have focused on a simple circle compactification, namely a 6d SCFT on $S^1 \times \mathbb{R}^{4,1}$. However, it is also possible to consider a circle compactification incorporated by a symmetry of a type IIA brane setup. In particular, we will consider a twisted circle compactification where the identification is done with a reflection. In other words, we put a 6d SCFT on $M \times \mathbb{R}^{3,1}$ where $M$ stands for a M\"obius strip with an infinite width. The geometry has a $S^1$ boundary with a cross-cap, and T-duality along the $S^1$ yields a pair of an $O7^-$-plane and an $O7^+$-plane \cite{Keurentjes:2000bs}. The resolution of the $O7^-$-plane yields new 5d gauge theories again.

The organization of the paper is as follows. In section \ref{sec:ONO5}, we first describe properties of an $ON^0$-plane in a 5-brane web in type IIB string theory. We propose a microscopic picture of an $ON^0$-plane incorporated in a 5-brane web and also conjecture a 5d duality associated to it. In section \ref{sec:O6}, we start with 6d SCFTs 
on M5-branes probing a D-type singularity. The system can be described by  O6-planes and D6-branes with multiplet half NS5-branes on them. Although its 5d gauge theory description has been known, we take a different route to obtain the same 5d theory on 5-branes webs involving two $ON^0$-planes, which make clear the role of $ON^0$-planes. 
In section \ref{sec:ON0}, we explore the simplest type IIA brane setup with an $ON^0$-plane. Namely we introduce an $ON^0$-plane at the end of 
 D6-branes split by NS5-branes.
In section \ref{sec:ON0O8O6}, we combine the system analyzed in section \ref{sec:O6} and section \ref{sec:ON0}. Namely, we consider the type IIA brane setup with $O6$-planes and an $ON^0$-plane. The presence of the two types of the orientifolds in fact implies the existence of an $O8^-$-plane. The circle compactification yields three types of 5d gauge theories, depending on either we resolve no $O7^-$-planes, we resolve two $O7^-$-planes or we resolve only one of the two $O7^-$-planes. Finally in section \ref{sec:O7pm}, we consider a twisted circle compactification of the type IIA brane setup with D6-branes and NS5-branes. 

\bigskip

\section{\texorpdfstring{\boldmath 5-brane webs with an $ON^0$-plane}{ON0}}
\label{sec:ONO5}
In this section, we first describe various properties of an $ON^0$-plane incorporated in 5-brane web diagrams in type IIB string theory. 

\subsection{A microscopic description of an $ON^0$-plane in a 5-brane web}
An $ON^0$-plane was originally formulated as an object which is S-dual to a D5-brane on top of an $O5^-$-plane and it arises as an orbifold fixed point in string compactifications \cite{Kutasov:1995te, Sen:1996na}. The $ON^0$-plane was further made use of as a building block for brane configurations in type II string theory \cite{Sen:1998rg, Sen:1998ii, Kapustin:1998fa, Hanany:1999sj}. We here particularly focus on the behavior of the $ON^0$-plane in a 5-brane web diagram in type IIB string theory and propose a microscopic description of the $ON^0$-plane as a combination of an NS5-brane and an $ON^-$-plane which we define as an object S-dual to an $O5^-$-plane. We will utilize this microscopic description in the later sections.

\begin{figure}[t]
\begin{tabular}{cc}
\begin{minipage}{0.5\hsize}
\begin{center}
\includegraphics[width=6cm]{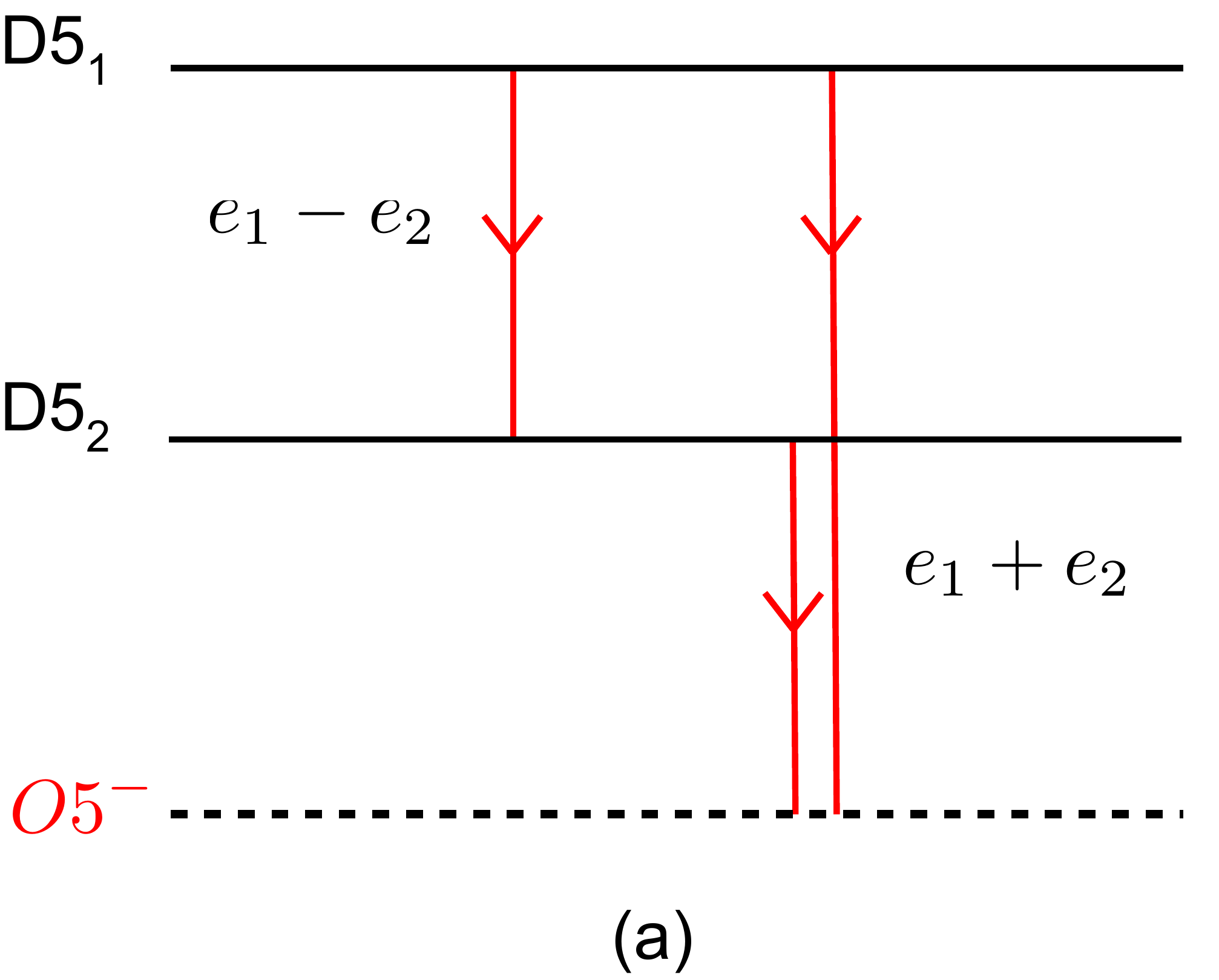}
\end{center}
\end{minipage}
\begin{minipage}{0.5\hsize}
\begin{center}
\includegraphics[width=4cm]{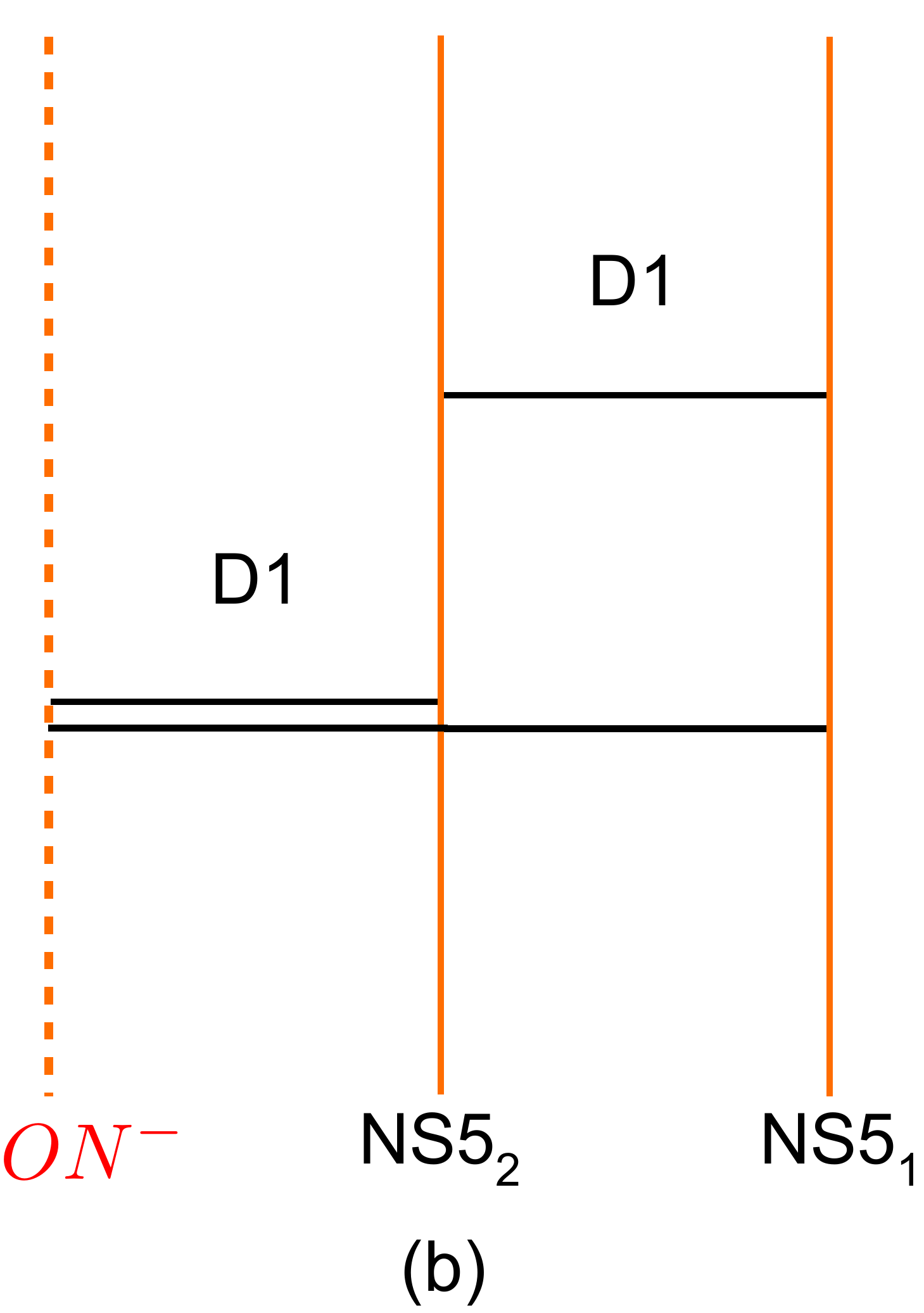}
\end{center}
\end{minipage}
\end{tabular}
\caption{(a): The brane configuration of two D5-branes and an $O5^-$-plane on a Coulomb branch of the $SO(4)$ gauge theory. The red lines denote the fundamental strings which yield gauge fields in the $SO(4)$ gauge theory. (b): The S-dual diagram to the figure (a). The orange lines denote NS5-branes. The D1-branes between the $ON^-$-plane and the NS5$_2$ looks separated pictorially but they are coincident. }
\label{fig:O5mONm}
\end{figure}

In order to understand the microscopic description of an $ON^0$-plane in a 5-brane web, let us first start from the behavior of fundamental strings between two D5-branes on top of an $O5^-$-plane. The two D5-branes on top an $O5^-$-plane give an $SO(4)$ gauge theory and the quantization of fundamental strings between the two D5-branes should give gauge fields in the adjoint representation of the $SO(4)$ gauge group. Let us denote the two D5-brane D5$_1$by D5$_1$ and D5$_2$. Fundamental strings which connect D5$_1$ directly to D5$_2$ yield gauge fields in roots $\pm(e_1-e_2)$ where $e_1$ and $e_2$ are the orthonormal basis in $\mathbb{R}^2$. There are also another type of fundamental strings which originate from D5$_1$, pass through D5$_2$ and also the $O5^-$-plane, and then end on the mirror D5$_2$ or vice versa. The fundamental strings yield gauge fields in roots $\pm(e_1+e_2)$. On the other hand, fundamental strings which connect D5$_1$ to the mirror D5$_2$ or fundamental strings which connect D5$_2$ to the mirror D5$_2$ are projected out by the orientifold action. Those fundamental strings supply all the roots of the $SO(4)$ Lie algebra. The configuration of the fundamental strings is drawn in Figure \ref{fig:O5mONm} (a).
The S-dual picture gives rise to D1-strings which connect two NS5-branes and an $ON^-$-plane but the configuration of the connections should be essentially the same as how the fundamental strings connect between the two D5-branes and the $O5^-$-plane. See Figure \ref{fig:O5mONm} (b).

Performing T-duality in directions transverse to D1-strings but along the NS5-branes, one finally obtains a 5-brane web with an $ON^-$-plane. 
\begin{figure}[t]
\begin{center}
\includegraphics[width=12cm]{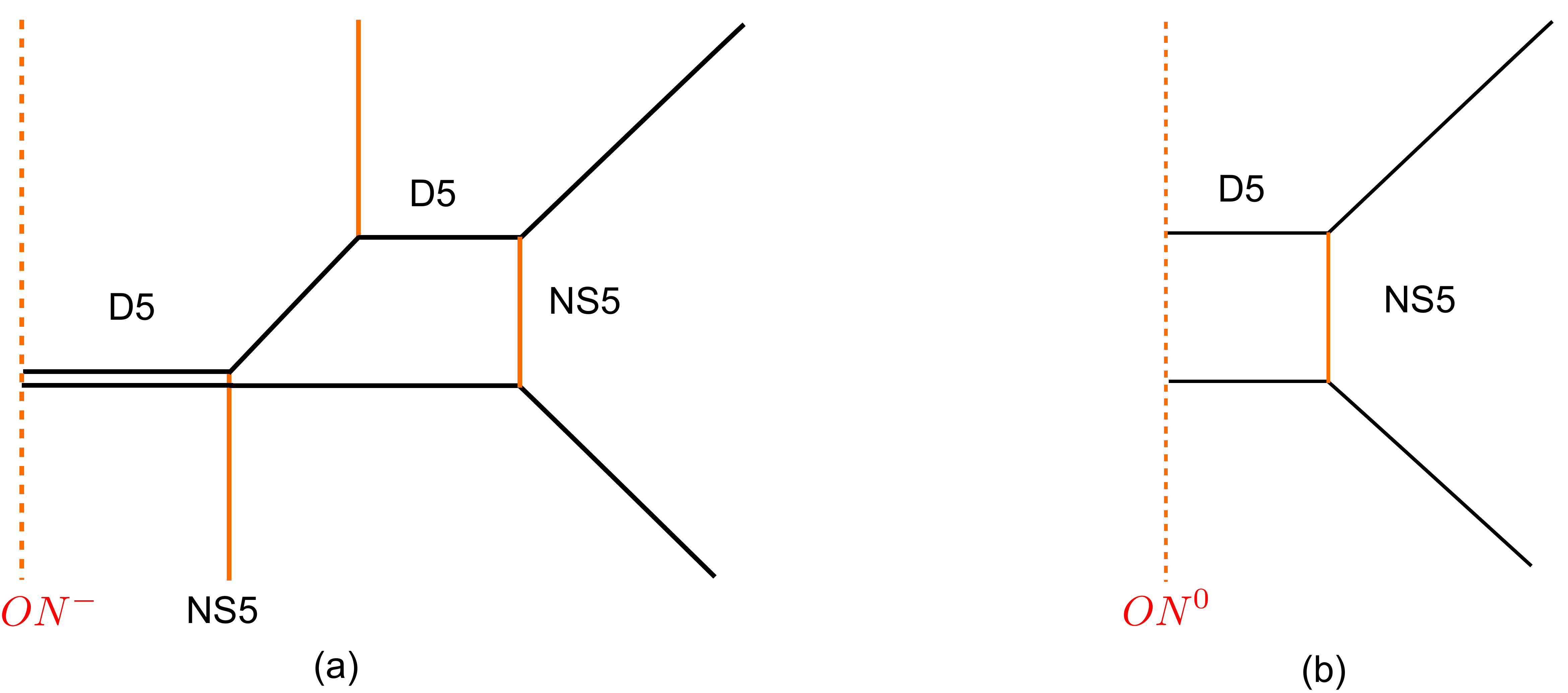}
\end{center}
\caption{(a): The proposal of the microscopic description of an $ON^0$-plane. (b): A shorthand picture of the figure (a). We often make use of this picture in the later sections for simplicity but it always means figure (a). }
\label{fig:microON0}
\end{figure}
We propose that the brane configuration in Figure \ref{fig:microON0} (a) is a microscopic description of an $ON^0$-plane in the system of 5-brane webs. The two coincident D5-branes separate the $ON^-$-plane into two pieces. Indeed, we have a pair of an $ON^-$-plane and an NS5-brane in the left part in Figure \ref{fig:microON0} (a) which may form an $ON^0$-plane after a suitable tuning. A shorthand way to write the brane configuration is depicted in Figure \ref{fig:microON0} (b). In the later sections, we will often make use of the picture in Figure \ref{fig:microON0} (b). Whenever we write Figure \ref{fig:microON0} (b), we always mean that the precise configuration is the one in Figure \ref{fig:microON0} (a). 

\begin{figure}[t]
\begin{center}
\includegraphics[width=6cm]{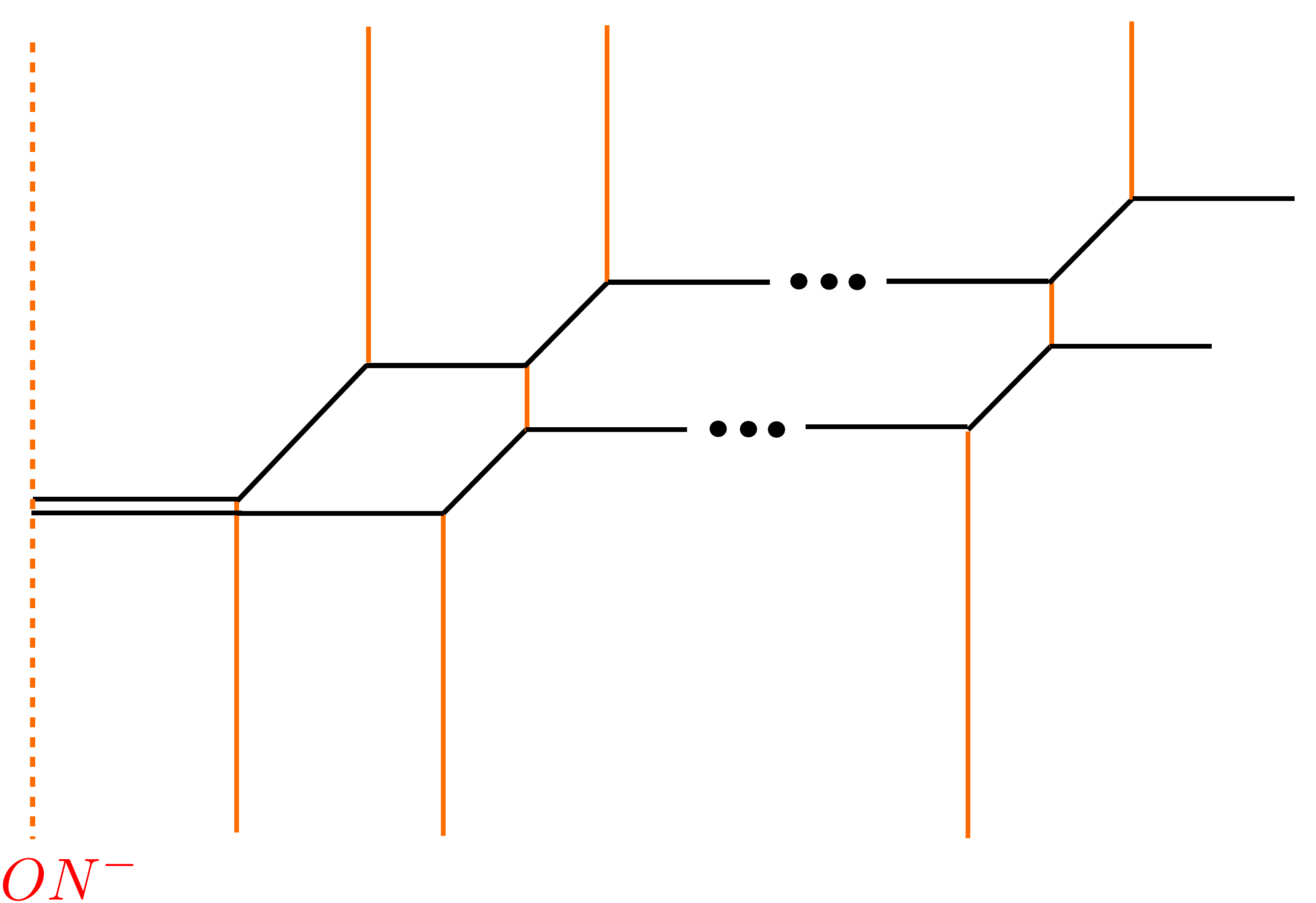}
\end{center}
\caption{A microscopic description of an $ON^0$-plane with multiple NS5-branes. We also add two semi-infinite D5-branes at the right end for later use. We will call this configuration split D5-branes on an $ON^0$-plane. 
}
\label{fig:ONmNS5}
\end{figure}

A simple generalization by adding multiple NS5-branes to Figure \ref{fig:microON0} is straightforward and given in Figure \ref{fig:ONmNS5} (a).  We also put two semi-infinite D5-branes at the right end of the brane configuration of Figure \ref{fig:ONmNS5} (a) since we will often encounter this case.

Now, we discuss the worldvolume theory realized on the D5-branes in this brane configuration.
As discussed in \cite{Hanany:1999sj}, $N$ D5-branes suspended between an NS5-brane and an $ON^0$-plane give $SU(N_1) \times SU(N_2)$ gauge group factor with $N_1+N_2=N$. 
Here, $N_1$ is the number of D5-branes which connect the NS5-brane adjacent to the $ON^0$-plane to the NS5-brane in the $ON^0$-plane from the right, while $N_2$ is the number of D5-branes which end on the NS5-brane next to the $ON^0$-plane, pass through the NS5-brane in the $ON^0$-plane, is reflected by the $ON^-$-plane inside the $ON^0$-plane and end on the NS5-brane in the $ON^0$-plane from the left. Both of the gauge group have bi-fundamental matter, which couples to the other adjacent gauge node instead of each other. In our case, $N_1=N_2=1$ and thus, we can formally write the 5d quiver gauge theory from this brane configuration with $N-1$ NS5-branes is
\begin{align}
\label{5dsu1}
5d~~
{\text{``$SU(1)$"}}-\underbrace{{\overset{\overset{\text{\normalsize``$SU(1)$''}}{~~~\textstyle\vert}}{SU(2)}}-
SU(2) - \cdots-SU(2)}_{N-2} - [2].
\end{align}
where we have $N-2$ $SU(2)$ gauge nodes and we denoted the $n$ hypermultiplets in the fundamental representation as $[n]$.
In \cite{Bergman:2014kza, Hayashi:2014hfa, Hayashi:2015fsa}, it is discussed that the ``$SU(1)$'' gauge node together with a bi-fundamental hypermultiplet between the ``$SU(1)$''' and the $SU(2)$ give two flavors for the $SU(2)$. Using this claim also to our  
case, we find that we can interpret that the above quiver gauge theory is interpreted as
\begin{equation}\label{SU2quiver}
5d~~[4] - \underbrace{SU(2) - \cdots - SU(2)}_{N-2} - [2].
\end{equation}

\subsection{A transition between $O5^{-}-O5^+$-planes and an $ON^0$-plane}

\begin{figure}[t]
\begin{center}
\includegraphics[width=13cm]{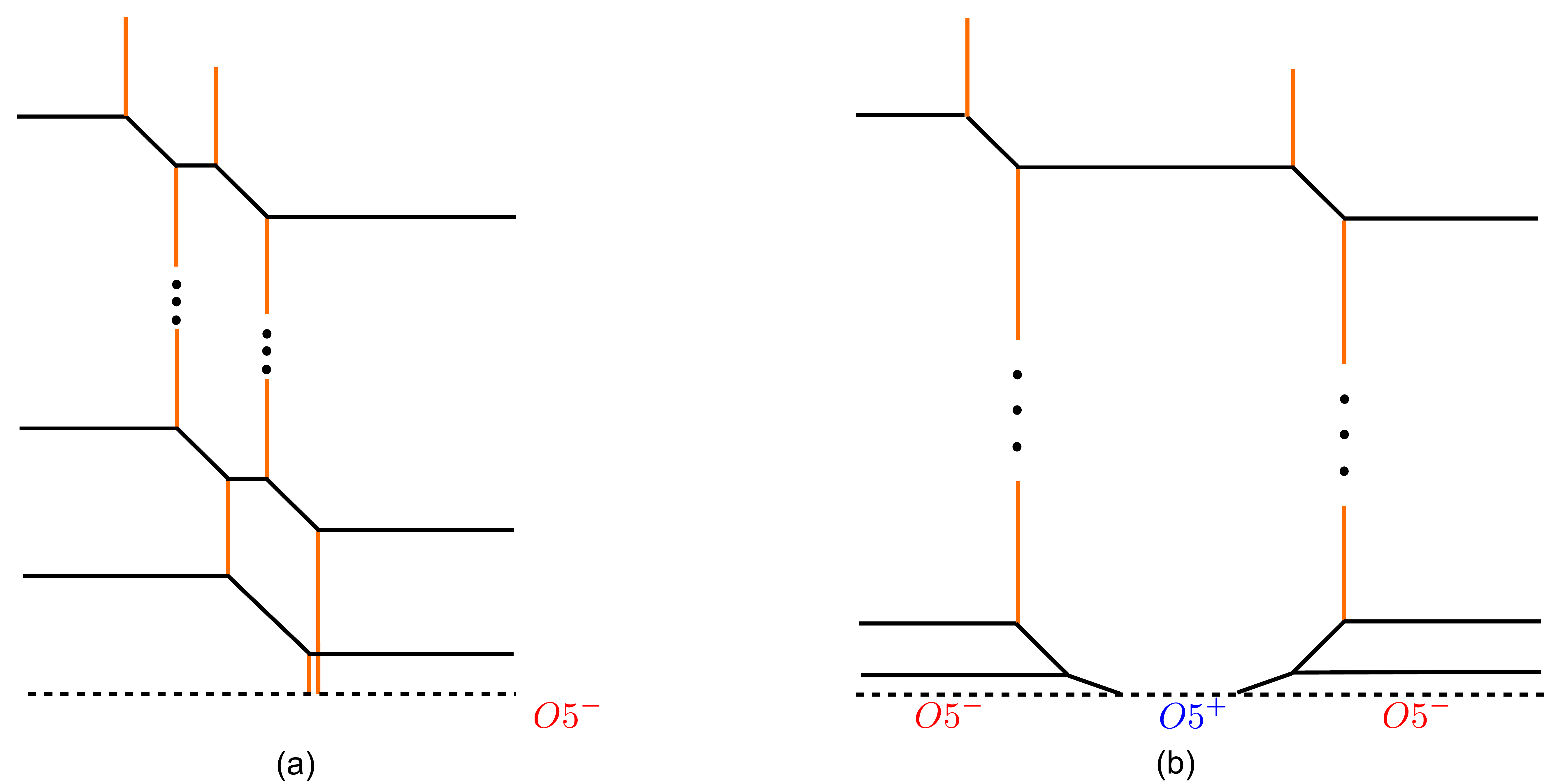}
\end{center}
\caption{(a): The S-dual to the Figure \ref{fig:ONmNS5}. (b): Another alternative brane configuration including an $O5^+$-plane.}
\label{fig:5ddualbase}
\end{figure}

Let us then consider the S-dual of the brane configuration in Figure \ref{fig:ONmNS5}.  The S-duality amounts to the $90^{\circ}$ rotation and one obtains a 5-brane web in Figure \ref{fig:5ddualbase} (a).
We conjecture here that the brane conifguration in Figure \ref{fig:5ddualbase} (a) is connected to the brane configuration in Figure \ref{fig:5ddualbase} (b) by a deformation of parameters of the worldvolume theory realized on the 5-branes. That is, we can deform the brane configuration from (a) to (b) in Figure \ref{fig:5ddualbase} without changing the UV fixed point of the theory. Hence we claim that the 5d theory realized on the 5-brane web in Figure \ref{fig:5ddualbase} (a) is dual to the 5d theory realized on the 5-brane web in Figure \ref{fig:5ddualbase} (b).

\begin{figure}[t]
\centering
\includegraphics[width=12cm]{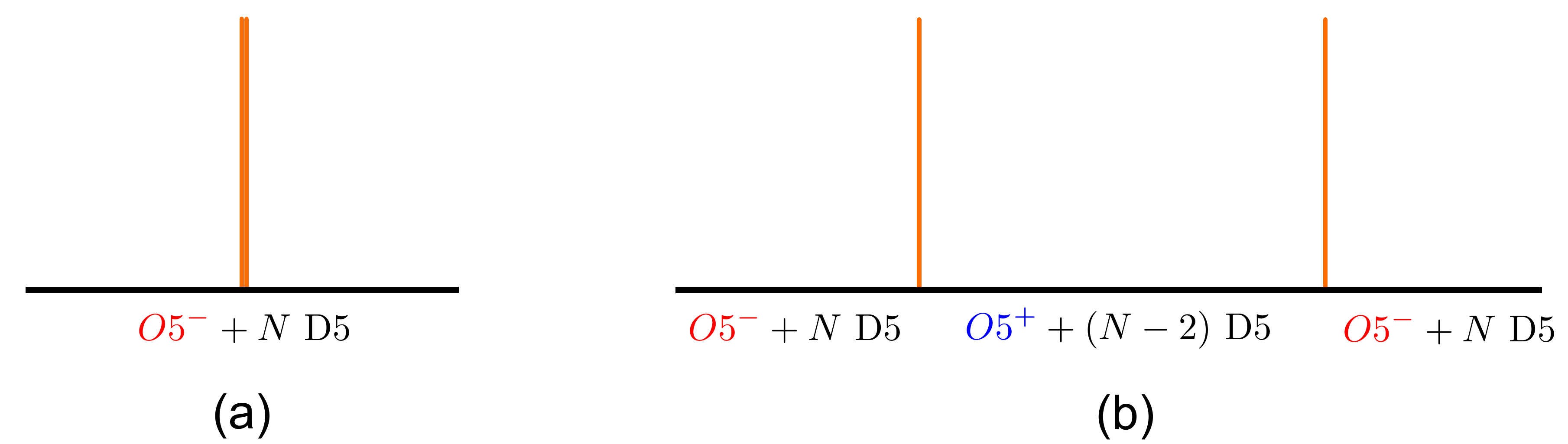}
\caption{
(a): The brane configuration with an unsplit NS5-brane attached to the $O5^-$-plane. This configuration is obtained by moving $N$ D5-branes to the place of $O5^-$ plane from Figure \ref{fig:5ddualbase} (a).
(b): An NS5-brane split on $O5^-$ plane into two fractional ones. An $O5^+$ plane are created between the two fractional NS5-branes. This configuration is obtained analogously from Figure \ref{fig:5ddualbase} (b).
}
\label{fig:equivtransition}
\end{figure}

The transition may look more natural when one moves all the D5-branes to the place of the $O5$-plane.
Then, (a) and (b) in Figure \ref{fig:5ddualbase} reduce to (a) and (b) in Figure \ref{fig:equivtransition}, respectively.
In Figure \ref{fig:equivtransition} (a), we see that an ``unsplit'' NS5-brane is attached to the $O5^-$-plane.
In fact, an``unsplit'' NS5-brane can be split into two fractional NS5-branes on the $O5^-$-plane without any cost of energy in certain cases. 
After the splitting, an $O5^+$-plane should be created since the NS5-brane changes the sign of the RR-charge of the $O5^-$-plane. 
In addition, the splitting can create certain number of D5-branes between the two fractional NS5-branes. 
It is argued in \cite{Bertoldi:2002nn}, the split configuration has the same energy as the 
unsplit confguration if the number of created D5-branes in between the two fractional NS5-branes is less than or equal to $N-2$. 
If we consider the case where the bound of the number of created D5-branes is saturated, we obtain Figure \ref{fig:equivtransition} (b).
Essentially, our conjecture is to claim that we obtain Figure \ref{fig:5ddualbase} (a) if we move all the D5-branes away from $O5^-$ plane in Figure \ref{fig:equivtransition} (a).

Another support for the conjecture comes from checking the duality of the realized 5d theories on the two configurations.
The Figure \ref{fig:5ddualbase} (b) gives a 5d $Sp(N-2)$ gauge theory with $N_f =2N$ flavors. On the other hand, the S-dual diagram to Figure \ref{fig:5ddualbase} (a), which is nothing but the brane web in Figure \ref{fig:ONmNS5}, gives the quiver gauge theory in \eqref{SU2quiver}. The latter theory is S-dual to a 5d $SU(N-1)$ gauge theory with $N_f=2N$ flavors and the Chern-Simons level $\pm 1$.
 In fact, the $Sp(N-2)$ gauge theory with $N_f = 2N$ flavors is dual to the $SU(N-1)_{\pm 1}$ gauge theory with $N_f = 2N$ flavors \cite{Gaiotto:2015una, Hayashi:2015zka},
\begin{align}
5d~~~SU(N-1)_{\pm 1} -[2N] \quad \longleftrightarrow \quad Sp(N-2) -[2N]  .
\end{align}
Therefore, the two brane configurations in Figure \ref{fig:5ddualbase} indeed give two dual 5d theories, which implies that the two brane webs in Figure \ref{fig:5ddualbase} are equivalent to each other. 

The above arguments strongly support the equivalence of the two 5-brane webs in Figure \ref{fig:5ddualbase}. By combinining the S-duality, the 5-brane web in Figure \ref{fig:5ddualbase} (b) is further dual to the 5-brane web in Figure \ref{fig:ONmNS5}.  Hence, if there is a  brane configuration of Figure \ref{fig:5ddualbase} (b) or  Figure \ref{fig:ONmNS5} in a local part of a 5-brane web, we can replace it with the one of Figure \ref{fig:ONmNS5} or the one of Figure \ref{fig:5ddualbase} (b) respectively since they are equivalent to each other.  More schematically, the S-duality relates the following two configurations
\begin{equation}
 (O5^{-}+ 2 {\rm D5}) - O5^+  \quad \longleftrightarrow \quad (ON^0+ {\rm NS5}) . \label{equivalence}
\end{equation}
Although we will often use the duality \eqref{equivalence} in later sections to modify 5-brane webs in a schematic way, the precise replacement is given by the the transition between Figure \ref{fig:ONmNS5} and Figure \ref{fig:5ddualbase} (b).

\subsection{Duality between a D-type $SU$ quiver and an $SO-Sp$ linear quiver}

\begin{figure}[t]
\begin{center}
\includegraphics[width=15cm]{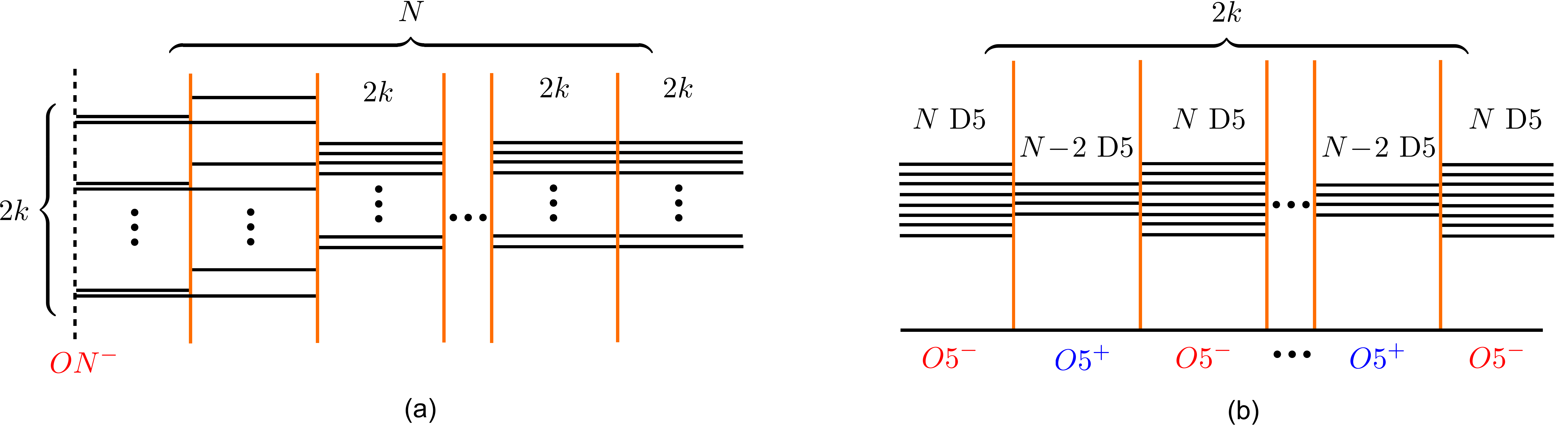}
\end{center}
\caption{(a): A 5-brane web configuration with an $ON^0$-plane and $N-1$ NS5-branes. (b): An alternative 5-brane web configuration with $O5^{\pm}$-planes.}
\label{fig:general5dduality}
\end{figure}

In order to demonstrate one of interesting implications of this conjecture, we consider a brane configuration depicted in Figure \ref{fig:general5dduality} (a).%
\footnote{Here and in the following sections, we often draw all the $(p,1)$ 5-branes by vertical lines for simplicity.}
This is interpreted as the following 5d quiver gauge theory,
\begin{align}
\label{5dDtypequiver}
5d~~
{SU(k)}-\underbrace{{\overset{\overset{\text{\normalsize$SU(k)$}}{~~~\textstyle\vert}}{SU(2k)}}-
SU(2k) - \cdots-SU(2k)}_{N-2} - [2k].
\end{align}
The case of $k=1$ reduces to \eqref{SU2quiver}. When we take S-dual of the brane configuration in Figure \ref{fig:general5dduality} (a) and locally use the deformation from (a) to (b) in Figure \ref{fig:5ddualbase}, we obtain Figure \ref{fig:general5dduality} (b). This brane configuration corresponds to a 5d quiver gauge theory
\begin{align}
\label{5dSOSpquiver}
5d~~
[N]-\underbrace{Sp(N-2) - SO(2N)-Sp(N-2) - SO(2N)-\cdots - Sp(N-2)}_{2k-1} - [N],
\end{align}
where we have $k$ $Sp(N-2)$ gauge nodes and $k-1$ $SO(2N)$ gauge nodes.
Therefore, we can expect that these two theories are dual to each other, which means, they have identical UV fixed point.

We can support the conjecture by comparing the dimension of the Coulomb branch moduli space and the number of parameters of both theories. The dimension of the Coulomb branch moduli space of the D-type quiver theory \eqref{5dDtypequiver} is
\begin{equation}\label{Coulomb5dDtypequiver}
\underbrace{(k-1)\times 2}_{2\;SU(k)'s} + \underbrace{(2k-1) \times (N-2)}_{N-2\;SU(2k)'s} = 2Nk - N - 2k.
\end{equation}
The dimension of the Coulomb branch moduli space of the $SO$-$Sp$ quiver theory \eqref{5dSOSpquiver} is
\begin{equation}
\underbrace{(N-2)\times k}_{k\;Sp(N-2)'s} + \underbrace{N \times (k-1)}_{k-1\;SO(2N)'s} = 2Nk - N - 2k,
\end{equation}
which precisely agrees with \eqref{Coulomb5dDtypequiver}. The number of the parameters of the D-type quiver theory \eqref{5dDtypequiver} is
\begin{equation}\label{parameter5dDtypequiver}
\underbrace{N}_{2\;SU(k)'s\;\text{and}\;N-2\;SU(2k)'s} + \underbrace{N-1}_{\text{bi-fund.}} + \underbrace{2k}_{2k\;\text{flavors}} = 2N+2k-1.
\end{equation}
The number of the parameters of the $SO$-$Sp$ quiver theory \eqref{5dSOSpquiver} is
\begin{equation}
\underbrace{2k-1}_{k\;Sp(N-2)'s\;\text{and}\;k-1\;SO(2N)'s} + \underbrace{2N}_{\text{two}\; N\; \text{flavors}} = 2N + 2k -1,
\end{equation}
which again exactly agrees with \eqref{parameter5dDtypequiver}.

In this case, one can see more refined support from the instanton operator analysis given in \cite{Tachikawa:2015mha, Yonekura:2015ksa}. All the $SU(N_c)$ gauge nodes\footnote{$N_c$ is either $k$ or $2k$ in this case.} of the 5d D-type quiver theory \eqref{5dDtypequiver} satisfy the condition $N_f = 2N_c$ where $N_f$ is the number of flavors coupled to the $SU(N_c)$ gauge node. Therefore, the one-instanton states create two Dynkin diagrams whose shape is exactly the quiver \cite{Yonekura:2015ksa}. Namely, the one-instanton states give rise to two $D_{N}$ Dynkin diagrams, which implies that the enhanced flavor symmetry is at least $SO(2N) \times SO(2N)$. The D-type quiver theory \eqref{5dDtypequiver} also has a flavor symmetry $SU(2k)$, hence the expected enhanced flavor symmetry at the 5d UV fixed point is $SO(2N) \times SO(2N) \times SU(2k)$. On the other hand, the $SO(2N) \times SO(2N)$ flavor symmetry is perturbatively seen in the 5d $SO$-$Sp$ quiver theory \eqref{5dSOSpquiver}. 
The remaining part of flavor symmetry, $SU(2k)$, is not clear from the one-instanton analysis but we can use 7-branes in the 5-brane web in Figure \ref{fig:general5dduality} (b). In general, one can think of attaching 7-branes to the ends of external 5-branes in a 5-brane web when the charge of the 7-branes is the same as that of the external 5-branes. In Figure \ref{fig:general5dduality}, we have $2k$ external NS5-branes and hence they can end on $2k$ $[0, 1]$ 7-branes. The global symmetry of the 5d theory may be understood from a symmetry realized on 7-branes in the corresponding 5-brane web. Therefore, the presence of the $2k$ $[0, 1]$ 7-branes implies that we have an enhanced flavor symmetry $SU(2k)$ from the instantons of the $SO$-$Sp$ quiver theory \eqref{5dSOSpquiver}. In summary, the 5d $SO$-$Sp$ quiver theory \eqref{5dSOSpquiver} is expected to have an enhanced flavor symmetry $SO(2N) \times SO(2N) \times SU(2k)$ at the 5d UV fixed point. The flavor symmetry precisely agrees with the enhanced flavor symmetry of the 5d D-type quiver theory \eqref{5dDtypequiver}, determined by the instanton operator analysis.

\bigskip

\section{\texorpdfstring{Type IIA brane setup with {\boldmath $O6$}-planes}{O6}
}\label{sec:O6}

\begin{figure}
\centering
\includegraphics[width=9cm]{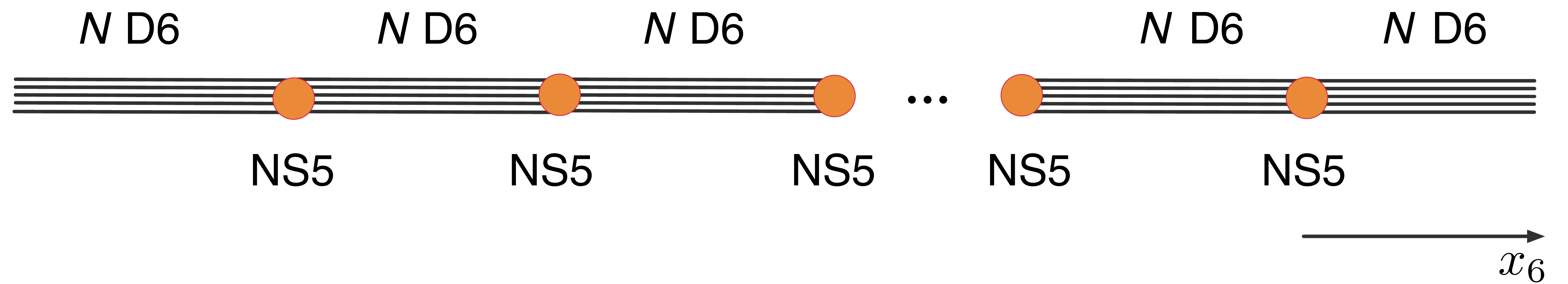}
\caption{The type IIA brane configuration realizing a 6d $SU(N)$ linear quiver theory. Each horizontal line represents an D6-branes and each orange circle represents an NS5-brane. We chose the horizontal direction as the $x_6$ direction.}
\label{fig:base}
\end{figure}

In type IIA string theory, 6d SCFTs are realized on D6-branes fractionated by NS5-branes. The D6-branes extend in the $x_0, x_1, \cdots, x_6$ directions whereas the NS5-branes extend in the $x_0, x_1, \cdots, x_5$ directions, which is summarized in Table \ref{tb:branebase}. 
\begin{table}[t]
\begin{center}
\begin{tabular}{c|cccccccccc}
\hline
\ &0 &1&2&3&4&5&6&7&8&9\\
\hline
D6& x& x& x& x& x& x& x& .& .& .\\
NS5& x& x& x& x& x& x& .& .& .& .\\
\hline
\end{tabular}
\end{center}
\caption{The basic brane setup in type IIA string theory which realizes a 6d linear $SU$ quiver theory.}
\label{tb:branebase}
\end{table}
Hence, the NS5-branes cut the D6-branes into pieces in the $x_6$ direction as in Figure \ref{fig:base}. 
When we consider $N$ D6-branes and $k$ NS5-branes, the resulting 6d SCFT on the tensor branch is 
\begin{equation}
6d~~[N] - SU(N) - \cdots - SU(N) - \cdots - SU(N) - [N] \label{6dAtype}
\end{equation}
where we have $k-1$ $SU(N)$ gauge nodes and the number in the bracket $[\cdot]$ represents the number of hypermultiplets in the fundamental representation of the $SU(N)$. The same 6d SCFT can be also realized on $k$ M5-branes probing an $A_{N-1}$ singularity. M5-branes extend in the $x_0, x_1, \cdots, x_5$ directions, and the $A_{N-1}$  
singularity is a point in the non-compact four-dimensional space spanned by $x_7, x_8, x_9, x_{11}$. This is the base setup for 6d SCFTs in the type IIA construction or M-theory.

It is possible to generalize this configuration by including an $O8^-$-plane \cite{Hanany:1997gh, Brunner:1997gf}. The 5d descriptions of such 6d SCFTs were extensively discussed in \cite{Hayashi:2015fsa, Zafrir:2015rga, Hayashi:2015zka, Ohmori:2015tka} by utilizing T-duality. The main aim of this paper is to consider yet another generalization by including different types of orientifolds and to determine corresponding 5d gauge theory descriptions. 

In this section, we first include $O6^{\pm}$-planes in the same direction as D6-branes in Table \ref{tb:branebase} where we have two semi-infinite $O6^-$-planes on the both sides in the $x_6$ direction. 
The same 6d SCFT can be also realized on $k$ M5-branes probing an $D_N$ singularity. After a circle compactification of the M-theory configuration, we obtain a system with an $O6^-$-plane on top of $N$ D6-branes with $k$ NS5-branes on top of each other. On a generic tensor branch of the theory, the $k$ full NS5-branes split into $2k$ factional NS5-branes due to the presence of the orientifold six-plane. The fractional NS5-branes can change the sign of the RR charge of the orientifold. Namely, when an $O6^-$-plane crosses the fractional NS5-brane, it changes into an $O6^+$-plane. Similarly, when the $O6^+$-plane crosses the fractional NS5-brane, it changes into an $O6^-$-plane. Furthermore, the condition of the cosmological constant of massive type IIA background implies that the number of D6-branes on top of the $O6^+$-plane must be $N-4$. 

\begin{figure}
\centering
\includegraphics[width=9cm]{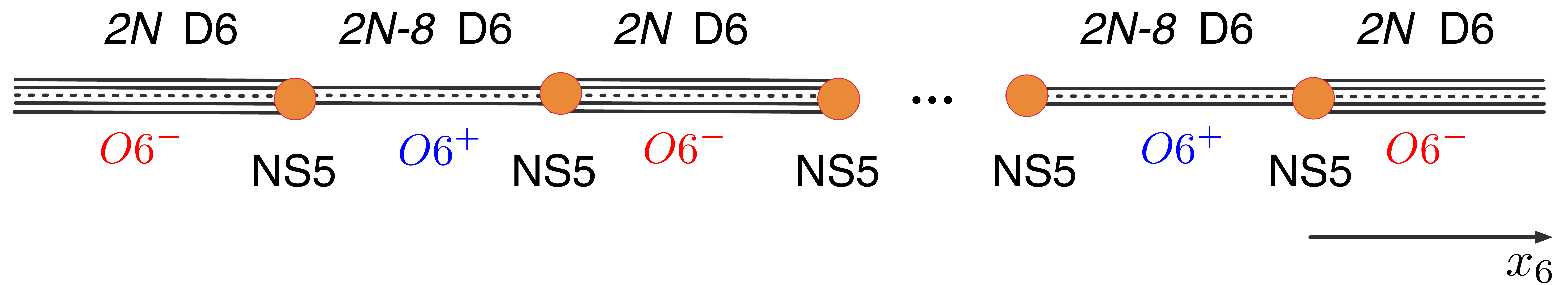}
\caption{The type IIA brane configuration realizing an 6d $SO-Sp$ alternating quiver theory. The horizontal dotted lines denote an $O6^-$-plane or an $O6^+$-plane. Each NS5-brane is a fractional NS5-brane. The number of D6-branes includes the mirror images.}
\label{fig:Dconformal}
\end{figure}

Let us understand this phenomenon in two steps. The original configuration is $k$ full NS5-branes which divides one $O6^-$-plane on $N$ D6-branes into two pieces in the $x_6$ direction. Since the $k$ NS5-branes are put on the $O6^-$-plane, we in fact have $2k$ fractional NS5-branes. It is possible to distribute the $k$ pairs of fractional NS5-branes along the $x_6$ direction. In this case, we only have $O6^-$-planes since the RR charge of the $O6^-$-plane does not change when it crosses two fractional NS5-branes. Finally, we split the pair of fractional NS5-branes into separated fractional NS5-branes on an $O6^-$-plane on top of D6-branes. In this process, some amount of D6-branes is also created between the fractional NS5-branes. The number of the created D6-branes is completely fixed by the condition on the cosmological constant or equivalently the anomaly cancellation condition of the corresponding 6d gauge theory with tensor multiplets. 
As a result, $N-4$ D6-branes on top of the $O6^+$-plane are created between two adjacent factional NS5-branes.
The brane configuration of the system is depicted in Figure \ref{fig:Dconformal}. Note that we have $2k$ factional NS5-branes in this case.

In the end, as a 6d SCFT on the tensor branch, we have a 6d alternating quiver theory 
\begin{equation}
6d~~[SO(2N)] - Sp(N-4) - SO(2N) - Sp(N-4) - \cdots - SO(2N) - Sp(N-4) - [SO(2N)] \label{Dconformal}
\end{equation}
where we have $k$ $Sp(N-4)$ gauge nodes and $k-1$ $SO(2N)$ gauge nodes. The two $[SO(2N)]$ at the two ends of the quiver represent the flavor symmetry. Instead of $O6^\pm$-planes, it is also possible to construct 6d SCFTs by using $\widetilde{O6}^{\pm}$-planes.\footnote{The number of D6-branes put on either $\widetilde{O6}^-$-planes or $\widetilde{O6}^+$-planes is determined by the condition of the cosmological constant. Here it is important to note that the $\widetilde{O6}^-$-plane exits only when the value of the cosmological constant is odd \cite{Hyakutake:2000mr}. For example, when one considers an 
 NS5-brane on $\widetilde{O6}^-$ which splits into two fractional NS5-branes,  the resulting 6d SCFT on the tensor branch is an $SO(2N+1)$ gauge theory with $2N-7$ hypermultiplets in the vector representation. With multiple fractional NS5-branes on D6/$\widetilde{O6}$ system, one exemplary  6d quiver theory with minimal odd cosmological constant is $6d ~[SO(2N+9)]-Sp(N)-SO(2N+7)-Sp(N-1)-[SO(2N+5)]$ in the tensor branch. }

We are interested in a 5d description of the 6d theory \eqref{Dconformal}. In fact, the 5d description has been known already and it can be obtained by considering a different $S^1$ compactification from the original M-theory picture. Namely, when we compactify, for example, the $x_5$ direction on $S^1$, then we can regard this $S^1$ as the M-theory circle. The circle compactification yields $k$ D4-branes sitting at the $D_N$ singularity. The 5d gauge theory description on the D4-branes is an affine D-type quiver theory \cite{Douglas:1996sw}
\begin{equation}
5d~~SU(k)-{\overset{\overset{\text{\normalsize$SU(k)$}}{\textstyle\vert}}{SU(2k)}}-SU(2k)- \cdots -SU(2k) -{\overset{\overset{\text{\normalsize$SU(k)$}}{\textstyle\vert}}{SU(2k)}} - SU(k), \label{5dDtype}
\end{equation}
where we have four $SU(k)$ gauge nodes and $N-3$ $SU(2k)$ gauge nodes.

In the remaining part of this section, 
we take a different route to arrive at the same 5d description \eqref{5dDtype}. This way in fact makes use of an $ON^0$ orientifold in \cite{Sen:1998rg, Sen:1998ii, Kapustin:1998fa, Hanany:1999sj}. This example also nicely illustrates a gauge theory description from a brane configuration involving the $ON^0$-plane, which we will use in the later sections.

\begin{figure}
\centering
\includegraphics[width=9cm]{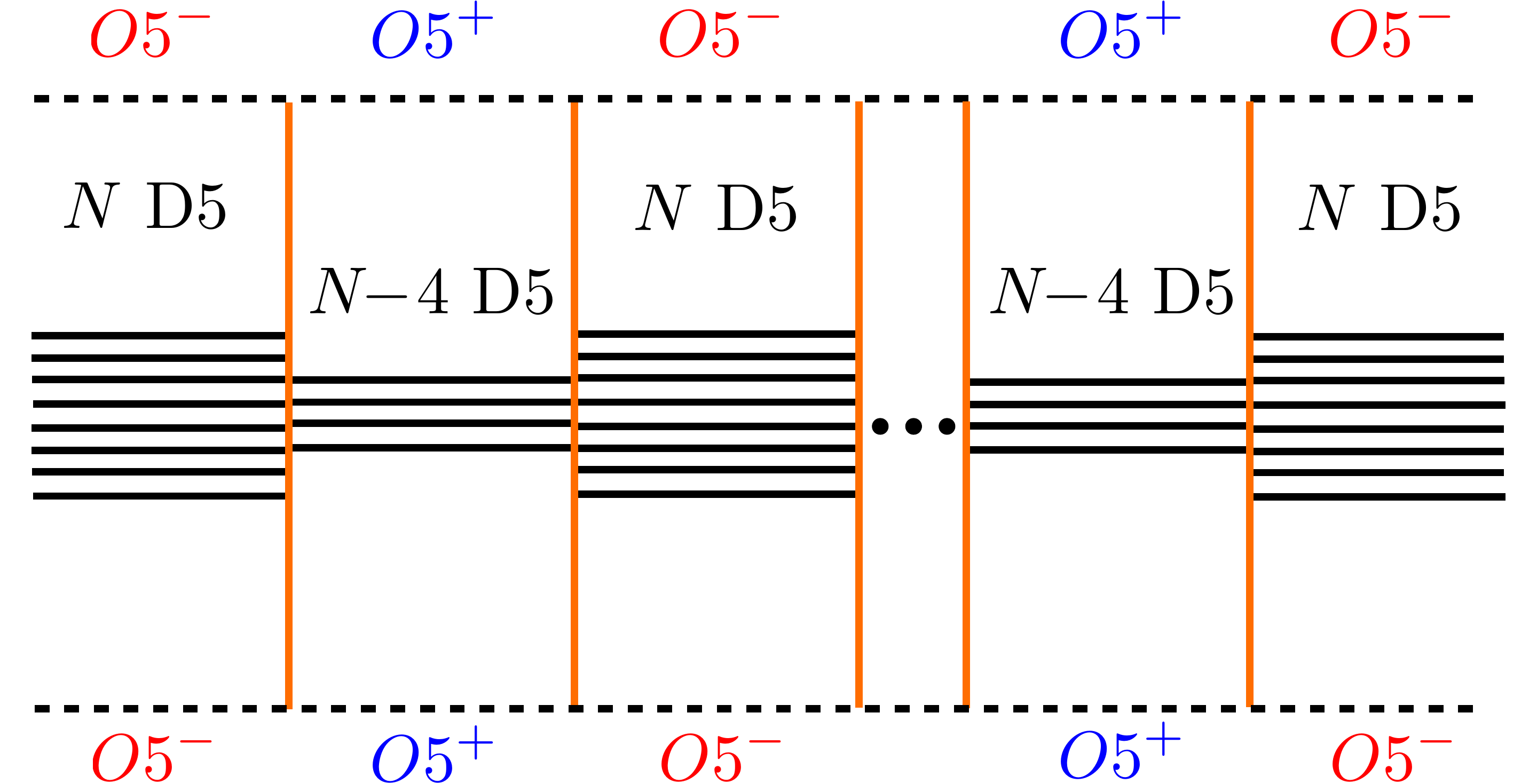}
\caption{The 5-brane web diagram which is T-dual to the circle compactification of the type IIA brane setup in Figure \ref{fig:Dconformal}. The number of the fractional NS5-branes is $2k$. The vertical direction is the $x_5$ direction and the horizontal direction is the $x_6$ direction}
\label{fig:O5mO5p}
\end{figure}

The starting point is the brane configuration in Figure \ref{fig:Dconformal} realizing the 6d theory of \eqref{Dconformal}. We compactify the $x_5$ direction on $S^1$ and then perform T-duality along the $S^1$. Then, we obtain a 5-brane diagram given in Figure \ref{fig:O5mO5p}. As for 5-brane web diagrams, we always write down the $(x_6, x_5)$ two-dimensional plane where the vertical direction is the $x_5$ direction and the horizontal direction is the $x_6$ direction.
Note that each $O6^-$-plane becomes two $O5^-$-planes, and each $O6^+$-plane becomes two $O5^+$-planes. In between the upper and lower $O5^-$-planes, we have $N$ D5-branes. In between the upper and lower $O5^+$-planes, we have $N-4$ D5-branes.

\begin{figure}
\centering
\includegraphics[width=6cm]{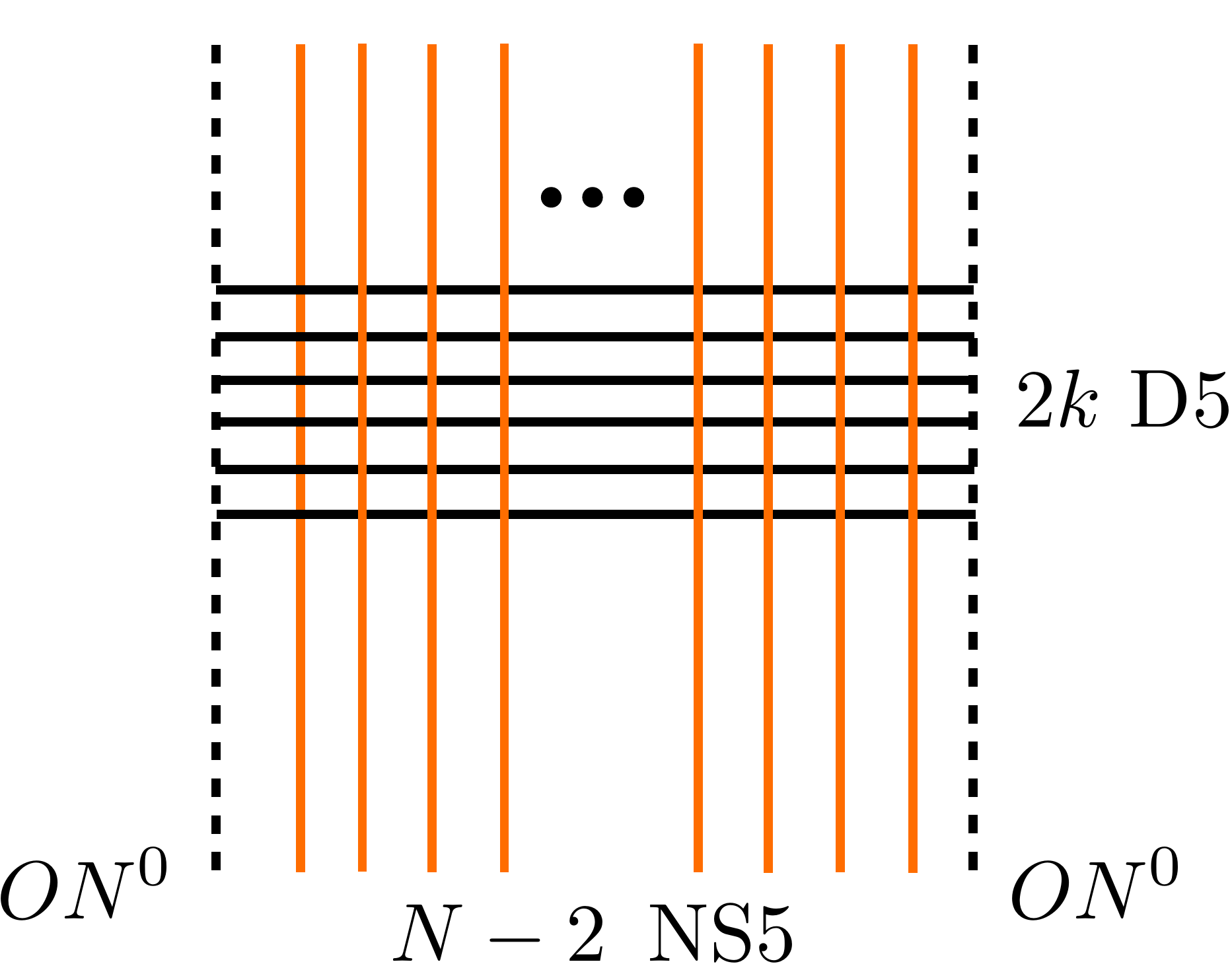}
\caption{The 5-brane web diagram which is obtained from the one in Figure \ref{fig:O5mO5p} after the transition (\ref{equivalence}).}
\label{fig:Sdual1}
\end{figure}
Here, we take S-duality taking into account the transition discussed at (\ref{equivalence}).
We then arrive at a configuration where we have two $ON^0$-planes at the both ends, $k$ pairs of fractional D5-branes in the horizontal direction and also $N-2$ NS5-branes in the vertical direction as in Figure \ref{fig:Sdual1}.
The 5d gauge theory description on the 5-brane configuration can be read off from the rules in \cite{Hanany:1999sj}, and it is exactly the same as \eqref{5dDtype}. For example, it is easy to see that we have $N-3$ $SU(2k)$ gauge nodes from $2k$ D5-branes between $N-2$ NS5-branes. The $2k$ fractional D5-branes between an $ON^0$-plane and an NS5-brane at the two ends give rise to the two-pronged $SU(k)$ gauge nodes at the two ends of the quiver. Hence, the (S- or/and) T-duality after the circle compactification of the original type IIA brane configuration given in Figure \ref{fig:Dconformal} gives a different way to show the 5d description \eqref{5dDtype} from the 6d $(D_N, D_N)$ conformal matter theory on $S^1$.

\bigskip

\section{\texorpdfstring{\boldmath Type IIA brane setup with an $ON^0$-plane}{ON0}
}\label{sec:ON0}

In the previous section, we included $O6$-planes in the D6--NS5-brane system and we found the 5d description of the 6d D-type conformal matter theory on $S^1$. 
The brane configuration in type IIB string theory with $ON^0$-orientifold, which yielded the 5d affine D-type quiver gauge theory, includes 
$ON^0$ orientifolds. The $ON^0$-plane appeared as an object which is S-dual to a combination $O5^- $-plane $+$ a full D5-brane in type IIB string theory. 

The existence of the $ON^0$-plane is not restricted in type IIB string theory. In fact, T-duality implies that an $ON^0$-plane also exists in type IIA string theory. 
Here we consider yet another case where we include an $ON^0$-plane in the brane configuration in type IIA string theory. Namely, we consider a type IIA brane configuration with an $ON^0$ orientifold in addition to D6-branes and NS5-branes \cite{Hanany:1999sj}. The $ON^0$-plane extends in the same direction as those of NS5-branes in Table \ref{tb:branebase} but is located at a different position in the $x_6$ direction.

\begin{figure}
\centering
\includegraphics[width=10cm]{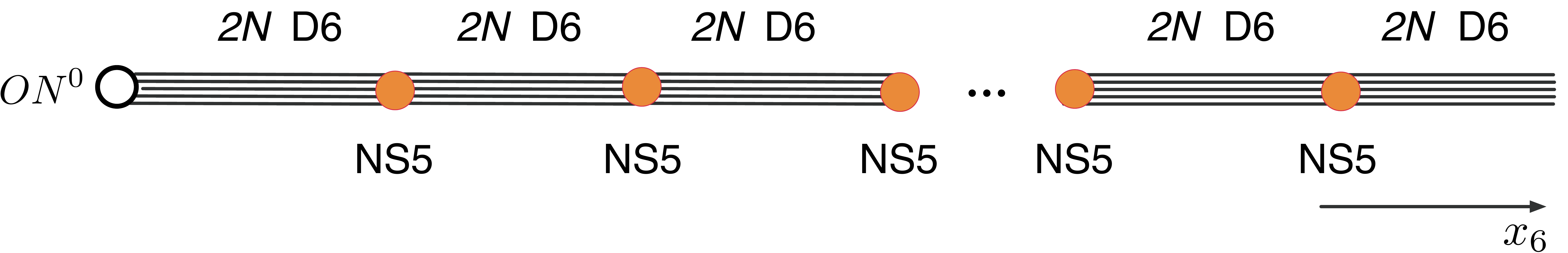}
\caption{Type IIA configuration with an $ON^0$-brane which is denoted by an empty circle. The presence of the $ON^0$-brane gives rise to a 6d $D$-type quiver gauge theory. $2N$ is the number of D6-branes and $k$ is the number of NS5-branes.} 
\label{fig:ON0}
\end{figure}

Let us focus on the brane configurations which do not involve other kinds of orientifolds. The IIA brane configuration with an $ON^0$-brane is given in Figure \ref{fig:ON0}.
We have $k$ NS5-branes, and also we have $2N$ D6-branes in between two NS5-branes, or in between one $ON^0$-plane and one NS5-brane. It is known that the presence of an $ON^0$-brane gives rise to a $D$-type quiver theory \cite{Hanany:1999sj} on a tensor branch of the corresponding 6d SCFT. The anomaly free 6d quiver theory is 
\begin{align}
6d~~SU(N)-{\overset{\overset{\text{\normalsize$SU(N)$}}{\textstyle\vert}}{SU(2N)}}-SU(2N)- \cdots -SU(2N)-[2N]. \label{6dDtype}
\end{align}
We have $2$ $SU(N)$ gauge nodes and $k-1$ $SU(2N)$ gauge nodes. The global symmetry of the 6d theory is $SU(2N)$. 

\begin{figure}[t]
\begin{center}
\includegraphics[width=7cm]{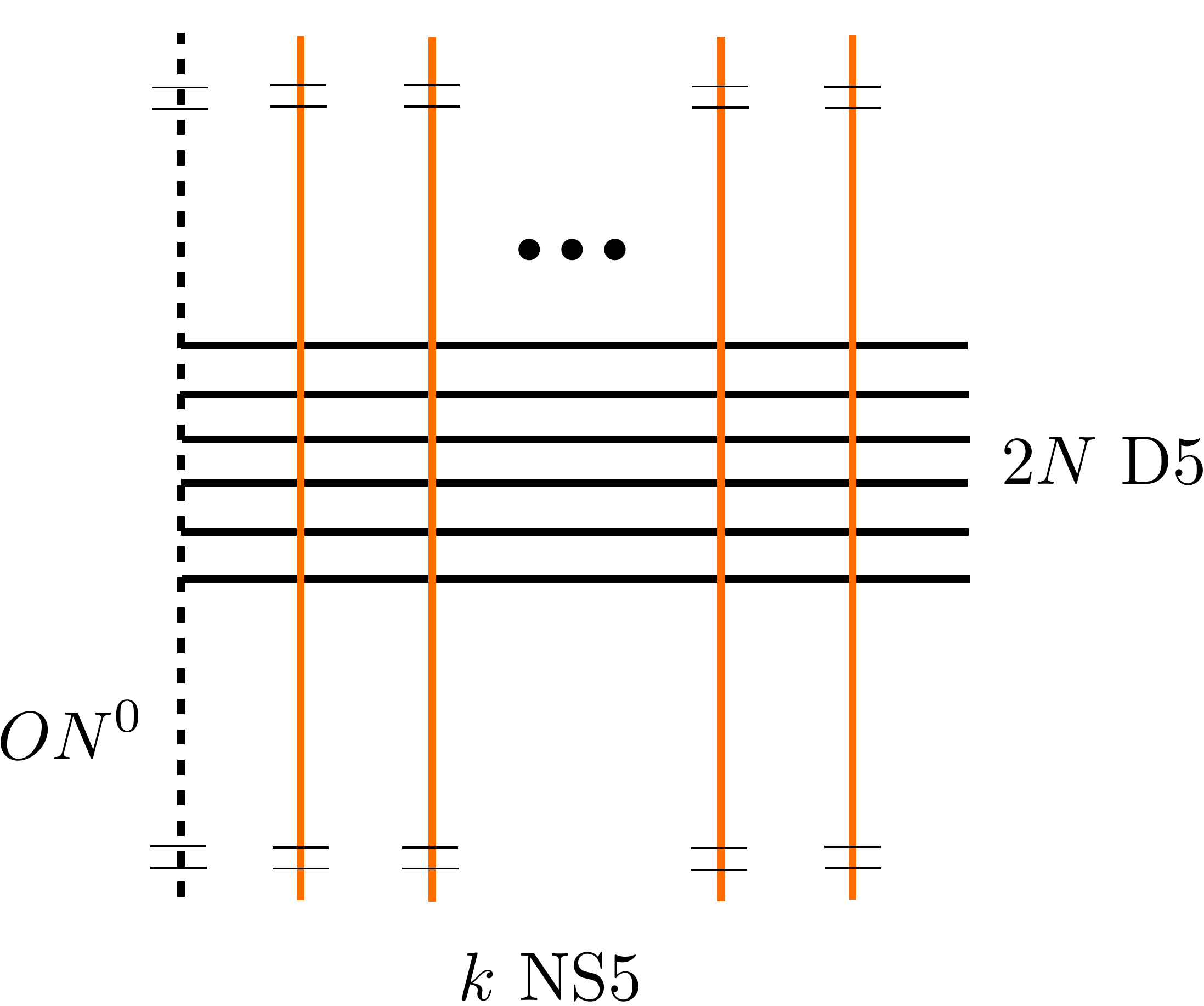}
\end{center}
\caption{
The T-dual description to the circle compactification of the type IIA brane setup in Figure \ref{fig:ON0}. The $ON^0$-plane and the NS5-branes are on $S^1$ in the $x_5$ direction and the identification is denoted by $||$. 
}
\label{fig:TdualON0}
\end{figure}

We now derive a 5d gauge theory description of the 6d D-type quiver theory \eqref{6dDtype} after a circle compactification along one of the worldvolume directions of NS5-branes. After the $S^1$ compactification and performing T-duality along it, we obtain a 5-brane web with an $ON^0$-plane on one end as well as $2N$ D5-branes and $k$ NS5-branes on $S^1$. 
The resulting 5-brane web diagram is depicted in Figure \ref{fig:TdualON0}.
\begin{figure}
\centering
\includegraphics[width=9cm]{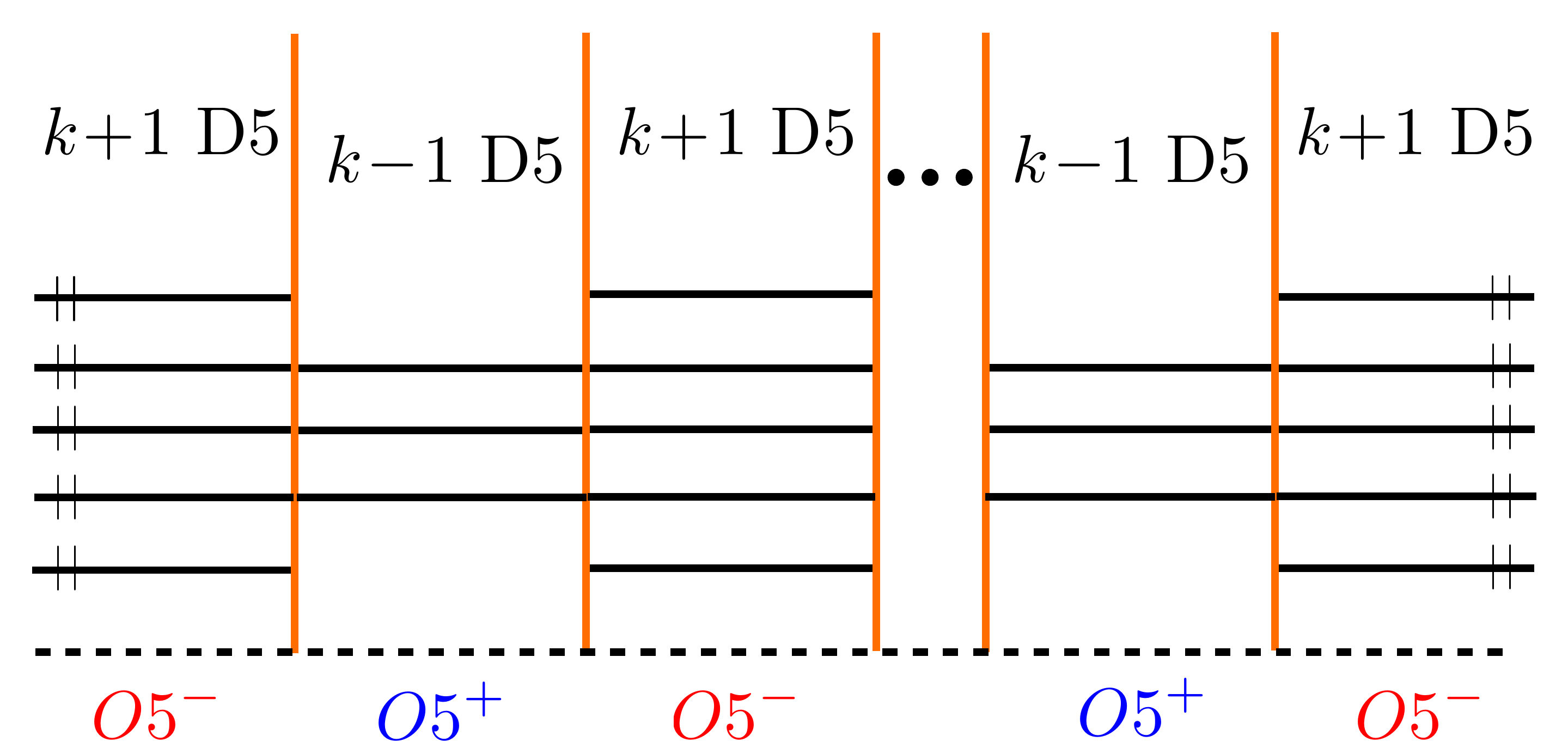}
\caption{A picture of the 5-brane web diagram obtained from Figure \ref{fig:TdualON0} by the transition (\ref{equivalence}). The leftmost D5-branes connect to the rightmost D5-branes and the identification is denoted by $||$. } 
\label{fig:ellipticSOSP}
\end{figure}

Here, we take S-duality taking into account the transition explained in (\ref{equivalence}).
Then, we have $k+1$ D5-branes on top of each $O5^-$-plane and $k-1$ D5-branes on top of each $O5^+$-plane. The total number of fractional NS5-branes is $2N$. The resulting 5d gauge theory realized by the 5-brane web in Figure \ref{fig:ellipticSOSP} is 
\begin{align}
5d~~{\overset{\overset{\text{\normalsize$Sp(k-1) - SO(2k+2)-\cdots-Sp(k-1) -SO(2k+2)$}}{\textstyle\vert\hspace{7cm}\textstyle\vert}}{SO(2k+2)-Sp(k-1)-\cdots -SO(2k+2)  - Sp(k-1)}},\label{ellipticSOSP}
\end{align}
where there are $N$ $Sp(k-1)$ gauge groups and $N$ $SO(2k+2)$ gauge groups. 
Hence, our claim is that when we circle compactify the 6d theory given by \eqref{6dDtype}, then the resulting 5d gauge theory description is the elliptic alternating quiver theory given by \eqref{ellipticSOSP}. This can be thought of as a natural generalization of the 5d elliptic quiver with $SU$ nodes which arises as a circle compactification of the simple D6-NS5-brane system.

We can support this claim by comparing the number of parameters in the 6d theory with those of the 5d theory. Let us count the parameters of the 6d theory given by \eqref{6dDtype}. We first count the number which become the number of the Coulomb branch moduli in the 5d theory after the circle compactification. In 6d, we count the sum of the number of tensor multiplets and the vector multiplets in the Cartan subalgebra. The number of the tensor multiplets corresponds to the number of the gauge nodes in \eqref{6dDtype}, which is $k+1$. From the brane picture in Figure \ref{fig:ON0}, they come from one $ON^0$-plane and $k$ NS5-branes. The two $SU(N)$ gauge nodes and $k-1$ $SU(2N)$ gauge nodes further supply the 6d vector multiplets in the Cartan subalgebra. Hence the sum finally becomes 
\begin{align}
\underbrace{k+1}_{\text{tensor multiplets}} + \underbrace{(N-1)\times 2}_{2\;SU(N)'s} + \underbrace{(2N-1)\times (k-1)}_{k-1\;SU(2N)'s} = 2Nk.\label{6dcounting}
\end{align} 
After the circle compactification, both the 6d tensor multiplets and the 6d vector multiplets in the Cartan subalgebra becomes 5d vector multiplets in the Cartan subalgebra. The number of the 5d vector multiplets in the Cartan subalgebra or the dimension of the Coulomb branch moduli space from the 5d elliptic alternating quiver theory given in \eqref{ellipticSOSP} is  
\begin{align}
\underbrace{(k+1)\times N}_{N\;SO(2k+2)'s} + \underbrace{(k-1)\times N}_{N\;Sp(k-1)'s} = 2Nk.
\end{align}
This precisely agrees with the expectation from the parameter counting of the 6d theory \eqref{6dcounting}.

Let us then look at the matching associated with the global symmetry. In 6d, the flavor symmetry is $SU(2N)$. The circle compactification will give the affine $A_{2N-1}$ structure whose finite part is $SU(2N) \times U(1)_I$ where $U(1)_I$ is a flavor symmetry associated with the Kaluza--Klein (KK) modes from the $S^1$ compactification. Therefore, the number of the parameters in the 5d theory should be $2N-1 + 1_I = 2N$ where $1_I$ denotes the parameter associated to the $U(1)_I$. In a 5d theory, parameters of the theory comes from the gauge couplings and also mass parameters of matter. The elliptic alternating theory given by \eqref{ellipticSOSP} has $2N$ gauge couplings from the $N$ $Sp(k-1)$ gauge groups and $N$ $SO(2k+2)$ gauge groups. It is also important to note that we cannot introduce mass parameters for the hypermultiplets which connect $Sp(k-1)$ gauge group and $SO(2k+2)$ gauge group. Therefore, the total number of the parameters is $2N$, which exactly agrees with the expectation. 

One can in fact see the global symmetry $SU(2N)$ directly from the web diagram in Figure \ref{fig:ellipticSOSP}. The 5-brane web has $2N$ external NS5-branes and a $[0, 1]$ 7-brane can be put on the ends of the NS5-branes. These $2N$ $[0, 1]$ 7-branes imply that the 5d theory have an $SU(2N)$ global symmetry.

\bigskip

\section{\texorpdfstring{Type IIA brane setup with \boldmath $ON^0-O6-O8^-$}{ON0O6O8} -planes
}\label{sec:ON0O8O6}
In section \ref{sec:O6} and \ref{sec:ON0}, we introduced $O6$-planes and an $ON^0$-plane respectively in the type IIA brane setup. As a generalization of the preceding sections, we introduce $O6$-planes and an $ON^0$-plane simultaneously into the type IIA brane configuration. 
The worldvolume configuration that we consider is shown in Table \ref{tb:ONO6O8}. 
\begin{table}[t]
\begin{center}
\begin{tabular}{c|cccccccccc}
\hline
\ &0 &1&2&3&4&5&6&7&8&9\\
\hline
$O8$/D8 & x& x& x& x&x& x& .& x& x& x\\
$O6$/D6 & x& x& x& x& x& x& x& .& .& .\\
$ON^0$ (NS5) & x& x& x& x& x& x& .& .& .& .\\
\hline
\end{tabular}
\end{center}
\caption{The type IIA brane configuration with $ON^0$, $O6$ and $O8$-planes.}
\label{tb:ONO6O8}
\end{table}

It was discussed in 
\cite{Hanany:1999sj} that if one introduces orientifold planes of different kinds among $ON^0$-, $O6^\pm$-, and $O8^\pm$-planes, one needs to consider all of the three orientifolds at the same time to describe the full system. It is also consistent from the point of view that the orientifold projections along the transverse directions with respect to each orientifold. We here consider brane configurations with an $O8$-plane,   
$O6$-planes and D6-branes 
divided by NS5-branes where an $ON^0$-plane is stuck at the intersection between the $O8$-plane and the $O6$-plane or D6-branes. It is worth noting that the $O8$-plane and $D8$-branes are 
fractionalized when they end on an $O6$-plane. 

There are four possible configurations in which an $ON^0$-plane can be included:
(i) $(O8^-,O6^-)$,
(ii) $(O8^-,O6^+)$,
(iii) $(O8^+,O6^-)$, and
(iv) $(O8^+,O6^+)$. Each type IIA brane configuration is anomaly free \cite{Hanany:1999sj} and thus leads to a 6d SCFT on a tensor branch. 
Each configuration gives different 6d quiver theories associated with a Dynkin diagram. Different types of quiver diagram depend on the relative sign of the RR-charge between an $O8$-plane and an $O6$-plane.  
For instance, if the relative sign between an $O8$-plane and an $O6$-plane is opposite, then 
the brane configuration leads to a $D$-type quiver. If the relative sign is same, on the other hand, the resulting configuration is an $A$-type quiver \cite{Hanany:1999sj}. 

In this section, we study possible 5d gauge theory descriptions of
these 6d theories via a circle compactification. As explained in \cite{Hayashi:2015fsa, Zafrir:2015rga, Hayashi:2015zka, Ohmori:2015tka} as well as the previous sections, the process that we propose to obtain a 5d description involves not only T-duality but also S-duality as well as the resolution of $O7$ orientifold planes into a pair of 7-branes. As resolving an $O7$-plane is possible only for an $O7^-$-plane \cite{Sen:1996vd}, obtaining a 5d description for the configuration with $O7^+$-planes is limited. It is also not clear 
whether an S-dual picture of an $O7^+$-plane leads to a sensible gauge theory description. 
Therefore, we will not discuss the brane configurations with an $O8^+$-plane among the possible IIA brane configurations above, as our proposal for obtaining 5d gauge theory description may not be fruitful for those configuration  
other than a natural IIB configuration having two $O7^+$-planes coming from T-duality. 
We will instead focus on the cases involving an $O8^-$-plane. In the following subsections, we study possible 5d gauge theory descriptions for  (i) $(O8^-,O6^-)$ and (ii) $(O8^-,O6^+)$ cases.

\subsection{\texorpdfstring{5d gauge theory descriptions for $ON^0-O6^--O8^-$ system}{O6plus}}\label{sec:ONO6minus}
\begin{figure}
\centering
\includegraphics[width=10cm]{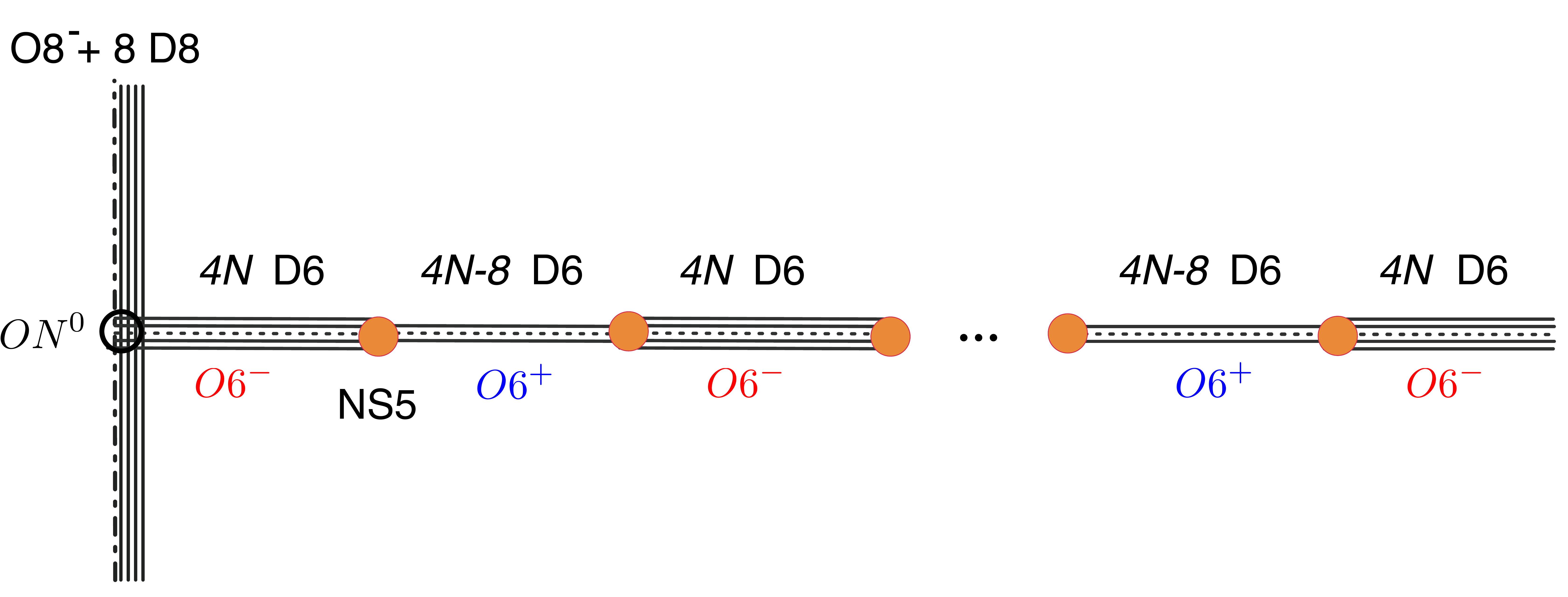}
\caption{The 6d brane configuration with an $O8^-$-plane, $O6$-planes 
and an $ON^0$-plane. A leftmost $O6$-plane is an $O6^-$-plane. The $O6^-$-plane is located on top of $2N$ 
physical D6-branes (The number of D6-branes in the figure includes its mirror images). We also have $2k$ fractional NS5-branes on top of the $O6$-plane and D6-branes. 
This leads to a 6d SCFT on a tensor branch 
$6d ~[8]-SU(2N) - Sp(2N-4) - SO(4N) - Sp(2N-4) -\cdots -Sp(2N-4)-[SO(4N)].$
 In the figure, 
an empty circle denotes an $ON^0$-plane and 
orange circles denote NS5-branes, and there are $k$ $O6^+$- and $(k+1)$ $O6^-$-planes.
}
\label{fig:6dO8mO6m}
\end{figure}

Let us first consider case (i), the configurations where an $O8^-$-plane, an $ON^0$-plane and an $O6^-$-plane are intersecting with each other as shown in Figure \ref{fig:6dO8mO6m}. For simplicity, we introduce eight $D8$-branes so that the net RR-charge is zero. We also 
consider the case of even number of D6-branes in the brane setup. These $2N$ D6-branes are split by $2k$ fractional NS5-branes. As discussed in earlier sections, due to the NS5-branes, the $O6$-planes split in such a way that $O6^-$- and $O6^+$-planes are alternating, which leads to $SO$ and $Sp$ gauge groups. 
A representative brane configuration is given in Figure \ref{fig:6dO8mO6m}.
\begin{figure}
\centering
\includegraphics[width=15cm]{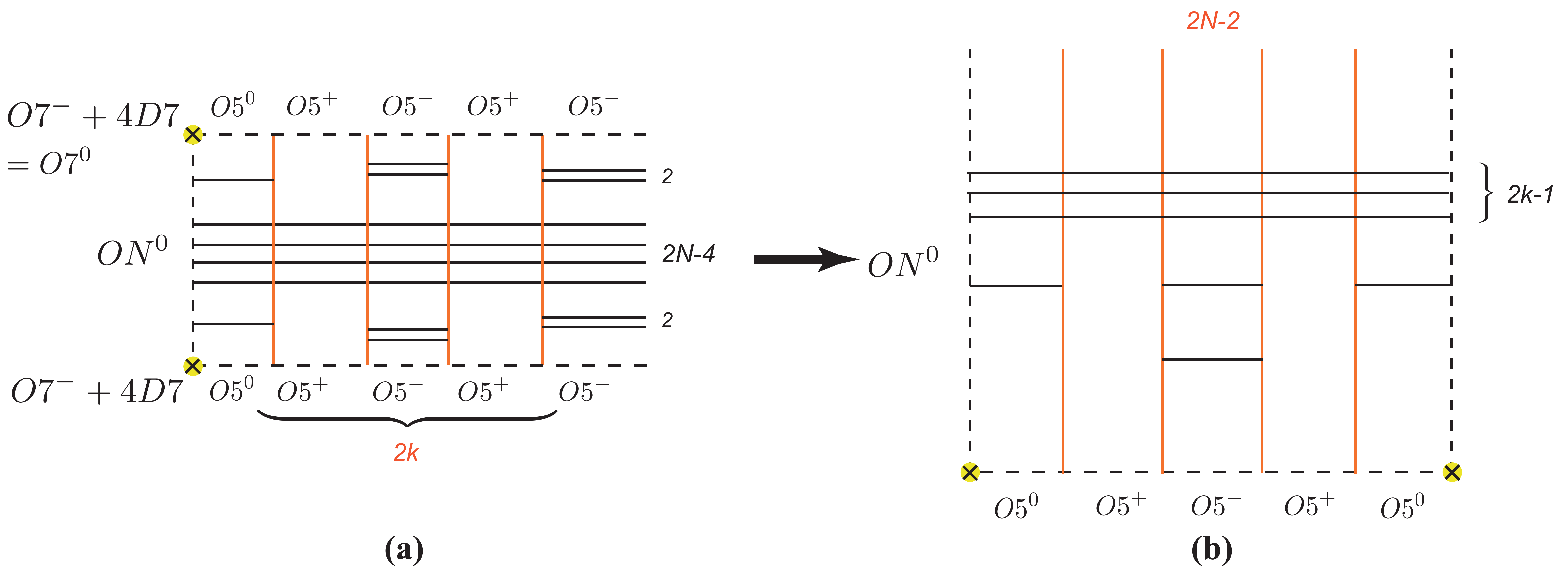}
\caption{A diagrammatic procedure for obtaining a 5d brane configuration for a 6d SCFT described by a brane setup involving $ON^0-O6^--O8^-$ intersection.  (a) Type IIB picture where the vertical direction is the circle compactification direction and the horizontal direction is the direction that D5-branes expanded. (b) The web diagram after S-duality with (\ref{equivalence}). Brane configuration contains two $ON^0$'s, $ (2N-2)$ NS5-branes giving rise to $(N-1)$ $O5^+$-planes and $N$ $O5^-$-planes.
} 
\label{fig:5dO8mO6m1}
\end{figure}
This $ON^0-O6^- - O8^-$-plane configuration yields the following 6d SCFT on a tensor branch:
\begin{align}\label{6dO8mO6m}
6d ~~
[8]
-SU(2N) - Sp(2N-4) -SO(4N) -
\cdots-Sp(2N-4)-
[SO(4N)],
\end{align}
where the number of $Sp$ gauge group is $k$, 
and that of $SO$ gauge group is $k-1$.

The number of the parameters associated in the theory is given as follows. The number of the 6d vector multiplets in the Cartan subalgebra and the 6d tensor multiplets is
\begin{align}\label{eq:coulombo8mo6m}
\underbrace{2N-1+1}_{SU(2N)+\rm tensor} +  \underbrace{(2N-4+1)k}_{Sp(2N-4)'s+\rm tensor}+\underbrace{(2N+1)(k-1)}_{SO(4N)'s+\rm tensor} = 4Nk-2k -1.
\end{align}
The global symmetry of the system is $SO(4N) \times SU(8) \times U(1)$ and hence the rank of the 
global symmetry is $2N+8$.
In the following subsections, we discuss three different ways of obtaining 5d gauge theory descriptions.

\subsubsection{S-dual picture in Type IIB setup}\label{sec:S-duality}
We now consider a type IIB brane configuration for this brane setup via a circle compactification and T-duality. 
As a result of T-dual action on the $O8^-$-plane, one has two $O7^-$-planes in the type IIB brane picture. 
As the $O8^-$-plane and eight D8-branes are fractional, the resulting two $O7^-$-planes and eight D7-branes  
are also fractional. 
With suitable Wilson lines, we allocate the $O7^-$-plane and four D7-branes on top of each other to make an $O7$/D7 combination whose net RR charge is zero, and we call it an $O7^0$-plane. (See Figure \ref{fig:5dO8mO6m1} (a).) The distance between the two $O7^-$-planes is inversely proportional to the radius of the compactifying circle. This is just a IIB description of the 6d theory on a circle. 

Note also that when one has the fractional $O7^0$-plane and the $ON^0$-plane at the same time, one must have another orientifold 5-plane \cite{Hanany:1999sj}. Namely, the fractional $O7^0$-plane has to be located at the intersection between the $ON^0$-plane and the $O5$-plane. The charge conservation further implies that the $O5$-plane has no net RR-charge, which should be a combination of an $O5^-$-plane and a D5-brane. We call it an $O5^0$-plane in this article.  Therefore, the fractional $O7^0$-plane is located at the intersection between the $ON^0$-plane and the $O5^0$-plane in the 5-brane web. From the T-duality, the $O5^0$-plane originates from the combination of the $O6^-$-plane and one D6-brane in the type IIA brane configuration. The RR-charge conservation implies that one $O6^-$-plane yields two $O5^-$-planes after the T-duality. Hence, we have one $O5^0$-plane at the top of the diagram and another $O5^0$-plane at the bottom of the diagram.  (See Figure \ref{fig:5dO8mO6m1} (a).)

In order to obtain a genuine 5d gauge theory description for the 6d theory, we utilize the S-duality as discussed in section \ref{sec:ONO5}.  
Due to the transition \eqref{equivalence},
the 5-brane web after S-duality is given in Figure \ref{fig:5dO8mO6m1} (b),
which corresponds to a 5d quiver gauge theory interpretation as 

\begin{align}\label{5dSpSOfromO8ONO6m}
5d~~[4] - SU(2k+1)-Sp(2k-1)-SO(4k+2)-\cdots -Sp(2k-1)-SU(2k+1)-[4],
\end{align}
where the number of $Sp$ gauge groups is $N-1$, while that of $SO$ gauge group is $N-2$.

Let us then count the number of parameters of the 5d gauge theory and see its comparison with the 6d computation. The dimension of the Coulomb branch moduli space is given by
\begin{align}\label{5dO5mp}
\underbrace{4k}_{{\rm two}~SU(2k+1)'s}+ \underbrace{\big(N-1\big)(2k-1)}_{Sp(2k-1)'s}+ \underbrace{\big(N-2\big)(2k+1)}_{SO(4k+2)'s} = 4Nk-2k-1,
\end{align}
which precisely matches with \eqref{eq:coulombo8mo6m}.
The number of the parameters is then
\begin{align}
\underbrace{4+4}_{\rm flavors}\,+ \,\underbrace{2+2N-3}_{\rm instantons}\,+\underbrace{2}_{\rm bi-fund} = 2N+8+1_I,
\end{align}
which is the expected number including the symmetry arising from the KK modes, $1_I$. Note that we can introduce a mass parameter only for the bi-fundamental hypermultiplet between the $Sp(2k-1)$ gauge group and the $SU(2k+1)$ gauge group. From the brane configuration and the parameter counting, \eqref{5dSpSOfromO8ONO6m} is a possible 5d description for the 6d
theory \eqref{6dO8mO6m}. 

\subsubsection{\texorpdfstring{The resolution of two $O7^-$ orientifolds}{two O7s}}\label{sec:twoO7s}
We now consider the resolution of an $O7^-$-plane \cite{Sen:1996vd,DeWolfe:1999hj} into a pair of two 7-branes of the charge $[1,-1]$ and $[1,-1]$ from the configuration in Figure \ref{fig:5dO8mO6m1} (a).
As for notation, we denote a D7-brane whose $[p,q]$ charge is $[1,0]$ by $\mathbf{A}$, and $[1,-1]$ 7-brane by $\mathbf{B}$, $[1,1]$ 7-brane by $\mathbf{C}$, and $[0,1]$ 7-brane by $\mathbf{N}$. It follows that the resolution of $O7^-$-plane means to replace $O7^-$ into a pair of ($\mathbf{B}$,$\mathbf{C}$) in a counterclockwise way\footnote{More precisely, when the branch cuts of 7-branes extend down there, an $O7^-$-plane can resolve into a $\mathbf{B}$ 7-brane and a $\mathbf{C}$ 7-brane from left to right. The monodromy matrix is 
\begin{eqnarray*}
M_{[p, q]} = \left(
\begin{tabular}{cc}
$1+pq$ & $-p^2$\\
$q^2$ & $1-pq$
\end{tabular}
\right),
\end{eqnarray*}
when a 7-brane or a 5-brane crosses the branch cut of a $[p, q]$ 7-brane in a counterclockwise way. }, and $O7^0$ becomes $\mathbf{A^2BCA^2}$ as a D7-brane and an $O7^-$-plane mutually commute with each other. It is straightforward to see from 7-brane monodromies that 
\begin{align}\label{7-braneresol}
\bf A^2BCA^2 = (ANA)(ANA).
\end{align}
In the current brane setup, the $O7^0$-plane is stuck at the intersection between an $ON^0$-plane and an $O5^0$-plane. As 
it is stuck at the intersection of the orientifold planes, 
the $O7^0$-plane becomes fractional. From the 7-brane configuration in \eqref{7-braneresol}, a half-$O7^0$-plane may be described effectively by the full $\bf ANA$ 7-branes. 
Since the half $O7^0$-plane consists of three full 7-branes, it can be detached away from the orientifolds. Therefore, we propose that a half $O7^0$-plane at the intersection between an $ON^0$-plane and an $O5^0$-plane can be resolved into the full $\bf ANA$ 7-branes.

\begin{figure}
\centering
\includegraphics[width=9cm]{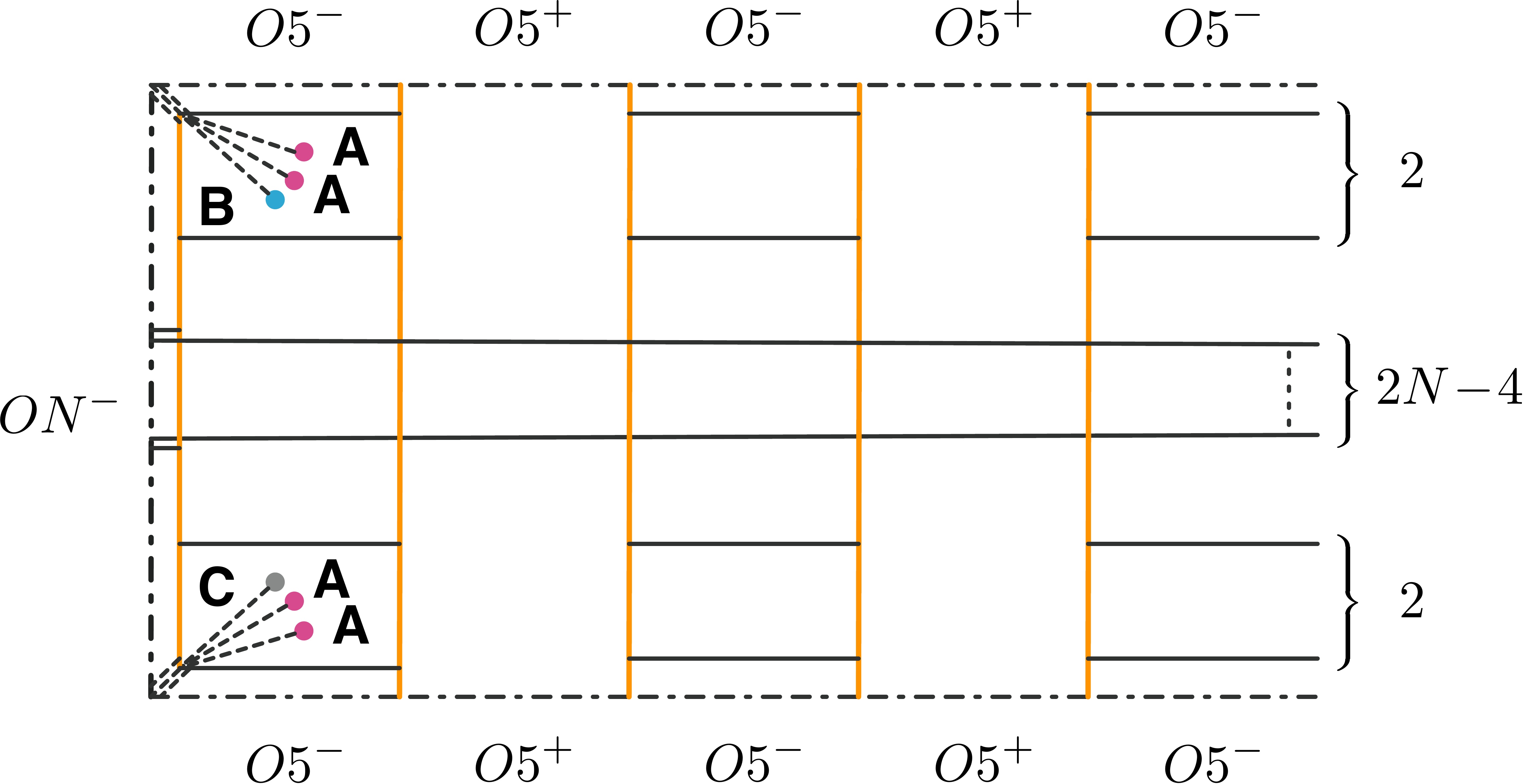}
\caption{A type IIB brane configuration with two resolved $O7^-$s. At the intersections of $ON^0$ and $O5^0$ on the top and the bottom, fractional $O7^-$ and four fractional D7-branes located. A full 7-brane combination ($\bf A,A,B $) or ($\bf C,A,A $) is pulled out from the intersection. The branch cuts of the 7-branes, denoted by the dotted line coming from 7-branes, are all pointing the intersection.
} 
\label{fig:5dONO57branes}
\end{figure}

This resolution is also natural from the point of view of the $(p, q)$ charge conservation at the interaction between an $ON^0$-plane and an $O5^0$-plane, involving the cuts due to $\bf ANA$ 7-branes. (See Figure \ref{fig:5dONO57branes}.) 
The monodromy created by the branch cuts of the $\bf ANA$ 7-branes is in fact exactly the S-duality transformation, Therefore, as the cuts of the $\bf ANA$ 7-branes are going through the intersection of the orientifolds, 
 an $O5^0$-plane 
is converted to an $ON^0$-plane, or vice versa.  
After the resolution of the $O7^0$-plane, a microscopic picture of the $ON^0$-plane and the $O5^0$-plane is that the branch cuts of the $\bf ANA(=AAB=CAA)$ 7-branes change an $ON^-$-plane (and an NS5-brane) into an $O5^-$-plane (and a D5-brane) respectively. 

\begin{figure}
\centering
\includegraphics[width=7cm]{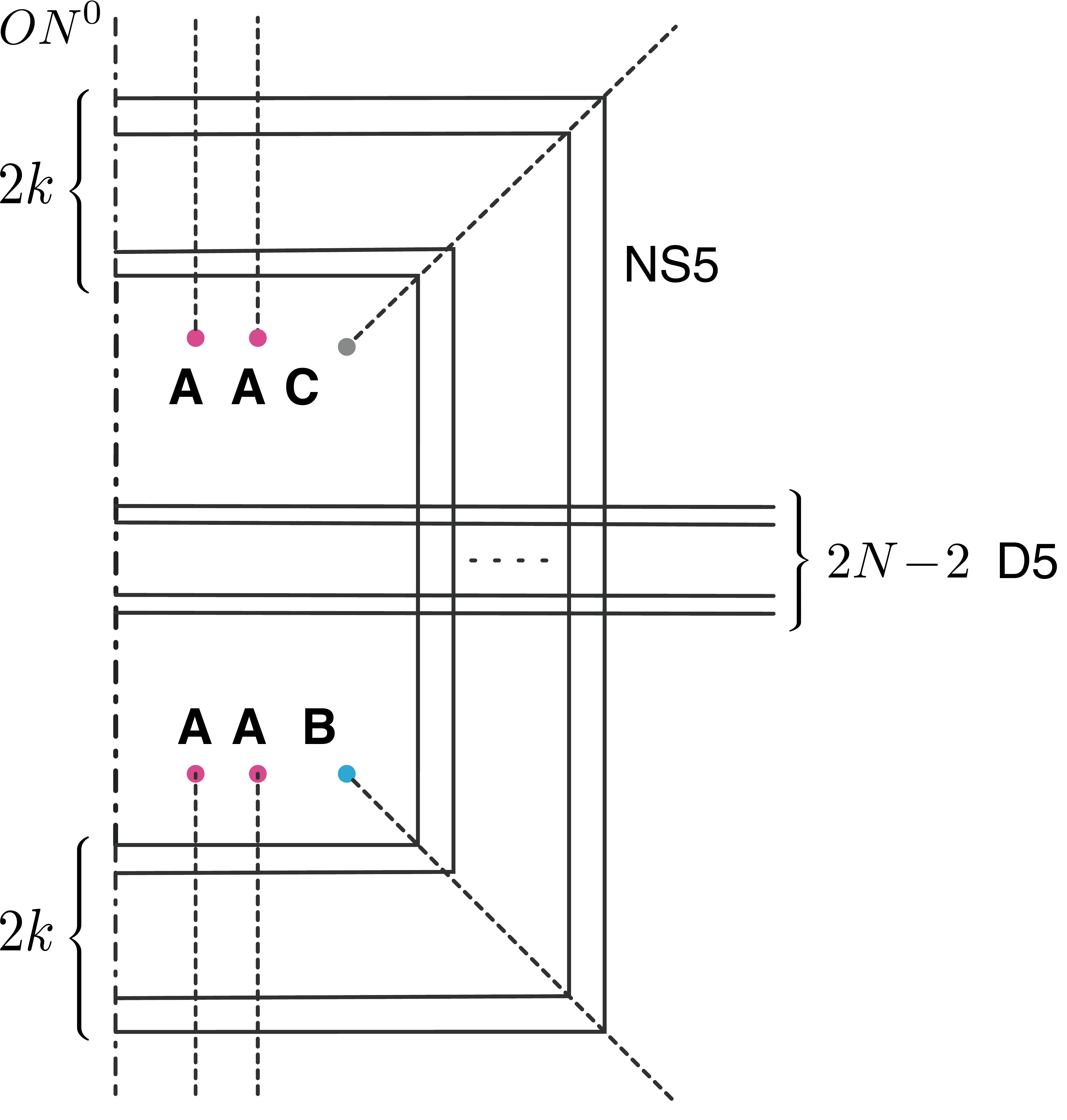}
\caption{A 5-brane configuration with an $ON^0$ and 7-branes coming from the resolutions of $O7^-$-planes. Here 7-branes are rearranged such that $O5^0$'s in Figure \ref{fig:5dONO57branes} turn into $ON^0$ as they across the cuts of $\bf B$ and $\bf C$ 7-branes.} 
\label{fig:ON7branes}
\end{figure}

We then turn around the cuts of 7-branes in a way that 
the alternating $O5^+$- and $O5^-$-planes go across these cuts. 
Since going across the cuts locally generate S-duality transformation
as discussed above, these $O5$-planes are converted into 
 $ON^0$-planes taking into account the transition (\ref{equivalence}).
The resulting web diagram is depicted in Figure \ref{fig:ON7branes}. 

\begin{figure}
\centering
\includegraphics[width=12cm]{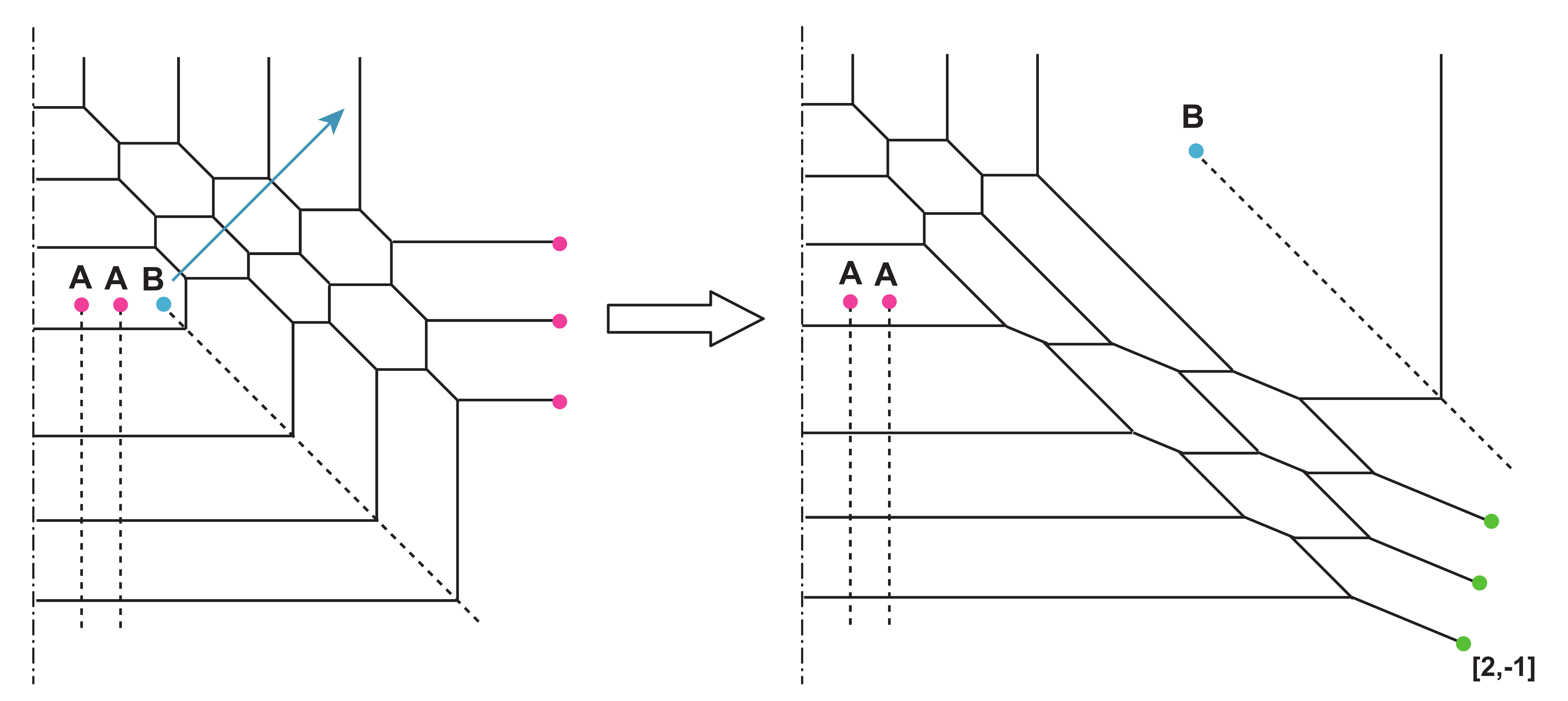}
\caption{
An example of 5-brane web with 7-branes which demonstrates how a $\bf B$ 7-brane moves through the $(1,-1)$ 5-branes without generating any new 5-branes. In so doing, some of D7-branes ($[1,0]$ 7-branes) become $[2,-1]$ 7-branes.} 
\label{fig:Bbranemoving}
\end{figure}

\begin{figure}
\centering
\includegraphics[width=15cm]{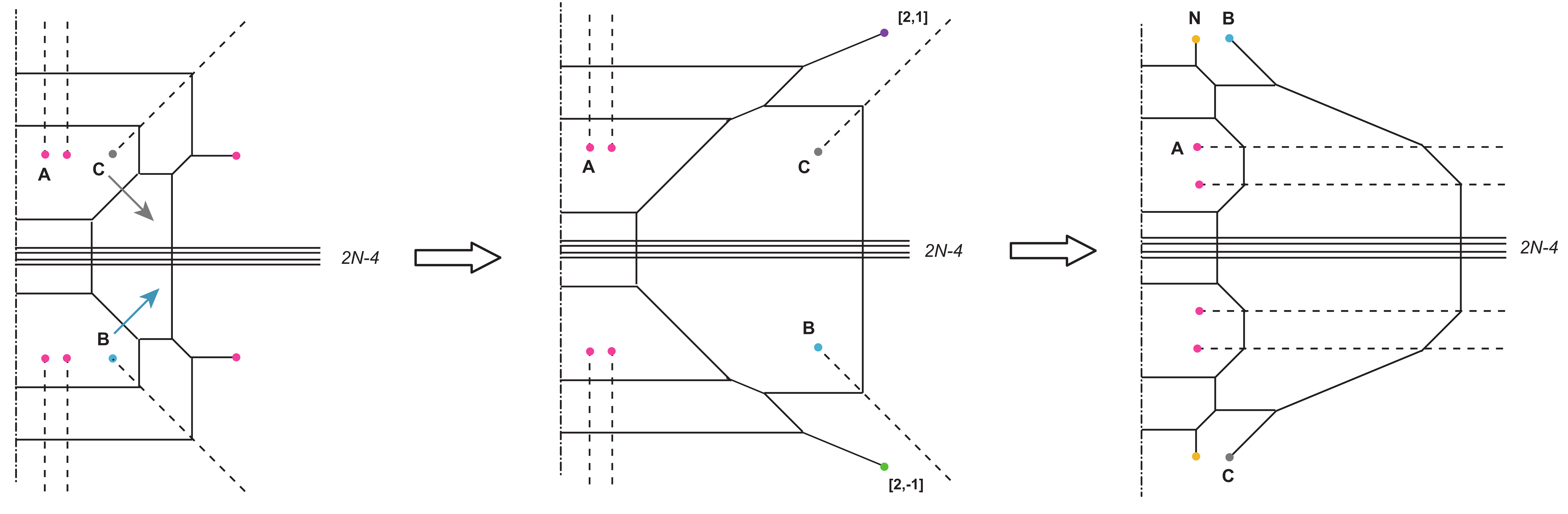}
\caption{A procedure leading to a 5d gauge theory from the web diagram after allocating 7-branes in suitable positions. By turning around the cuts of 7-branes, one can see a 5d gauge theory description more clearly.
} 
\label{fig:ON0BC}
\end{figure}

As done in \cite{Hayashi:2015zka}, we can pull out the $\bf B, C$ 7-branes in such a way to yield a 5d gauge theory interpretation, which is to move the $\bf B, C$ 7-branes through the 5-branes of the same $(p,q)$ charge so that no 5-branes are created in so doing. 
For example, take the $\bf B$ 7-brane and move it through the $(1,-1)$ 5-brane as shown in Figure \ref{fig:Bbranemoving}. We move the $\bf B, C$ 7-branes until they are put as close as possible in Coulomb branch moduli chambers. This leads to 
the following two possible configurations for which one can give a 5d gauge theory interpretation.\\
\noindent $\bullet$ 
The first case is $N\ge 2k$. 
In this case, as depicted in Figure \ref{fig:ON0BC} for $k=1$, 
the $\bf B$ and $\bf C$ 7-brane can move outside 
without meeting each other in a single Coulomb branch moduli chamber.
This yields the following $D$-type quiver theory
\begin{align}\label{5dNge2k}
5d~~
{[4]-SU(N+2k-1)}-\underbrace{{\overset{\overset{\text{\normalsize$[4]-SU(N+2k-1)$}}{~~~\textstyle\vert}}{SU(2N+4k-6)}}-
\cdots-SU(2N-4k+2)}_{2k-1} - [2N-4k].
\end{align}
Here we put all four $\bf A$ 7-branes back to the $ON^0$-plane so that they are again fractionalized, giving each four flavors.

It is straightforward to see the parameters of theory precisely match with those of the 6d theory \eqref{eq:coulombo8mo6m}. The dimension of the Coulomb branch moduli space is given by
\begin{align}
(N+2k-2)\times 2 +\sum^{2k-1}_{i=1} (2N+4k-7-4(i-1)) = 4Nk-2k-1, 
\end{align}
and the number of the parameters of the global symmetry is 
\begin{align}
\underbrace{4+4}_{\rm left~flavors}+ \underbrace{2 + (2k-1)}_{\rm instantons}+ \underbrace{2+2k-2}_{\rm bi-fund.}+\underbrace{2N-4k}_{\rm right~flavors}
= 2N+8+1_I,
\end{align}
where it includes the $U(1)$ part coming from the KK mode, $1_I$, as expected.

In this case, we can further see the consistency with the enhancement of the global symmetry by the one-instanton analysis given in \cite{Yonekura:2015ksa}. As for the D-type Dynkin quiver \eqref{5dNge2k}, almost all the gauge nodes except for the rightmost one satisfy $N_f = 2N_c$ for a $SU(N_c)$ gauge node where $N_f$ represents the effective number of flavors attached to the $SU(N_c)$ gauge node. On the other hand, the rightmost gauge node satisfies $N_f = 2N_c + 2$. The one-instanton analysis then implies that, in the case where $N_f = 2N_c$, the instanton states give rise to two Dynkin diagrams whose shapes are equal to the quiver diagram. In the case where $N_f = 2N_c + 2$, the two nodes given by the one-instanton states not only participate in the two Dynkin diagrams but also are the fundamental representation and the antifundamental representation of the flavor symmetry attached to the gauge node. Therefore, the one-instanton analysis of the D-type quiver theory \eqref{5dNge2k} at least yields the flavor symmetry whose structure is described by the affine $D_{2N}$ Dynkin diagram. The affine $D_{2N}$ Dynkin diagram implies that the UV completion is a 6d theory with a flavor symmetry $SO(4N)$, which is consistent with the flavor symmetry of the 6d theory \eqref{6dO8mO6m}.

\begin{figure}
\centering
\includegraphics[width=8cm]{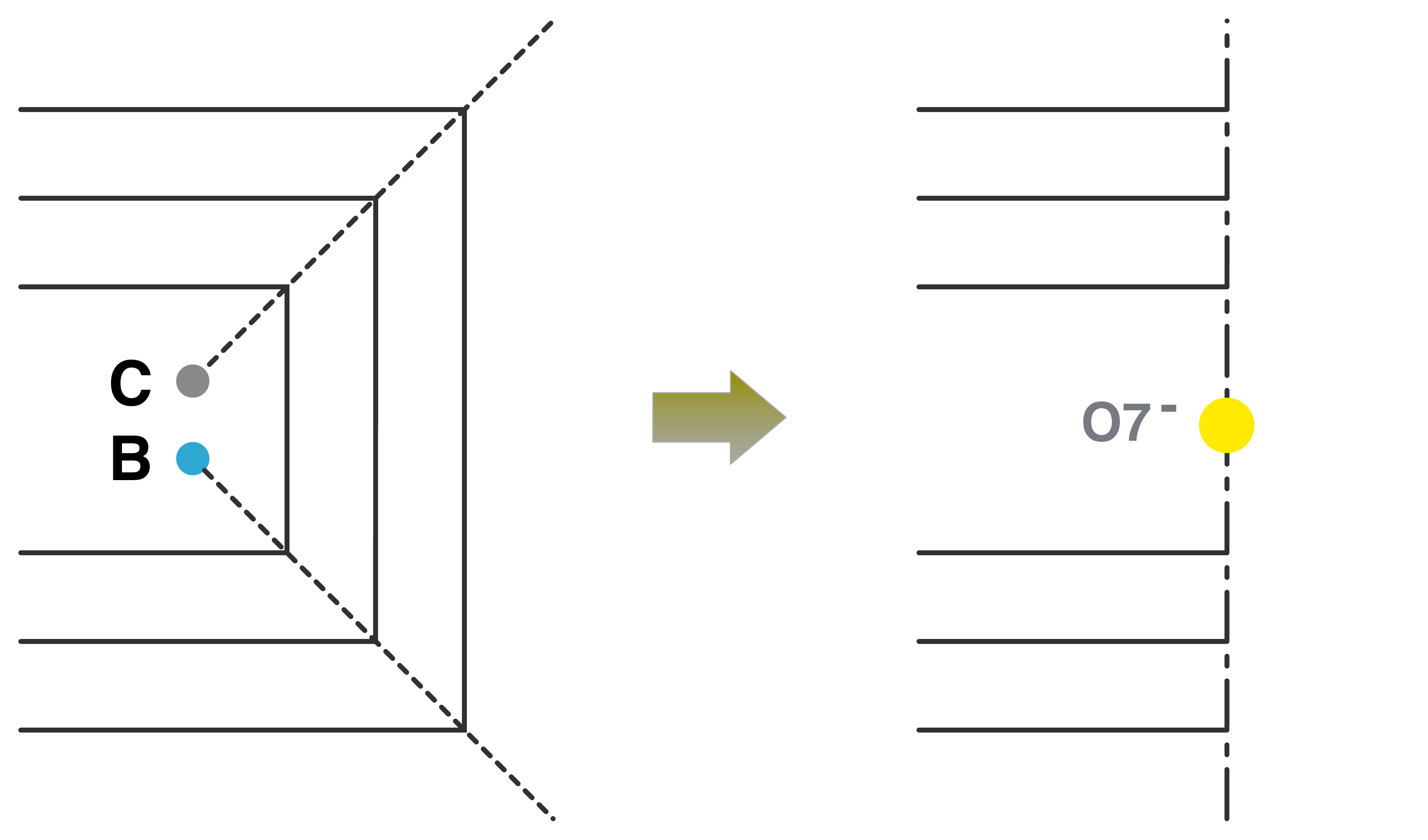}
\caption{A 7-brane pair $(\bf B,C)$ of a resolved $O7^-$-plane forming an $O7^-$-plane again. In so doing, the cut of $O7^-$ appears, in this case, vertically, giving rise to an $SO$ gauge group.
} 
\label{fig:BC2O7}
\end{figure}

\vskip 0.5cm
\noindent $\bullet$ The other case is when $N < 2k$. In this case, the $\bf B, C$ 7-branes are put together in a single Coulomb branch moduli chamber. By doing the trick converting $\bf B, C$ 7-branes back to an $O7^-$-plane, as depicted in Figure \ref{fig:BC2O7},
one finds that the corresponding 5d gauge theory of a $D$-type quiver is given by
\begin{align}
5d~~
{[4]-SU(2k+N-1)}-\underbrace{{\overset{\overset{\text{\normalsize$[4]-SU(2k+N-1)$}}{~~~\textstyle\vert}}{SU(4k+2N-6)}}-
\cdots-SU(4k-2N+6)}_{N-2} - Sp(2k-N+1).
\end{align}

It is again straightforward to check that the parameters of the theory match with those of the 6d theory \eqref{eq:coulombo8mo6m}. The dimension of the Coulomb branch moduli space reads
\begin{align}
(2k+N-2)\times 2 +\sum^{N-2}_{n=1} (4k+2N-7-4(n-1))+ 2k-N+1 = 4Nk-2k-1, 
\end{align}
and the number of the parameters of the global symmetry is given by 
\begin{align}
\underbrace{4+4}_{\rm left~flavors}+ \underbrace{2 + (N-2)+1}_{\rm instantons}+ \underbrace{2+N-3+1}_{\rm bi-fund.}
= 2N+8+1_I,
\end{align}
which includes the $U(1)$ part coming from the KK mode, as expected.

\subsubsection{\texorpdfstring{The resolution of only one $O7^-$ orientifold}{one O7}}\label{sec:oneO7s}
\begin{figure}
\centering
\includegraphics[width=8cm]{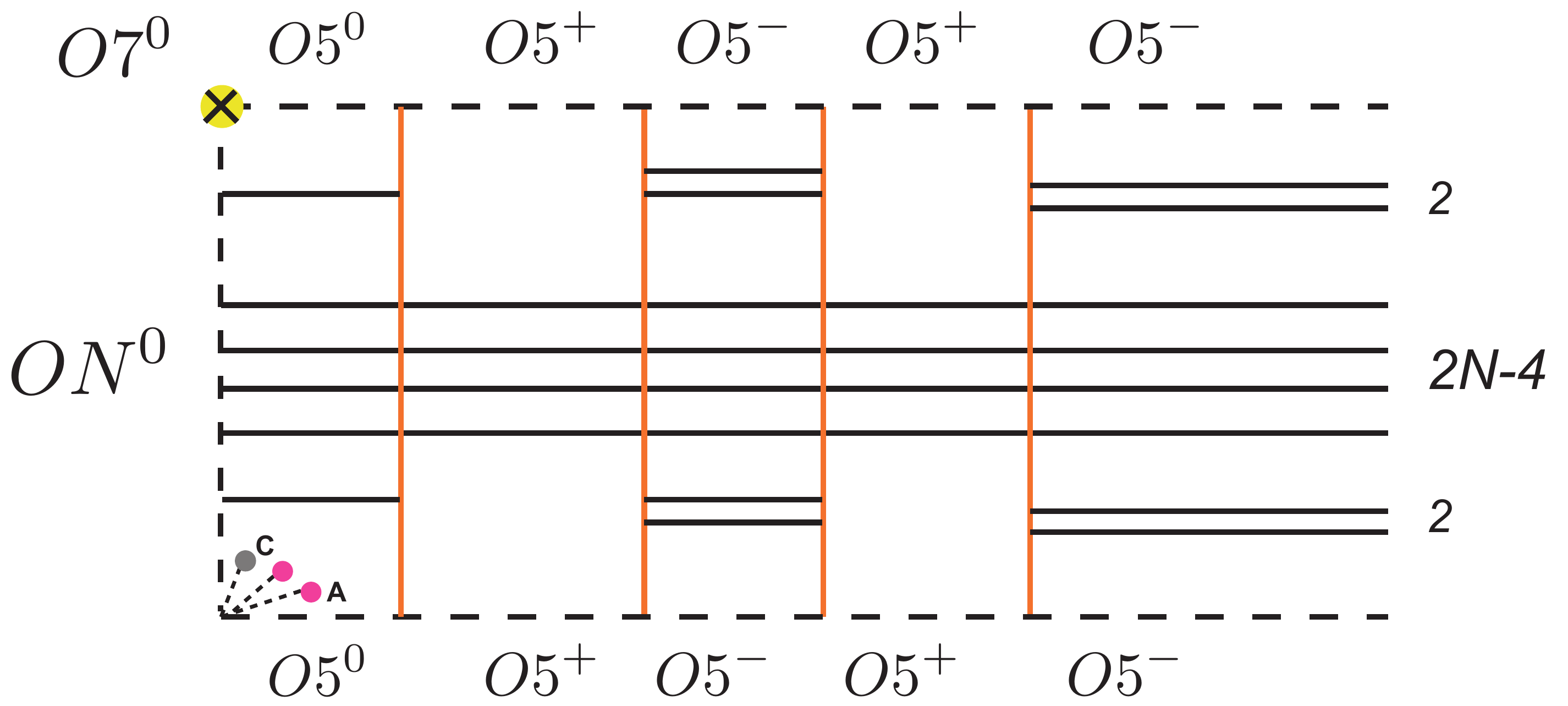}
\caption{LEFT: Only one resolved $O7^-$ and $4$ D7s at the intersection of $ON^0$ and $O5^0$ at the bottom. $O7^-$ and $4$ D7 branes are all fractional and a combination ($\bf C,A,A $) of full 7-branes are pulled out. The branch cuts of the 7-branes are all pointing the intersection.~~~  RIGHT: A brane configuration leads to a 5d gauge description with one resolved $O7^-$ for $2N\ge 2k+1$. The 7-branes ($\bf C,A,A $) rearrange themselves to be ($\bf A,A,B $), and $\bf B$ 7-brane moves across $(2k-1)$ 5-branes of the $(1,1)$ charge. By pulling $\bf B$ 7-brane along the its charge, one then sees a brane configuration yielding a 5d gauge description given in \eqref{1O7Nge2k}.} 
\label{Fig:oneO7}
\end{figure}

Regarding the resolution of the $O7^-$ orientifold planes, one can consider the case where only one out of the two $O7^-$-planes is resolved, which leads to an interesting UV duality in 5d. For instance, it was discussed in \cite{Hayashi:2015zka} that there are two seemingly different 5d descriptions whose UV completion is the same 6d conformal matter theory.  Depending on whether one resolves one or two $O7^-$-planes which come from the T-dual action on an $O8^-$-plane in the  
type IIA brane setup,  
one has an $SU$ gauge theory description for resolving two $O7^-$-planes, while one has an $Sp$ gauge theory description for resolving only one $O7^-$-plane. In this subsection, as a generalization, it is interesting to study what dual description we get when we resolve only one $O7^-$-plane leaving the other $O7^-$ as it is, like Figure \ref{Fig:oneO7}.
As in the previous subsection, possible brane configurations depend on the number of D5-branes and NS5-branes.

\begin{figure}
\centering
\includegraphics[width=13cm]{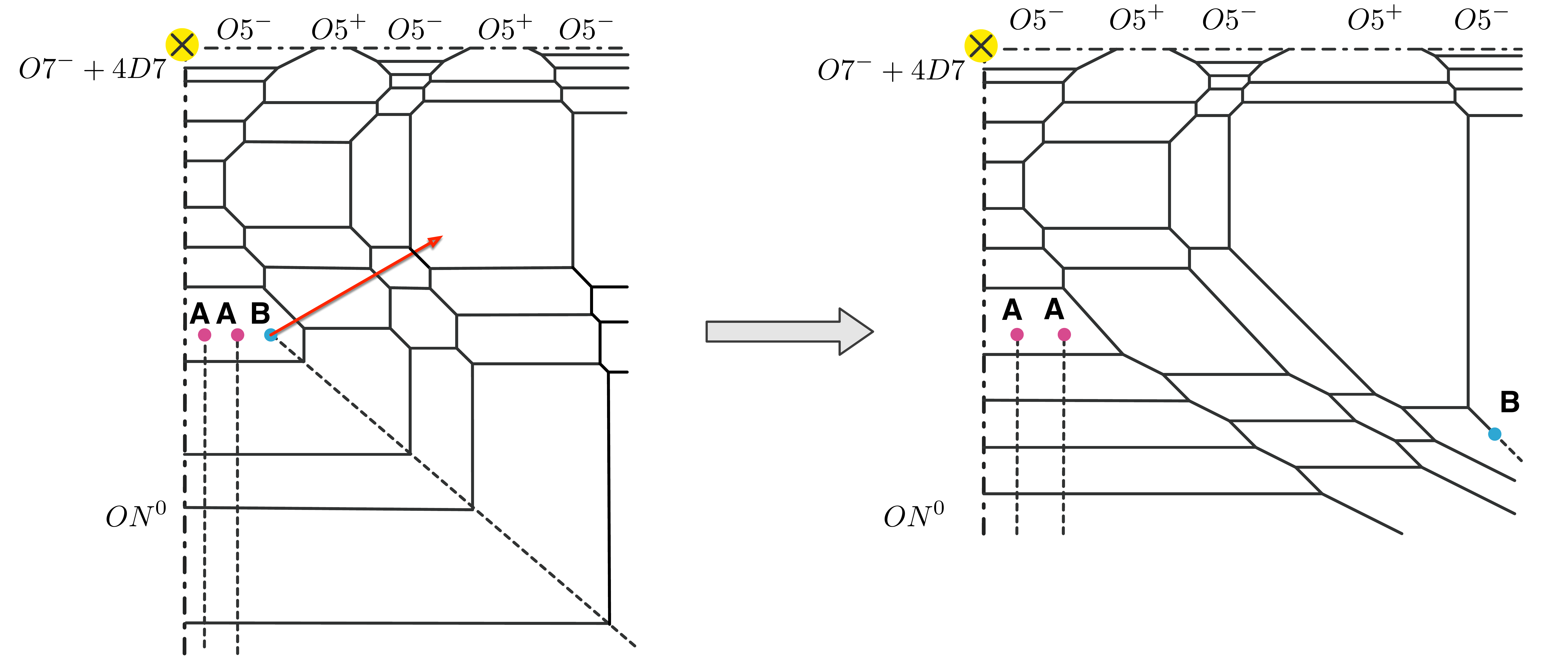}
\caption{A brane configuration for $N=4,k=2$ with only one $O7^-$ being resolved.  The fractional $O7^-+4 D7$'s are pulled out as $\bf A, A, B$ 7-branes (Left figure). The $\bf B$ 7-brane is moved through chambers of the Coulomb moduli spaces, in the direction where the arrow indicates. In order to see a 5d gauge theory, one then pulls out the $\bf B$ 7-brane along the direction of its charge $[1,-1]$, which leads to the 5d quiver theory, $[8]-SU(11)-Sp(7)-SO(14)-Sp(3)-[4]$.} 
\label{fig:1O7N4k2}
\end{figure}

We first consider the case when $2N\ge 2k+1$. After resolving only one $O7^-$-plane, say the bottom $O7^-$-plane in Figure \ref{fig:5dO8mO6m1} (a), we 
repeat the procedure of moving $\bf B$ 7-brane through the chambers of 5-brane webs
as explained in Figure \ref{fig:Bbranemoving}.
For definiteness, an example of brane configuration of one $O7^-$ resolution for $N=4$ and $k=2$ is shown in Figure \ref{fig:1O7N4k2}.
Alternating $O5^+$-$O5^-$-plane gives alternating $Sp-SO$ gauge groups. 
The resulting 5d gauge theory is given by
\begin{align}\label{1O7Nge2k}
5d~~
&[8]-SU(2N+2k-1)-Sp(2N+2k-5)-SO(4N+4k-10)-\cdots\cr
&-Sp(2N+2k-1-4i)-SO(4N+4k-2-8i)-\cdots\cr
&-Sp(2N-2k+3)-SO(4N-4k+6)-Sp(2N-2k-1)-[2N-2k],
\end{align}
where there are $2k$ gauge nodes.

We now check the parameters of this 5d quiver theory and compare them with the 6d theory that we started with. There are one $SU$ gauge group, $k$ $Sp$ gauge groups, and $k-1$ $SO$ gauge groups which lead to the total dimension of the Coulomb branch moduli space to be
\begin{align}
&2N+2k-2 +\sum^k_{i=1} \big(2N+2k-5-4(i-1)\big)+\sum^{k-1}_{i=1} \big(2N+2k-5-4(i-1)\big) \cr
&= 4Nk-2k-1,
\end{align}
which agrees with \eqref{eq:coulombo8mo6m}.
The number of the global symmetry parameters of the 5d theory is 
\begin{align}
\underbrace{8}_{\rm left~ flavors}\;+\underbrace{2k}_{\rm instantons}\;+\; \underbrace{2N-2k}_{\rm right~ flavors}\, +\underbrace{1}_
{\rm bi-fund.}
=2N+8+1_{I},
\end{align}
where we have $2k$ instanton fugacities from the $2k$ gauge nodes, and one bi-fundamental field connecting the first $SU$ gauge group and the adjacent $Sp$ gauge group. Again it agrees with the 6d parameters as expected.

When $2N< 2k+1$, one may repeat the same procedure, which leads to the case where the $\bf B$ 7-brane will eventually hit the upper $O5^0$-plane. In this case, however, it is not clear whether there is a well-defined gauge theory description. To obtain a proper gauge theory description, we might need to convert the 7-branes to a full or fractional $O7^-$-plane. In order for one to have a fractional $O7^-$-plane of a consistent configuration with the $(p,q)$ charge conservation, one needs a $\bf C$ 7-brane to hit the $O5^0$-plane. Here, on the other hand, we have the $\bf B$ 7-brane hitting the $O5^0$-plane, which does not lead to a gauge theory description.

\subsection{\texorpdfstring{5d gauge theory descriptions for $ON^0- O6^+-O8^-$ system}{O6plus}}\label{sec:ONO6plus}
\begin{figure}
\centering
\includegraphics[width=12cm]{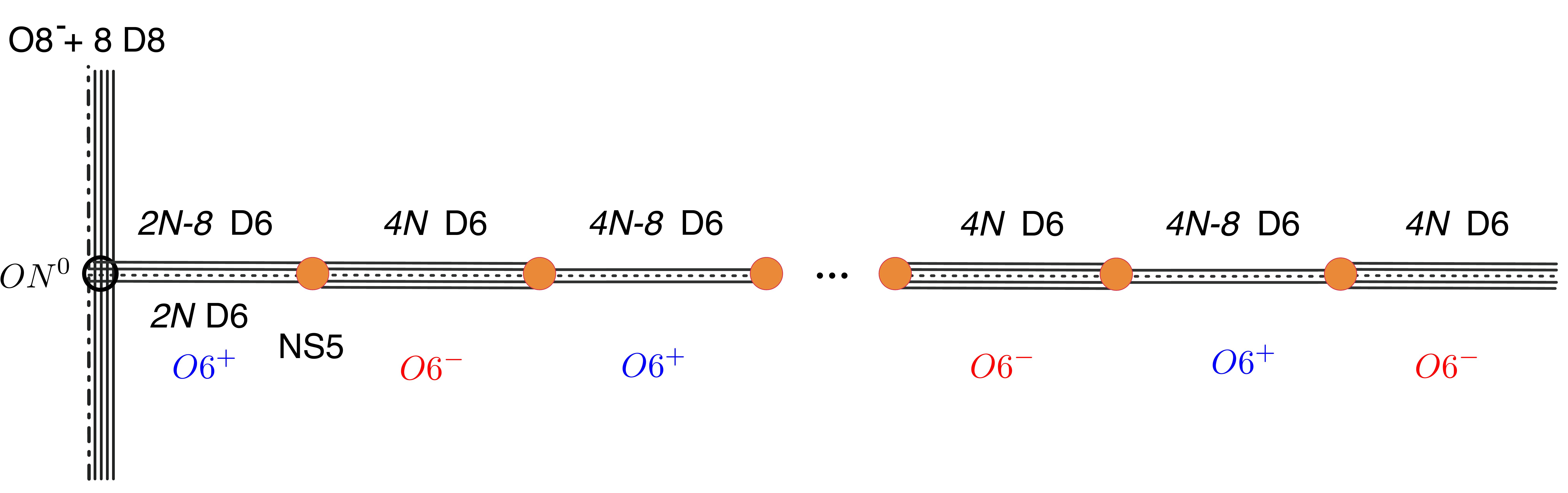}
\caption{A type IIA brane setup involving the intersection of $ON^0,O6^+$, and $O8^-$-planes. There are $2k-1$ fractional NS5-branes and the number of D6-branes includes their mirror pairs.} 
\label{Fig:O8-O6p-ON0}
\end{figure}
We now consider the case (ii) $(O8^-, O6^+)$ which is a 6d SCFT which can be described by a type IIA brane setup involving an $ON^0-O6^+-O8^-$ system.  
On top of a fractional $O8^-$-plane with $8$ D8-branes, an $ON^0$-plane is located, and D6-branes/$O6^+$-planes are extended along the $x_6$ direction as shown in Figure \ref{Fig:O8-O6p-ON0}. Like the previous cases with an $O6^-$-plane, D6-branes and an $O6$-plane are suspended between fractional NS5-branes, leading to a configuration with alternating $O6^-$- and $O6^+$-planes. 

In Figure \ref{Fig:O8-O6p-ON0}, 
there are two types of configurations for D6-branes between the $ON^0$-plane and the first NS5-brane in this case \cite{Hanany:1999sj}.
The first configuration is 
that $(2N-8)$ D6-branes are suspended between the first NS5-brane and the NS5-brane inside the $ON^0$. 
The other one is 
that $2N$ D6-branes are 
extended from the first NS5-brane and jump over the NS5-brane,  
passing through
the $ON^-$-plane  
and then attach back to the NS5-brane or the mirror NS5-brane inside the $ON^0$. 
These two types of D6-branes contribute to different $Sp$ gauge groups. 
The numbers of the D6-branes in Figure \ref{Fig:O8-O6p-ON0} are fixed by the cosmological constant condition, or equivalently the anomaly free condition.  

The resulting brane configuration then yields a $D$-type quiver with two-pronged $Sp$ gauge nodes of different ranks. 
It follows then that the 6d SCFT on a tensor branch is given by
\begin{align}\label{6dO8mO6p1}
6d~~[8]-Sp(N)-
\underbrace{ 
{\overset{\overset{\text{\large$Sp(N-4)$}}{\textstyle\vert}}{SO(4N)}}-Sp(2N-4)-SO(4N) -\cdots-Sp(2N-4)}_{2k-2} - [2N].
\end{align}
The number of the 6d vector multiplets in the Cartan subalgebra plus the number of the tensor multiplets are given by
\begin{align} \label{6dCoulomb4O6plus}
\underbrace{N-3+N+1}_{{\rm two~split}~Sp's\rm+tensors} +\underbrace{(2N+1)(k-1)}_{SO~\rm groups+tensors} + \underbrace{(2N-3)(k-1)}_{Sp~\rm groups+tensors}=
4Nk-2N-2k.
\end{align}
The global symmetry is $SO(4N) \times SO(16)$ 
and thus the rank of the global symmetry is $2N+8$.

In general, the number of tensor multiplets is equal to the number of gauge nodes. 
Since we have two-pronged $Sp$ gauge nodes in the left part of the quiver \eqref{6dO8mO6p1},  
the distance between the NS5-brane and the $ON^-$-plane,
which form the original $ON^0$ plane, 
is also related to the tensor branch moduli of one of these gauge groups.


\begin{figure}
\centering
\includegraphics[width=7cm]{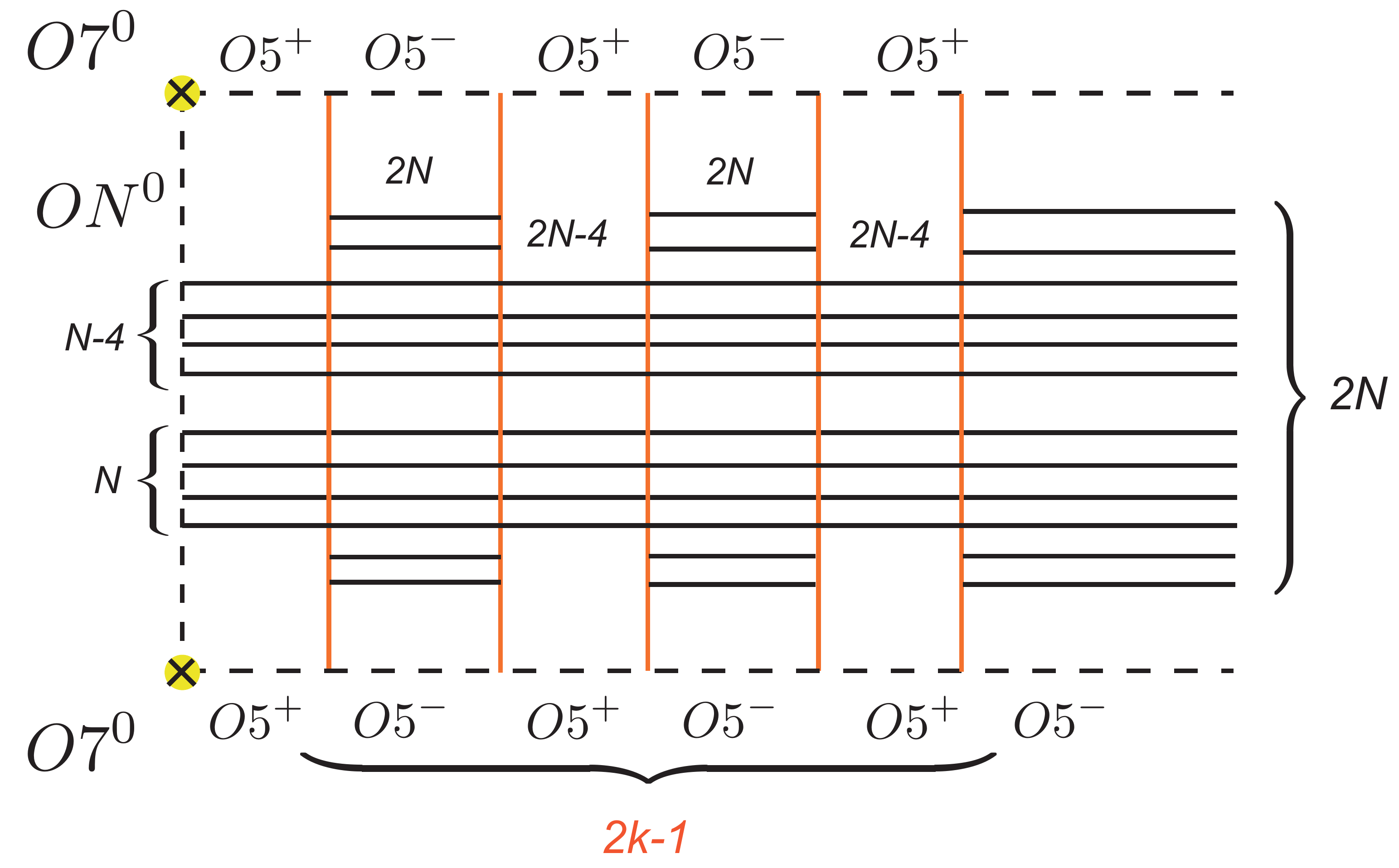}
\caption{A IIB version of Figure \ref{Fig:O8-O6p-ON0} after a circle compactification and T-duality. The vertical direction is the compactification direction, where two fractional $O7^0(=O7^-+4D7{\rm 's})$s are located at the two corners. 
} 
\label{Fig:sec4O5plus}
\end{figure}

\subsubsection{S-dual picture in Type IIB setup}\label{sec:S-dualONminus}
We now discuss a 5d description for this 6d SCFT on a tensor branch, by compactifying it on a circle and taking a T-duality. A straightforward type IIB setup is given in Figure \ref{Fig:sec4O5plus}.
\begin{figure}
\centering
\includegraphics[width=15cm]{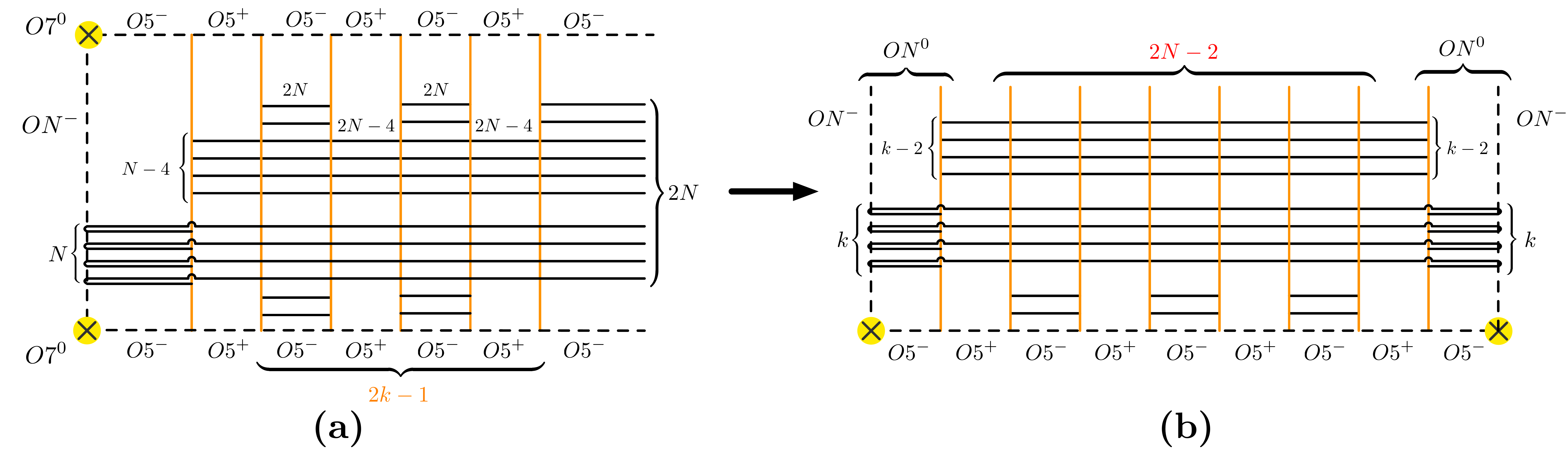}
\caption{A procedure leading to a 5d gauge theory description for the IIB brane configuration given in Figure \ref{Fig:sec4O5plus}. (a) A $ON^0$ is refined to be an $ON^-$-plane and an NS5-brane. 
(b) Taking (\ref{equivalence}) into account, we obtain S-dual description.
} 
\label{Fig:O5plusO5split}
\end{figure}
As mentioned above, 
one NS5-brane is away from the $ON^-$-plane and one fractional $O7^0$-plane is located at the intersection between the $ON^-$-plane and an $O5^-$-plane. 
This picture gives rise to a 5-brane configuration that D5-branes jump over this NS5-brane and reconnect to the same NS5-brane after passing through the $ON^-$-plane. See Figure \ref{Fig:O5plusO5split} (a).
The more precise configuration is the one explained in section \ref{sec:ONO5}.


After the S-duality, taking into account the transition (\ref{equivalence}), we 
obtain a 5-brane configuration with two $ON^-$-planes on the left and on the right, 
which comes from the S-dual of the $O5$-plane. 
See Figure \ref{Fig:O5plusO5split} (b).
This may lead to a 5d affine $D$-type quiver gauge theory,
\begin{align}\label{O5plussdual}
  5d~~[4]-Sp(k)-{\overset{\overset{\text{\normalsize$Sp(k-2)$}}{\textstyle\vert}}{SO(4k)}}-Sp(2k-2)- \cdots -{\overset{\overset{\text{\normalsize$Sp(k-2)$}}{\textstyle\vert}}{SO(4k)}}-Sp(k)-[4],
\end{align}
where there are two $Sp(k-2)$ as well as two $Sp(k)$ gauge groups in each end of the affine $D$-type quiver, and $(N-1)$ $SO(4k)$ gauge groups and $(N-2)$ $Sp(2k-2)$ gauge groups in the middle of the affine $D$-type quiver.

The dimension of the Coulomb branch moduli space of the proposed 5d quiver gauge theory \eqref{O5plussdual} is given by
\begin{align}
\underbrace{(k-2)\times 2}_{Sp(k-2)'s} +\underbrace{k\times 2}_{Sp(k)'s} + \underbrace{2k \times (N-1)}_{SO(4k)'s} + 
\underbrace{(2k-2)\times (N-2)}_{Sp(2k-2)'s} =
4Nk-2N-2k,
\end{align}
which matches  with the 6d parameter number \eqref{6dCoulomb4O6plus}.
The number of parameters associated with the global symmetry is
\begin{align}
\underbrace{4+4}_{\rm flavors} + \underbrace{N+2}_{Sp~{\rm instantons}}+\underbrace{N-1}_{SO~{\rm instantons}} = 2N+8+1_{I},
\end{align}
which also agrees with the number of the rank of the flavor symmetry of the 6d theory as expected.

\subsubsection{\texorpdfstring{The resolution of two $O7^-$ orientifolds}{two O7s}}\label{sec:twoO7s4O5plus}
As done in subsections \ref{sec:twoO7s} and \ref{sec:oneO7s}, we also consider the resolutions of $O7^-$-planes into a pair of 7-branes for either both or one $O7^-$-plane(s). The procedure of pulling out the $\bf B$ and $\bf C$ 7-branes together with flavor 7-branes from the orientifolds and then allocating these 7-branes throughout 5-branes is essentially same as that explained in subsections \ref{sec:twoO7s} and \ref{sec:oneO7s}. Here we hence discuss subtle issues or differences in obtaining 5d descriptions, compared with subsections \ref{sec:twoO7s} and \ref{sec:oneO7s}.

\begin{figure}
\centering
\includegraphics[width=12cm]{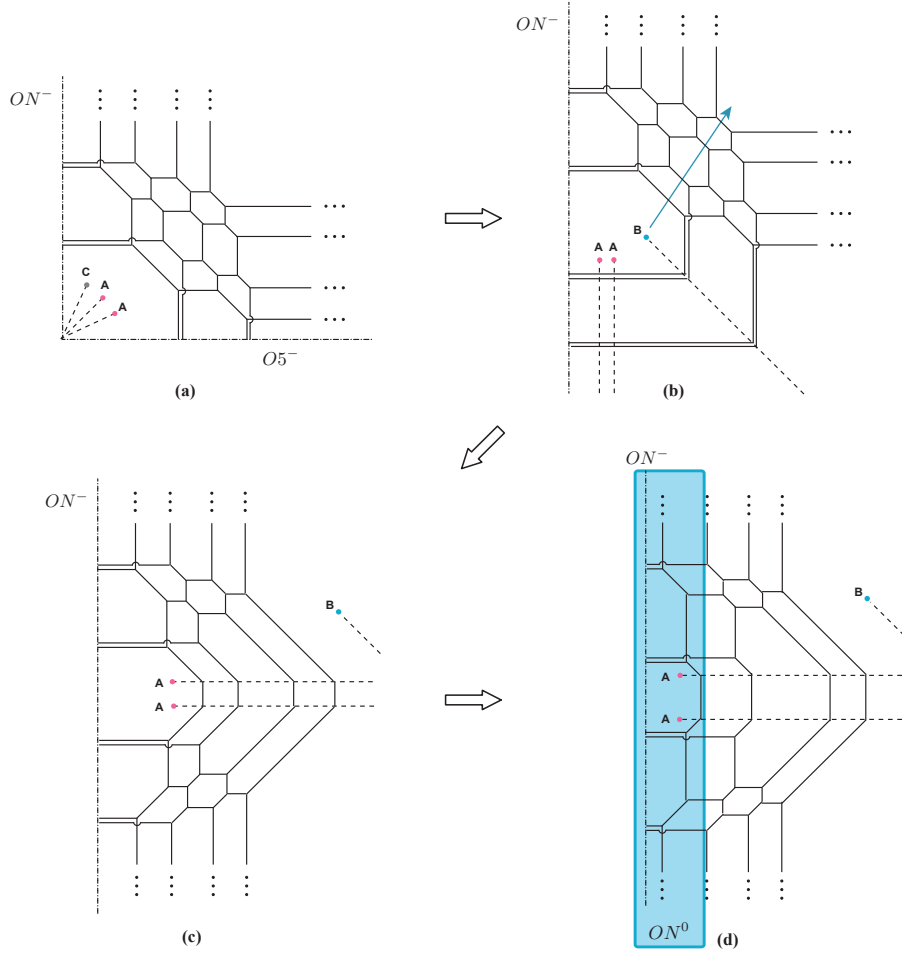}
\caption{A procedure of turning 5-brane configuration of $ON^-$, $O5^-$, and $O7^-$-planes into a 5-brane configuration of an $ON^0$ after resolving an $O7^-$ into a pair of two 7-branes.} 
\label{Fig:afterresolving}
\end{figure}
Let us consider resolutions of two $O7^-$-planes of the IIB brane configuration in Figure \ref{Fig:sec4O5plus} or more precisely in Figure \ref{Fig:O5plusO5split} (a). We focus on the configuration near one of the two $O7^-$-planes. 
Consider a lower half corner of Figure \ref{Fig:O5plusO5split} (a), where 
where we resolve $O7^-$-plane of the $O7^0$-plane to be
$\bf B$ and $\bf C$ 7-branes. Together with four D7-branes, they are initially stuck at the intersection between $ON^-$ and $O5^-$-planes and thus factional branes.  As explained in section \ref{sec:twoO7s}, these branes can leave the intersection as a combination of full 7-branes, for instance, $(\bf C, A,A)$ 7-branes. Recalling 5-brane configurations associated with $ON^-$-planes discussed in section \ref{sec:ONO5}, one easily sees the proper 5-brane configuration of intersecting $ON^-$ and $O5^-$-planes is given by a configuration shown in Figure \ref{Fig:afterresolving} (a). 
D5/NS5-branes jump over NS5/D5-branes to pass through the orientifolds and reconnect to the NS5/D5-branes. Using 7-brane monodoromies, we can rearrange the 7-branes to be $(\bf A, A, B)$ so as to convert the $O5^-$-plane into the $ON^-$-plane as shown in Figure \ref{Fig:afterresolving} (b). Then as done before, the $\bf B$ 7-brane can be allocated by crossing $(1,-1)$ 5-branes, with proper changes of 5-branes charges when the cut of $\bf B$ passes through. We also turn around the cuts of the two $\bf A$ 7-branes, which makes the nearest 5-brane from the $ON^-$-plane a generic NS5-brane, as shown in Figure \ref{Fig:afterresolving} (c). We then consider this NS5-brane as the 
 NS5-brane to form an $ON^0$-plane so that the resulting 5-brane configuration is of a 5-brane configuration with the $ON^0$-plane. See Figure \ref{Fig:afterresolving} (d).

In the same fashion, one can repeat the procedure for the remaining $O7^-$-plane in the other corner of the intersection between an $ON^-$-, an $O5^-$-, and an $O7^-$-plane. Putting together two resolved 7-branes at both corners gives rise to a 5d gauge theory description. We note that as in section \ref{sec:twoO7s}, two 5d gauge theory descriptions are possible depending on 
the relation between $N$ and $k$.

\begin{figure}
\begin{tabular}{cc}
\begin{minipage}{0.5\hsize}
\centering
\includegraphics[width=7cm]{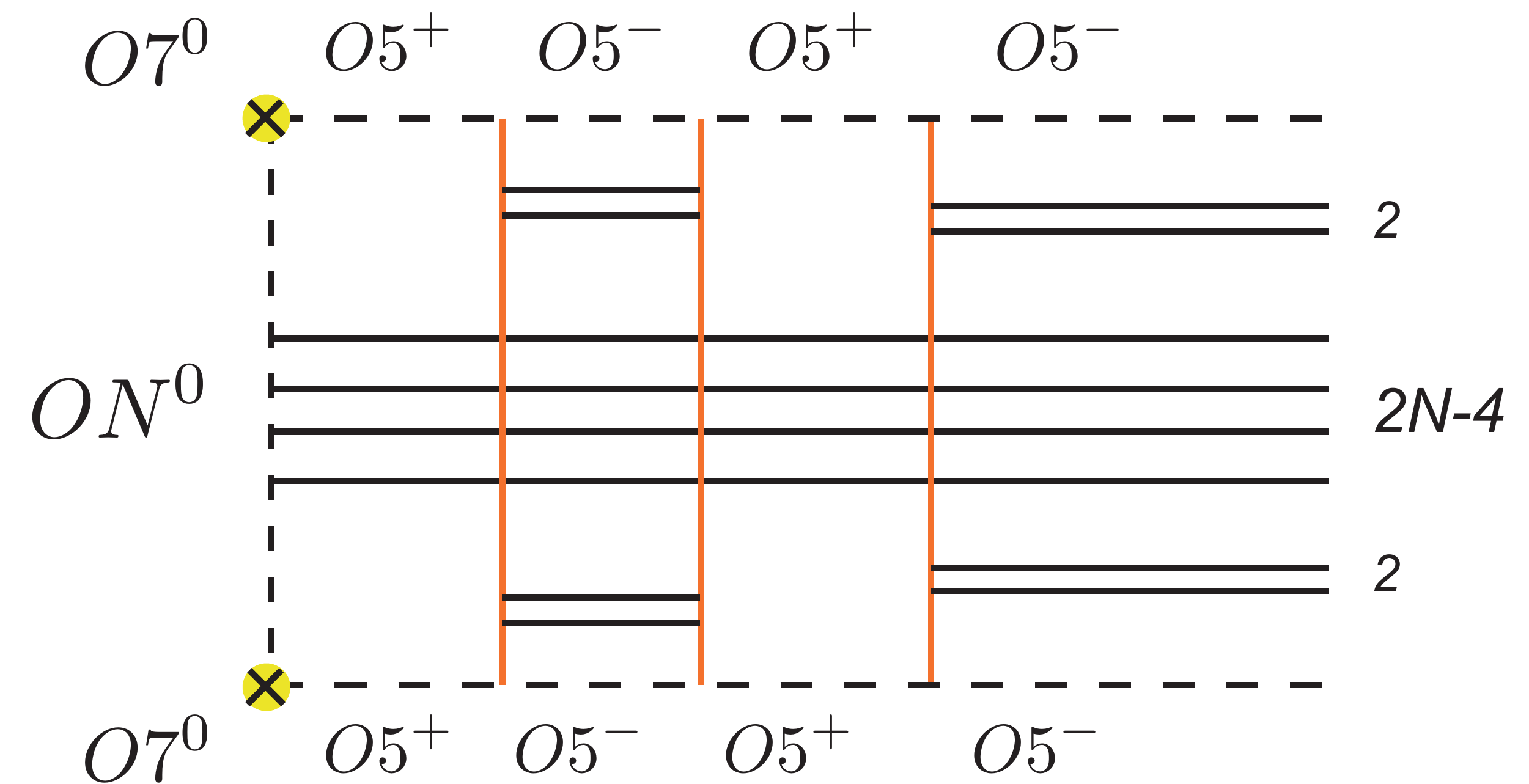}
\end{minipage}
\begin{minipage}{0.5\hsize}
\centering
\includegraphics[width=5cm]{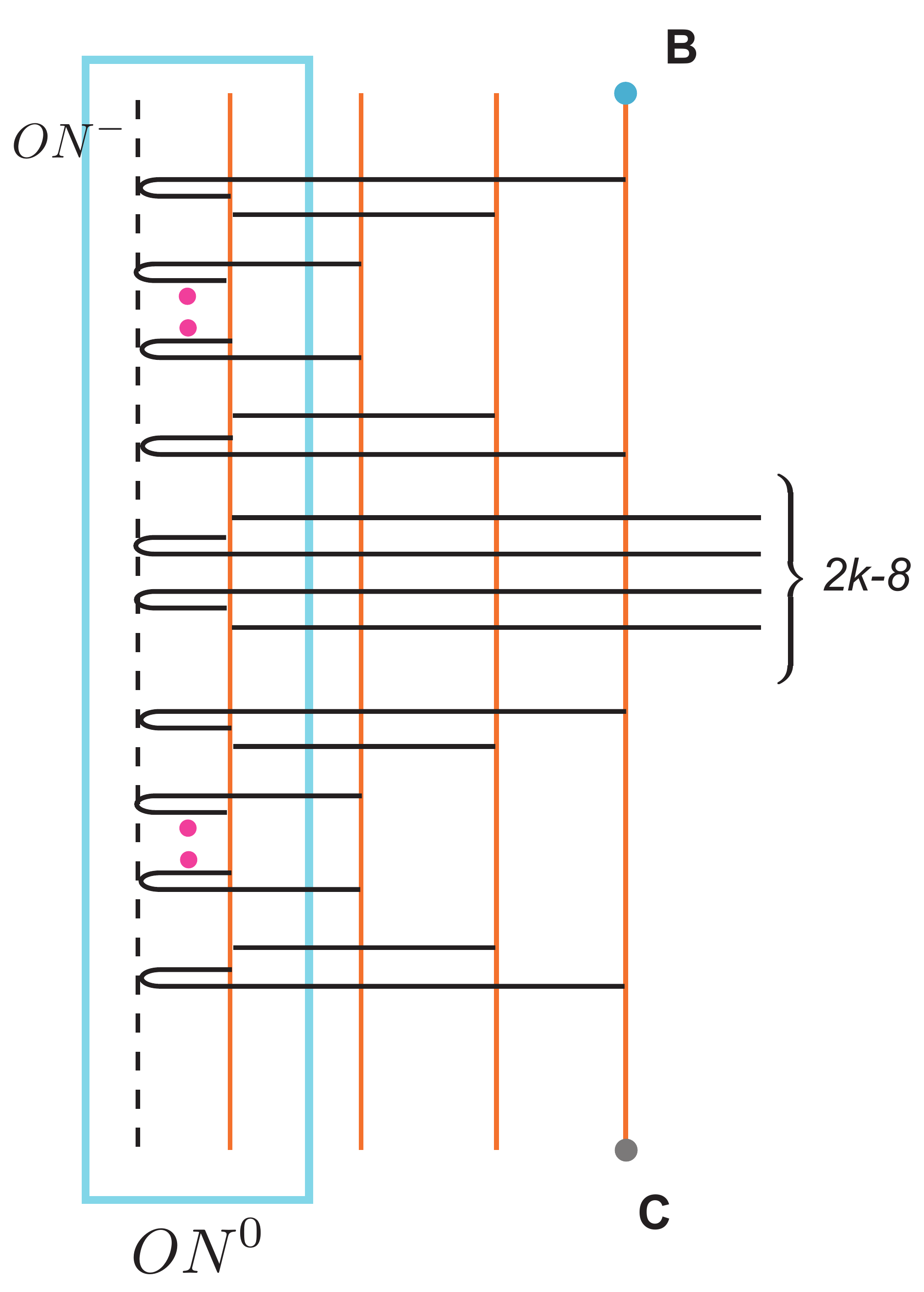}
\end{minipage}
\end{tabular}
\caption{Left: A Type IIB brane configuration with $ON^0$, $O5^+$, and two $O7^0$.
Right:A resulting 5d gauge theory description after resolving two $O7^-$-planes.
Here the brane configuration assumes that the number of D5-branes ($N$) and that of NS5-branes ($k-1$) satisfy $N\ge 2k-1$.}  
\label{Fig:O5plusfigure2}
\end{figure}
\noindent $\bullet$ When $N \ge 2k-1$, the corresponding 5d gauge theory description is given by
\begin{align}\label{O5plus5dNge2k}
5d~~
{[8]-SU(N+2k)}-{\overset{\overset{\text{\normalsize$SU(N+2k-4)$}}{~~~\textstyle\vert}}{SU(2N+4k-8)}}&-SU(2N+4k-12)-
\cdots-\cr
-\cdots-& SU(2N-4k+4) - [2N-4k+2],
\end{align}
where there $2k$ gauge nodes in the quiver. To compare to the 6d theory that we started with, we count the parameters of the resulting 5d theory. 
The dimension of the Coulomb branch moduli space is given by 
\begin{align}
2N+4k-6
+\sum^{2k-2}_{i=1} \big(2N+4k-9-4(i-1)\big) = 4Nk-2N-2k, 
\end{align}
which matches with \eqref{6dCoulomb4O6plus}. 
The number of the parameters of the global symmetry is 
\begin{align}
\underbrace{8}_{\rm left~flavors}+ \underbrace{2 + (2k-2)}_{\rm instantons}+ \underbrace{2+2k-3}_{\rm bi-fund.}+\underbrace{2N-4k+2}_{\rm right~flavors}
= 2N+8+1_I,
\end{align}
which again agrees with the number of the rank of the flavor symmetry of the 6d theory plus an extra parameter accounting for the $U(1)_I$ KK instanton. See Figure \ref{Fig:O5plusfigure2}.

It is also possible to perform the instanton operator analysis to see a subgroup of the enhanced flavor symmetry at the UV fixed point. In the 5d quiver \eqref{O5plus5dNge2k}, most of the gauge nodes satisfy $N_f = 2N_c$ whereas only one gauge node at the right end satisfies $N_f = 2N_c + 2$. The rule in \cite{Yonekura:2015ksa} implies that the one-instanton states create a Dynkin diagrams where two of the quiver diagrams of \eqref{O5plus5dNge2k} are connected by the Dynkin diagram of $SU(2N-4k+2)$. The resulting Dynkin diagram is an affine $D_{2N}$ Dynkin diagram. This agrees with the fact that the original 6d theory \eqref{6dO8mO6p1} has the $SO(4N)$ flavor symmetry. 

\begin{figure}
\centering
\includegraphics[width=4cm]{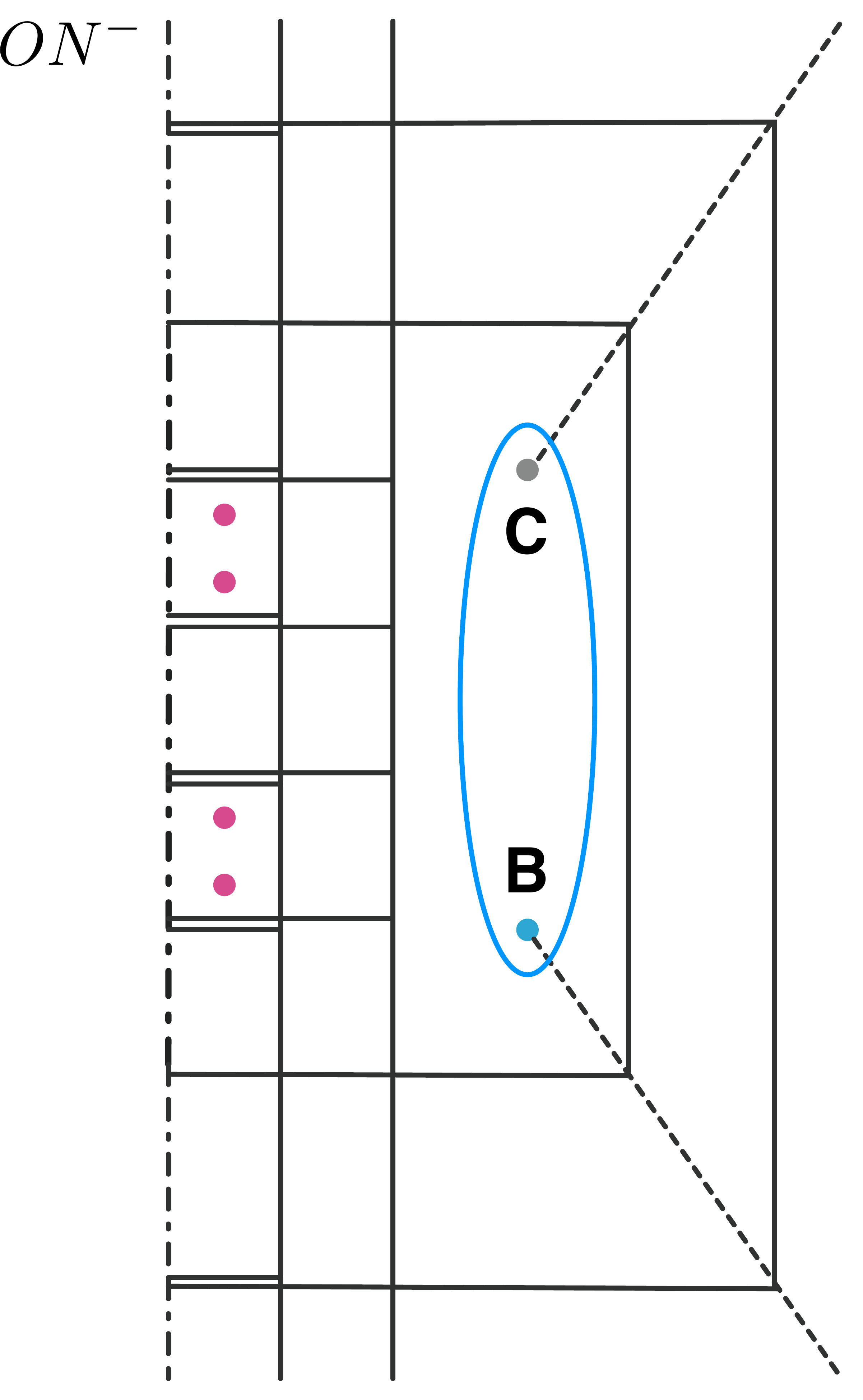}
\caption{A 5d gauge theory description for Type IIB brane configuration with $ON^0$ as a result of the resolution of two $O7^-$-planes in the Type IIB brane configuration with $ON^0$, $O5^+$, and two $O7^0$ on the left of Figure \ref{Fig:O5plusfigure2}. 
Here the brane configuration assumes that the number of D5-branes ($N$) and that of NS5-branes ($k-1$) satisfy $N\le 2k-1$, where $\bf B, C$ 7-branes are allocated in the same chamber of the Coulomb branch moduli space and in turn can form an $O7^-$ yielding an $Sp$ gauge theory associated with the chamber.} 
\label{Fig:O5plusfigure3}
\end{figure}
\vskip 0.5cm
\noindent $\bullet$ When $N < 2k-1$, we find that $\bf (B,C)$ 7-branes are located in a chamber of the Coulomb branch moduli space, which we can recombine to form an $O7^-$-plane, as shown in Figure \ref{Fig:O5plusfigure3}. Hence, the corresponding theory is given by an $D$-type quiver with two-pronged $SU$ gauge nodes of different ranks in one end, and an $Sp$ gauge node in the other end of the quiver
\begin{align}\label{O5plus5dNle2k}
5d~~
{[8]-SU(2k+N)}-{\overset{\overset{\text{\normalsize$SU(2k+N-4)$}}{~~~\textstyle\vert}}{SU(4k+2N-8)}}&-SU(4k+2N-12)-\cdots-\cr
-\cdots-&SU(4k-2N+4) - Sp(2k-N),
\end{align}
where there are $N$ $SU$-type gauge nodes in the quiver. 
The dimension of the Coulomb branch moduli space reads
\begin{align}
4k+2N-6 
+\sum^{N-2}_{i=1} \big(4k+2N-9-4(i-1)\big)+ 2k-N = 4Nk-2N-2k, 
\end{align}
and the number of the parameters associated with the global symmetry is given by 
\begin{align}
\underbrace{8}_{\rm left~flavors}+ \underbrace{2 + (N-2)+1}_{\rm instantons}+ \underbrace{2+N-3+1}_{\rm bi-fund.}
= 2N+8+1_I,
\end{align}
which includes the $U(1)$ part coming from the KK mode, as expected.

\subsubsection{\texorpdfstring{The resolution of only one $O7^-$ orientifold}{one O7}}\label{sec:oneO7O5plus}

\begin{figure}
\begin{tabular}{cc}
\begin{minipage}{0.5\hsize}
\centering
\includegraphics[width=7cm]{sec5221.pdf}
\end{minipage}
\begin{minipage}{0.5\hsize}
\centering
\includegraphics[width=5cm]{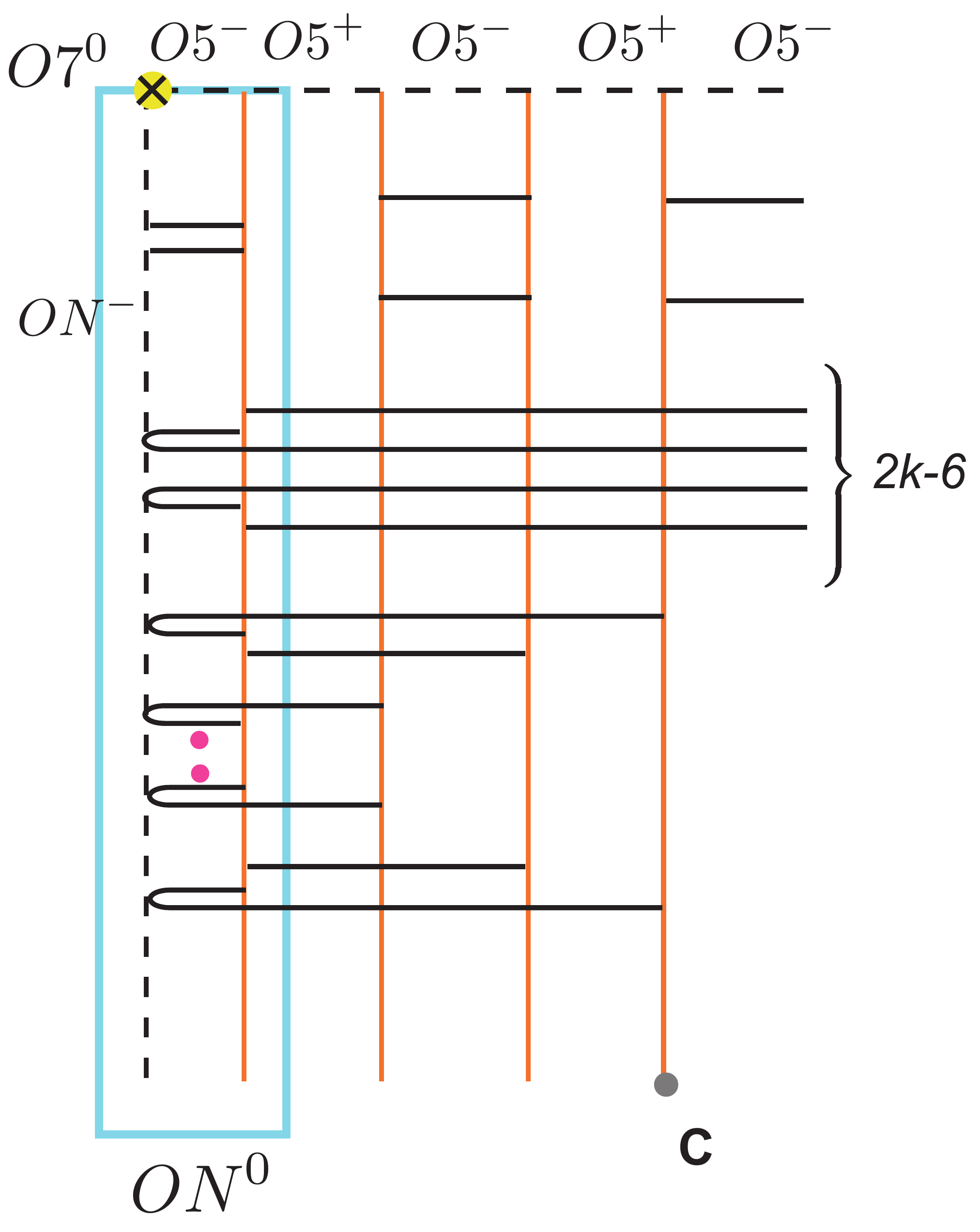}
\end{minipage}
\end{tabular}
\caption{Left: A type IIB brane configuration with $ON^0$, $O5^+$, and two $O7^0$. Right: A resulting 5d gauge theory description after resolving only one $O7^-$-plane.
Here the brane configuration assumes that the number of D5-branes ($N$) and that of NS5-branes ($k-1$) satisfy $2N\ge 2k-1$.}  
\label{Fig:O5plusfigure4}
\end{figure}

We now consider the case where we resolve only one $O7^-$-plane out of the two $O7^-$-planes. 
The way that one can obtain a gauge theory description is similar to section \ref{sec:twoO7s4O5plus}, except that the unresolved $O7^-$-plane is located at the intersection of $ON^-$ and $O5^-$-planes. See Figure \ref{Fig:O5plusfigure4}.
As discussed in section \ref{sec:oneO7s}, 5d gauge theory description is possible 
when $2N\ge 2k-1$, and is given by an $D$-type quiver with two-pronged $Sp$ gauge nodes of different ranks,
\begin{align}\label{O5plus1O7Nge2k}
5d~~
{[8]-Sp(N+k)}&-{\overset{\overset{\text{\normalsize$Sp(N+k-4)$}}{~~~\textstyle\vert}}{SO(4N+4k-8)}}-Sp(2N+2k-8)-\cr
&-SO(4N+4k-16)-Sp(2N+2k-12)-\cdots-\cr
&-SO(4N+4k-8i)-Sp(2N+2k-4-4i)-\cdots-\cr
&-SO(4N-4k+8)-Sp(2N-2k)-[2N-2k+1],
\end{align}
where there are $2k$ gauge nodes ($i=1,2,\cdots, k-1$). 
The dimension of the Coulomb branch moduli space for this quiver is given by
\begin{align}
&2N+2k-4 + \sum^{k-1}_{i=1} \big(2N+2k -4 -4(i-1)\big)+ \sum^{k-1}_{i=1} \big(2N+2k -8 -4(i-1)\big)\cr
& = 4Nk -2N-2k,
\end{align}
which exactly matches with \eqref{6dO8mO6p1}. The number of the parameters associated with global symmetry of the 5d theory \eqref{O5plus1O7Nge2k}
\begin{align}
\underbrace{8}_{\rm left~ flavors}\;+\underbrace{2k}_{\rm instantons}\;+\; \underbrace{2N-2k+1}_{\rm right~ flavors}\, 
=2N+8+1_{I},
\end{align}
where the $2k$ instanton factors are from the $2k$ gauge nodes. Again it agrees with the rank of the flavor symmetry of the 6d theory as expected. 

\bigskip

\section{Beyond circle compactification }\label{sec:O7pm}

So far, we have considered a simple circle compactification of 6d SCFTs and determined their 5d gauge theory description. In this section, we consider another one-dimensional compactification of a 6d SCFT. In this case also, it turns out that the different one-dimensional compactification still yields a 5d gauge theory.

The starting setup is a 6d $SU(N)$ gauge theory with $N_f = 2N$ flavors. This 6d theory can be realized either on $N$ D6-branes with two NS5-branes which fractionate the $N$ D6-branes in the $x_6$ direction in type IIA string theory or on an M5-brane probing an $A_N$-type singularity in M-theory. It is well-known that the standard circle compactification yields a 5d elliptic quiver theory where we have $N$ $SU(2)$ gauge nodes connected by bi-fundamental hypermultiplets. 

\begin{figure}
\centering
\includegraphics[width=6cm]{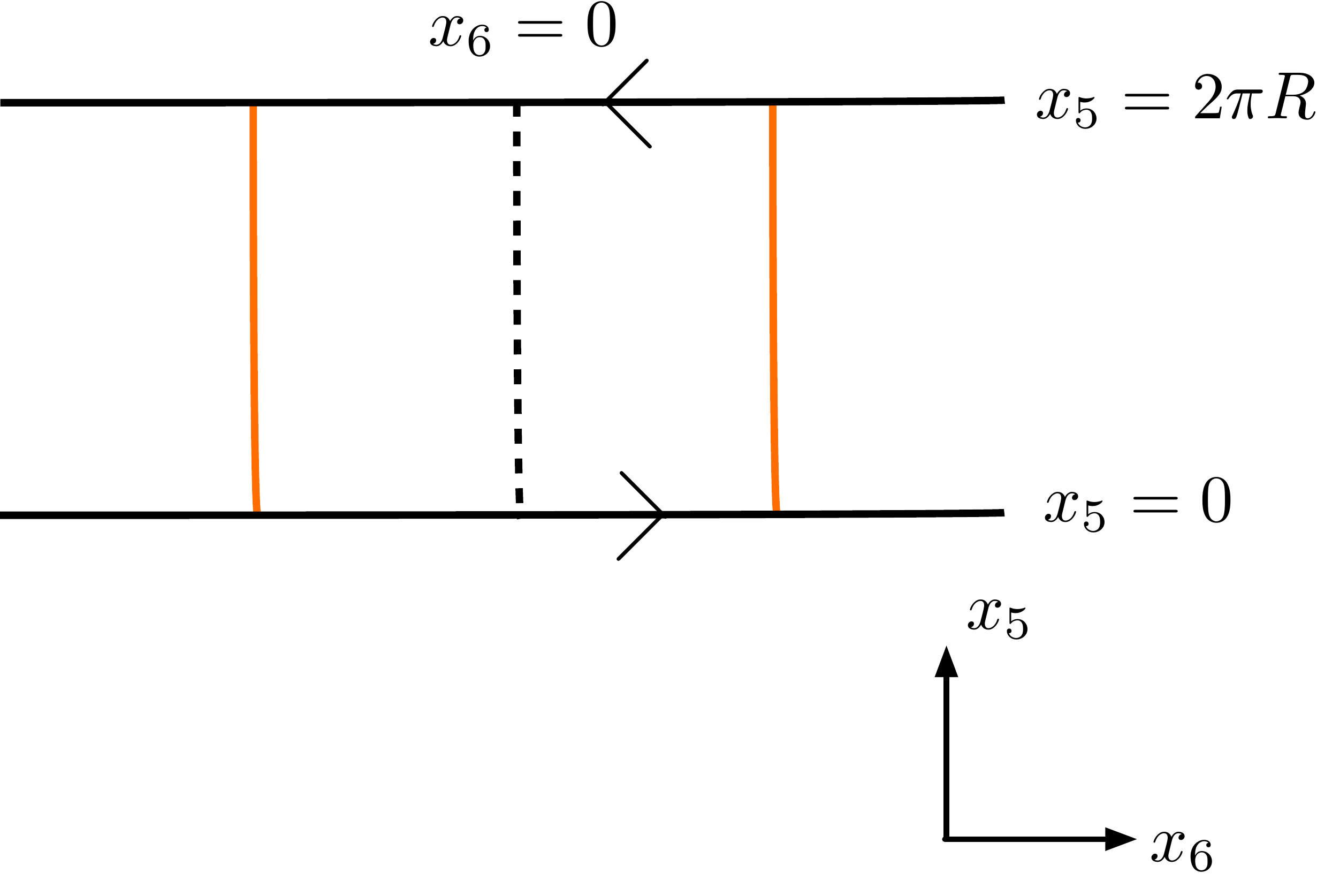}
\caption{The circle compactification with a reflection. The lower horizontal line is identified with the upper horizontal line after the reflection with respect to the axis $x_6=0$. The dotted line represents the reflection axis. The orange lines are NS5-branes. Due to the reflection, we have only singly connected NS5-brane. The $N/2$ D6-branes fill the space between the $x_5=0$ and $x_5=2\pi R$.}
\label{fig:reflection}
\end{figure}

Let us consider a different one-dimensional compactification. We start from the configuration of the $N$ D6-branes fractionated by the two NS5-branes in type IIA string theory, which is a special case with two NS5-branes in Figure \ref{fig:base}. Then, we compactify the brane setup along the $x_5$ direction. In the case of the circle compactification, we identify the configuration at $x_5=0$ with that at $x_5=2\pi R$ where $R$ is the radius of the $S^1$. This time we consider an identification with a reflection with respect to an axis along $x_6 = 0$. Here we choose $x_6=0$ as the middle point between the two NS5-branes, and hence it is a symmetry of the brane configuration. We call this compactification a twisted circle compactification. The schematic picture of the geometry in the $x_6$-$x_5$-plane is depicted in Figure \ref{fig:reflection}.  In other words, the compactification geometry in the $x_5$-$x_6$-plane is a M\"obius strip with an infinite width. 
Due to the identification, the original two NS5-branes are connected to each other and become a single NS5-brane.

\begin{figure}
\centering
\includegraphics[width=6cm]{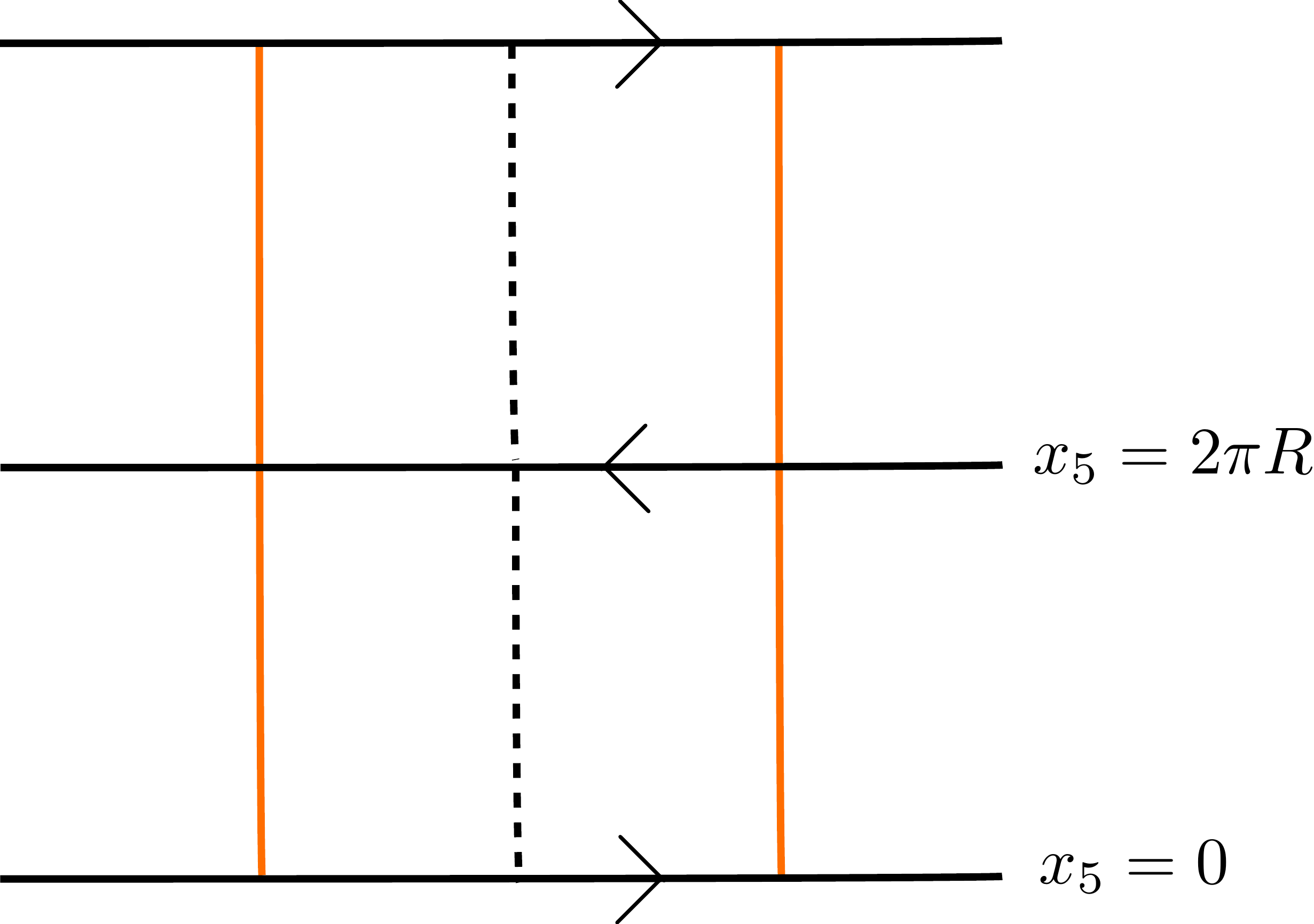}
\caption{The double cover of the configuration in Figure \ref{fig:reflection}.}
\label{fig:doublecover}
\end{figure}

\begin{figure}
\centering
\includegraphics[width=6cm]{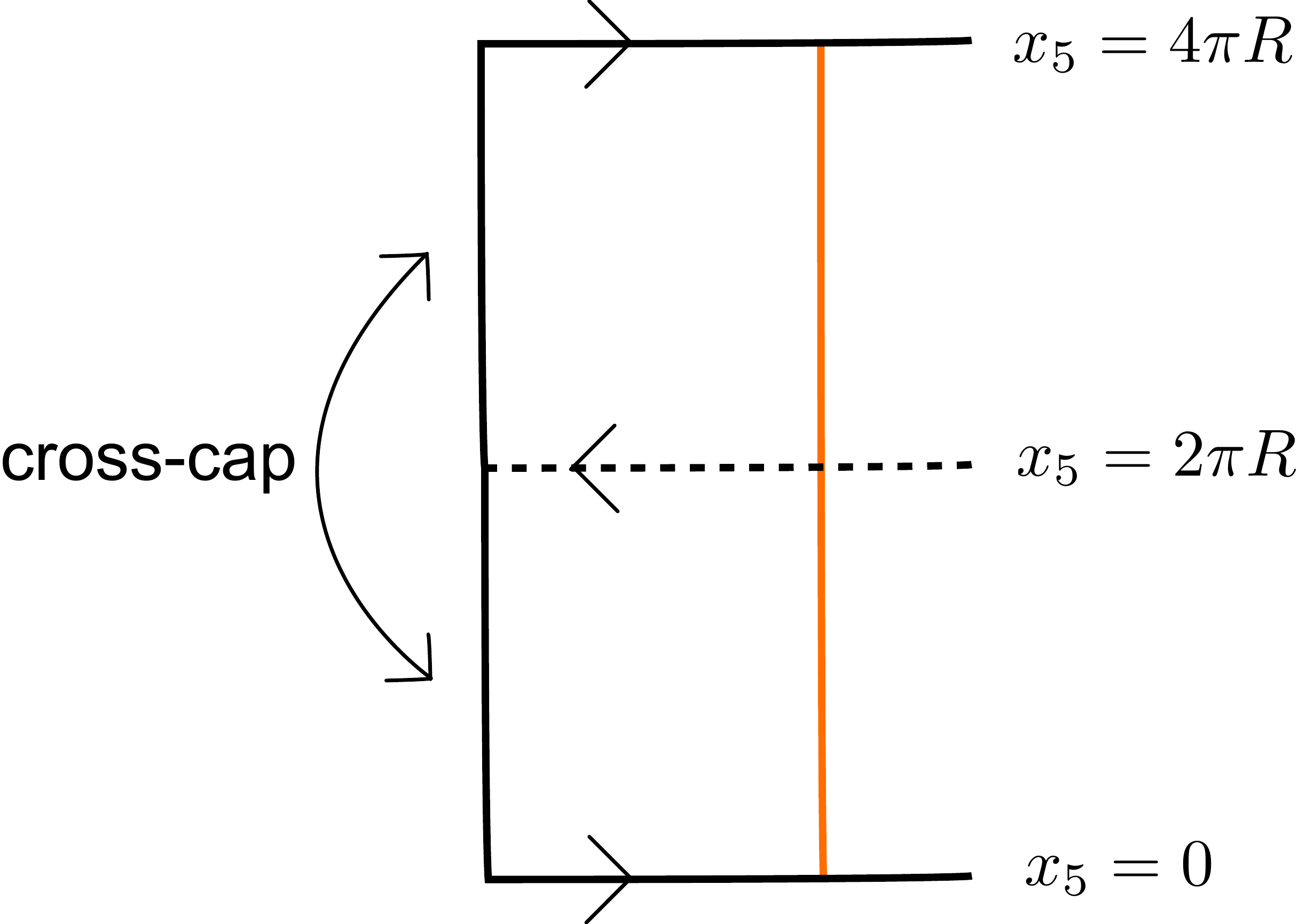}
\caption{Taking the right half of the double cover in Figure \ref{fig:doublecover} as a fundamental region.}
\label{fig:crosscap}
\end{figure}

The geometry of the compactification in the $x_6$-$x_5$-plane can be seen in another way. Let us consider a double cover of the geometry as in Figure \ref{fig:doublecover}. 
The original configuration in Figure \ref{fig:reflection} is obtained by taking the lower half of the double cover in Figure \ref{fig:doublecover} as a fundamental region. It is also possible to take a different fundamental region. Let us choose the right half of the double cover. Then, it can be regarded as a semi-infinite cylinder with a cross-cap at one boundary as in Figure \ref{fig:crosscap}.

\begin{figure}
\centering
\includegraphics[width=5cm]{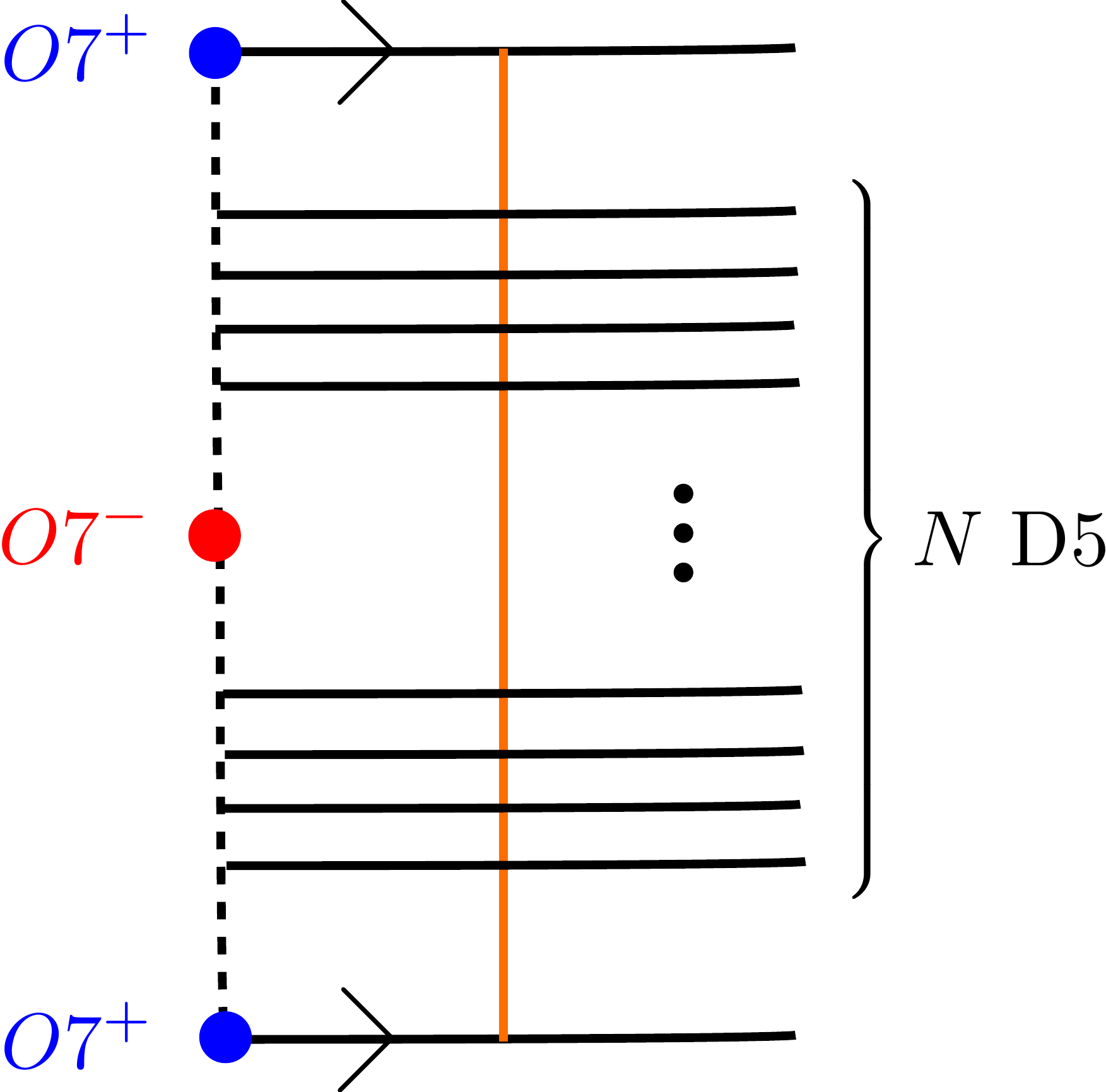}
\caption{The configuration after performing the T-duality along the $x_5$ direction to the configuration in Figure \ref{fig:crosscap}.}
\label{fig:crosscapTdual}
\end{figure}

Now we perform T-duality along the $x_5$ direction to go to a 5d description. In fact, it is known that the T-duality along the $S^1$ with a cross-cap produces a pair of an $O^-$-plane and an $O^+$-plane \cite{Keurentjes:2000bs}. Therefore, the brane configuration after the T-duality is that we have a pair of an $O7^-$-plane and an $O7^+$-plane and $N$ fractional D5-branes and also a single NS5-brane. The schematic picture is depicted in Figure \ref{fig:crosscapTdual}. 

\begin{figure}
\centering
\includegraphics[width=7cm]{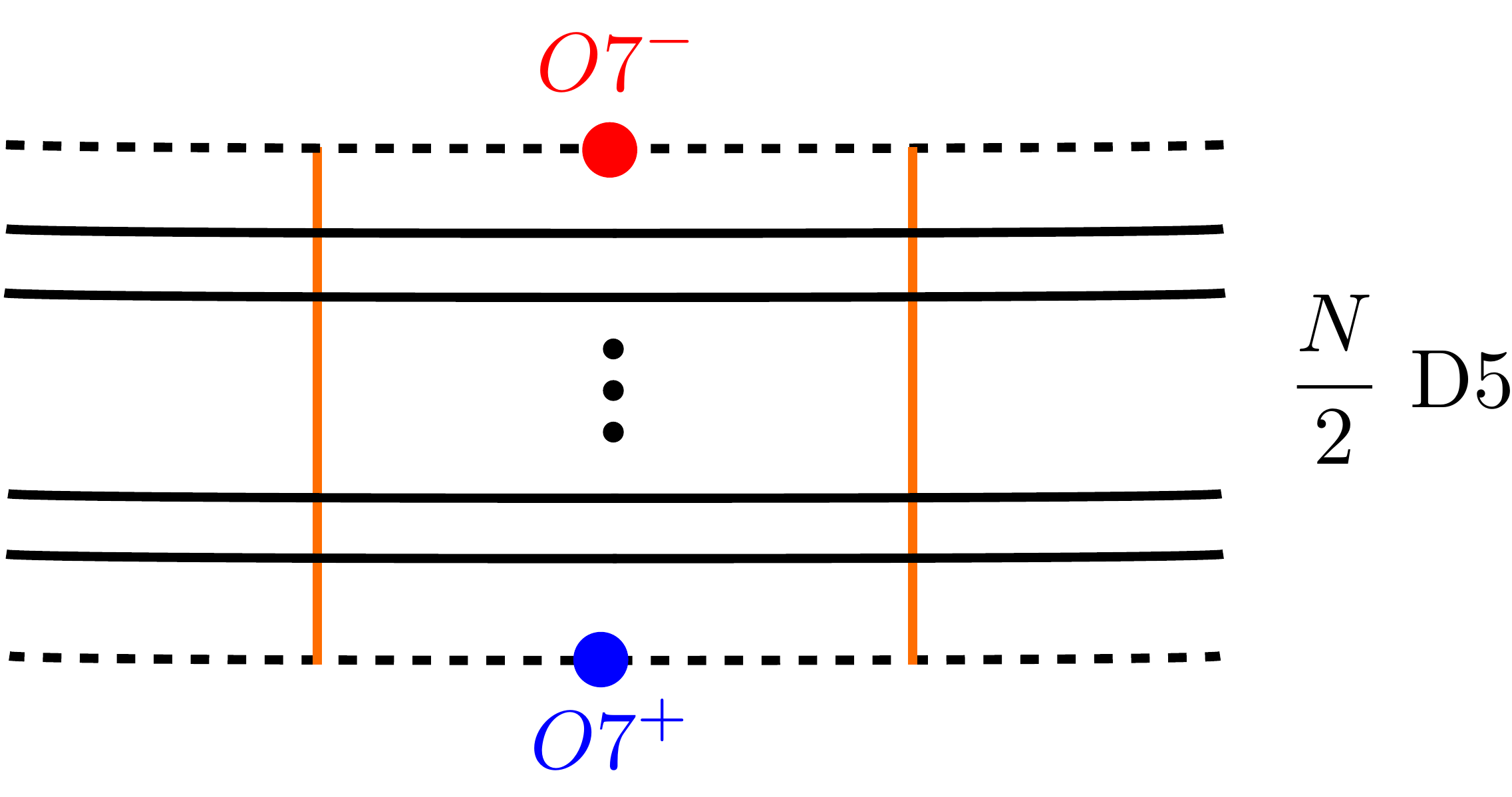}
\caption{The same 5-brane web diagram as the one in Figure \ref{fig:crosscapTdual} but we take a different fundamental region. When $N$ is odd, we have one fractional color D5-brane stuck on the $O7^+$-plane. The total number of the flavor D5-branes is always $N$. }
\label{fig:5dSO}
\end{figure}

In order to read off the gauge theory content from the diagram, we choose a different fundamental region as in Figure \ref{fig:5dSO}. 
When $N$ is even, then we have $\frac{N}{2}$ D5-branes between two NS5-branes. When $N$ is odd, then one of the color D5-brane is stuck on the $O7^+$-plane and we have $\frac{N-1}{2}$ D5-branes away from the $O7^+$-plane between the two NS5-branes. However, we have always $N$ flavor D5-branes.

\begin{figure}[t]
\begin{tabular}{cc}
\begin{minipage}{0.5\hsize}
\begin{center}
\includegraphics[width=6cm]{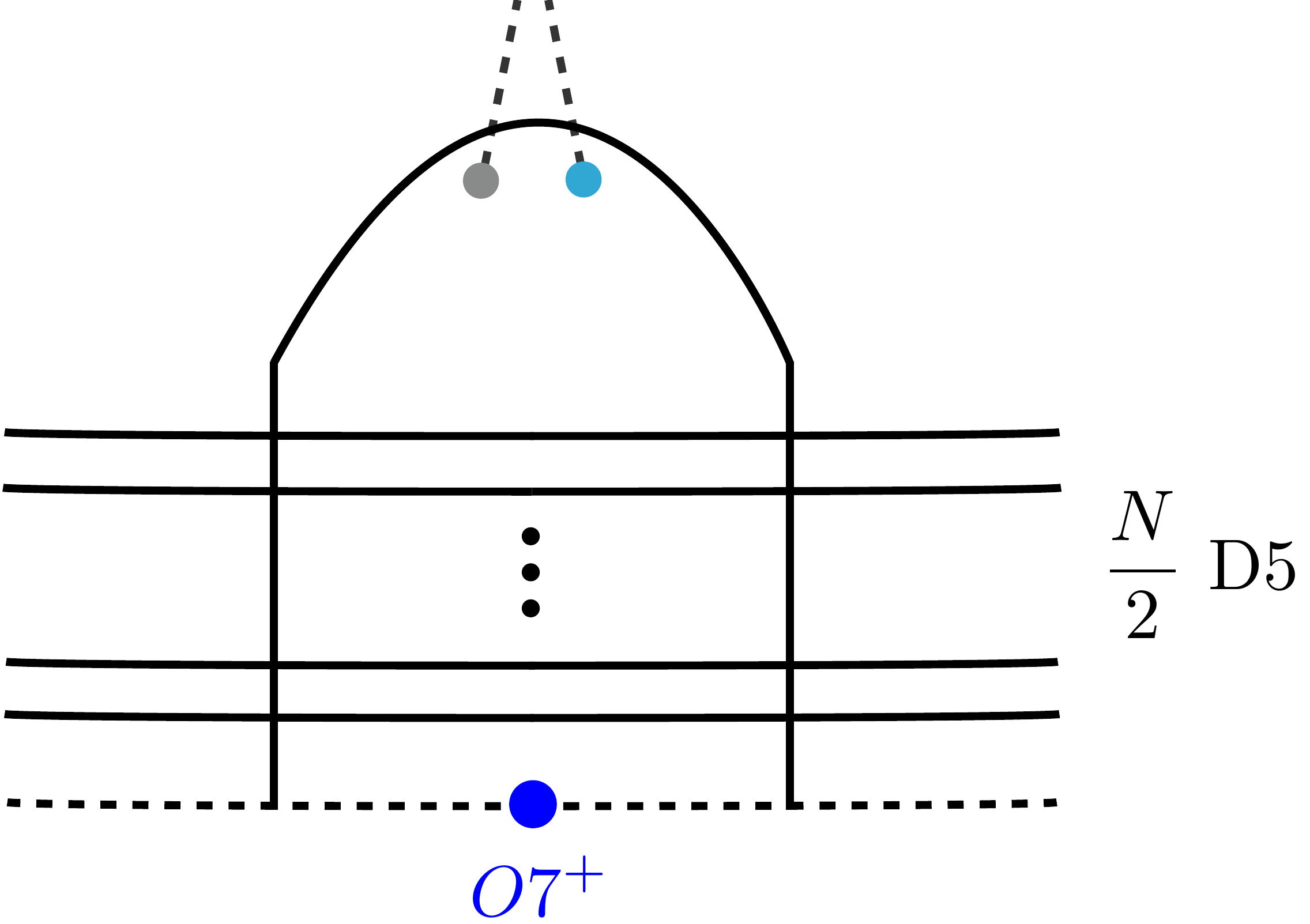}
\end{center}
\end{minipage}
\begin{minipage}{0.5\hsize}
\begin{center}
\includegraphics[width=6cm]{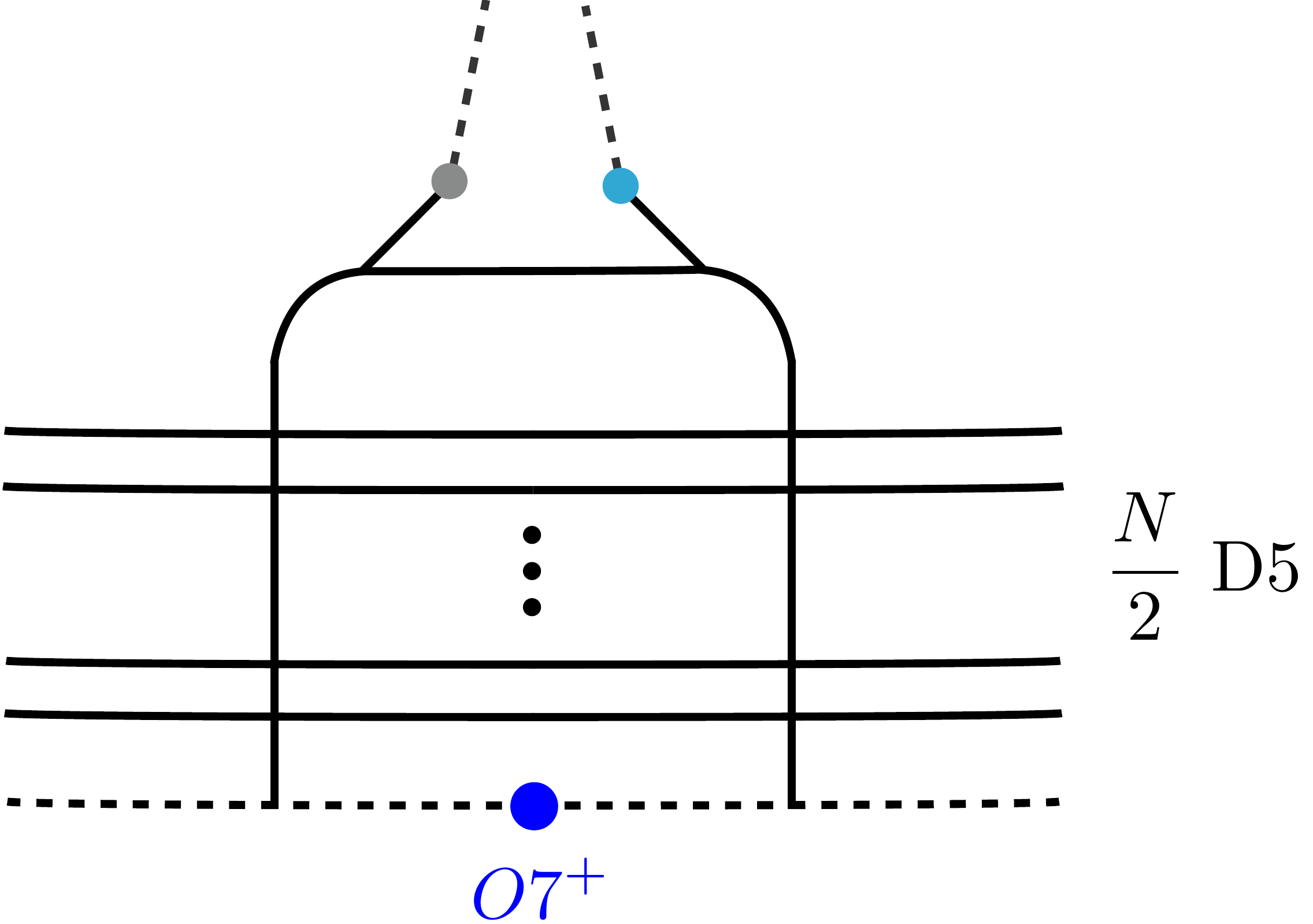}
\end{center}
\end{minipage}
\end{tabular}
\caption{Left: The brane configuration after splitting the $O7^-$-plane in Figure \ref{fig:5dSO}. When $N$ is odd, we have one fractional color D5-brane stuck on the $O7^+$-plane. Right: The brane configuration after pulling out the two 7-branes in the left figure. The 5-brane web yields a 5d $SO(N+2)$ theory with $N_v=N$ where there are $N+2$ color D5-branes including the mirror images and $N$ flavor D5-branes.}
\label{fig:5braneSO}
\end{figure}

By the strong coupling effect of the string coupling, an $O7^-$-plane splits into two 7-branes. After the resolution of the $O7^-$-
plane, we obtain a 5-brane web configuration with one $O7^+$-plane. The splitting process is depicted in Figure \ref{fig:5braneSO}. 

We have $N$ semi-infinite D5-branes and $\frac{N}{2}+1$ parallel finite D5-branes. This is nothing but a 5-brane web that yields a 5d $SO(N+2)$ gauge theory with $N_v = N$ hypermultiplets in the vector representation. Therefore, we propose 
 that the UV completion of the 5d $SO(N+2)$ gauge theory with $N_v=N$ hypermultiplets in the vector representation is the 6d $SU(N)$ gauge theory with $N_f = 2N$ flavors. In other words, the twisted circle compactification of the 6d $SU(N)$ gauge theory with $N_f=2N$ flavors yields a 5d $SO(N+2)$ gauge theory with $N_v=N$ hypermultiplets in the vector representation. 

It was originally discussed that the 5d $SO(M)$ gauge theory theory has 5d fixed point when the number of the hypermultiplets in the vector representation satisfies $N_v\le M-4$, based on the the requirement that the effective gauge coupling is non-negative in all the region of the Coulomb branch moduli space \cite{Intriligator:1997pq}. 
However, it was recently noticed that 5-brane web configurations predict the existence of more 5d theories with a 5d fixed point. Indeed it was pointed out that for the 5d $SO(M)$ theories, there exists a brane configuration involving an $O7^+$-plane with $N_v=M-3$ \cite{Bergman:2015dpa}, which suggests a new 5d fixed point beyond the bound from \cite{Intriligator:1997pq}. Our analysis shows that when one adds one more hypermultiplet in the vector representation of the $SO(M)$, the UV completion of the 5d theory is a 6d SCFT, namely the 6d $(A_{M-3}, A_{M-3})$ minimal conformal matter.

One can see another support for the claim from the flavor symmetry of the system. Before the twisted circle compactification, the flavor symmetry is $SU(2N) = A_{2N-1}$. The standard circle compactification yields the affine $A_{2N-1}$ algebra. Therefore, a natural expectation after the twisted circle compactification is that we obtain the twisted affine algebra $A_{2N-1}^{(2)}$. This in fact completely agrees with the flavor symmetry realized in the 5-brane web configuration.

For a 5d theory on a 5-brane web, the flavor symmetry is realized on 7-branes. We have now $N$ D7-branes attached to the $N$ semi-infinite D5-branes and also a pair of an $O7^-$-plane and an $O7^+$-plane. When $N$ D7-branes are on top of the $O7^-$-plane, we obtain an $SO(2N)$ symmetry. On the other hand, when $N$ D7-branes are on top of the $O7^+$-plane, we obtain an $Sp(N)$ symmetry. Hence, the full structure of the flavor symmetry should include both the $SO(2N)$ and the $Sp(N)$ symmetry. There is no finite group which includes both and the symmetry structure is in fact the twisted affine algebra $A_{2N-1}^{(2)}$. Indeed, it has been known that the $N$ Dp-branes with a pair of an $Op^-$-plane and an $Op^+$-plane yields the twisted affine algebra $A_{2N-1}^{(2)}$ \cite{Hanany:2001iy}. This also agrees with the one-instanton analysis in \cite{Zafrir:2015uaa}.

\begin{figure}
\centering
\includegraphics[width=15cm]{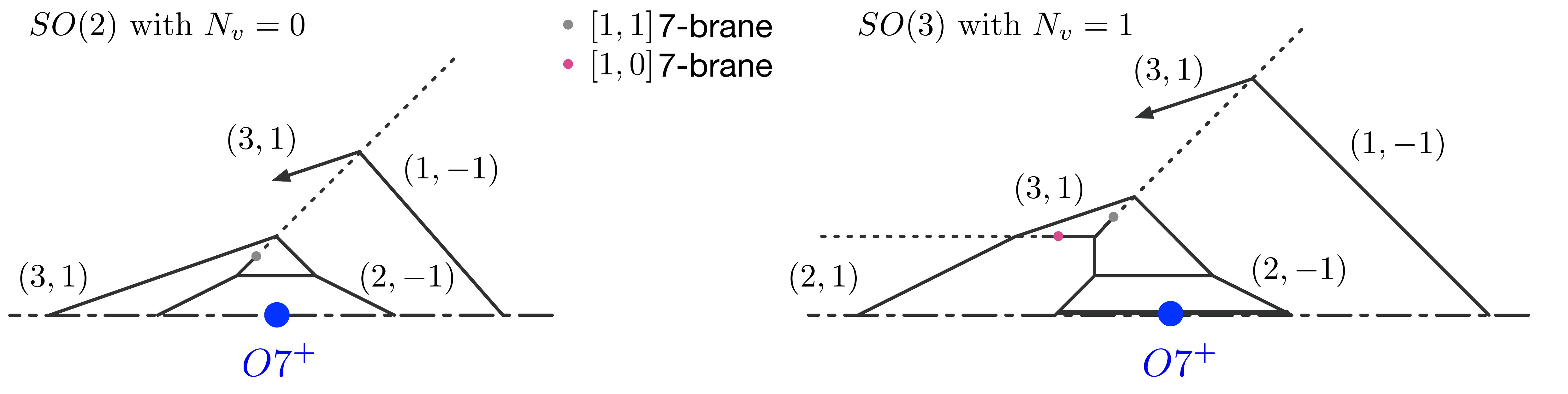}
\caption{Left: A Tao brane configuration for $SO(2)$ theory with $N_v=0$. 
As a rank one theory, there is only one D5 brane above the $O7^+$-plane. One takes a branch cut of a $[1,1]$ 7-brane, and let us pull out a $[1,-1]$ 7-brane along the direction of its $(p,q)$ charge. This 7-brane then must pass through the cut of the $[1,1]$ 7-brane, which then turns to a $[3,1]$ 7-brane. When passing the bottom of the fundamental region, it appears as again a $[1,-1]$ 7-brane on the right hand side of the $O7^+$ due to the action of $O7^+$ which will pass through the cut off the $[1,1]$ 7-brane then becomes again a $[3,1]$ 7-brane. A repeated application of pulling 7-branes leads to a Tao configuration with a constant period. ~Right: A Tao diagram of the brane configuration for $SO(3)$ theory with $N_v=1$. In this case, a factional D5-brane is stuck at the orientifold and one can introduce the cut of a $[1,0]$ 7-brane in addition to the cut of a $[1,1]$ 7-brane. In the same way, by pulling out a $[1,1]$ 7-brane which experiences the 7-brane cuts, one see that it also makes a Tao configuration. 
the $[1,0]$ 7-branes} 
\label{fig:SO2and3Tao}
\end{figure}

We note that, following \cite{Kim:2015jba,Hayashi:2015fsa}, one sees that the web diagram for $SO(2)$ with $N_v=0$ and $SO(3)$ with $N_v=1$ shows a rotating spiral shape (Tao sturcture), as shown in Figure \ref{fig:SO2and3Tao}.
A straightforward generalization for generic $SO(M)$ does not seem to work, but it does not mean that there is no Tao structure, as Tao diagram may appear depending on the number of 7-brane cuts which other 7-branes are passing through.

\begin{figure}
\centering
\includegraphics[width=6cm]{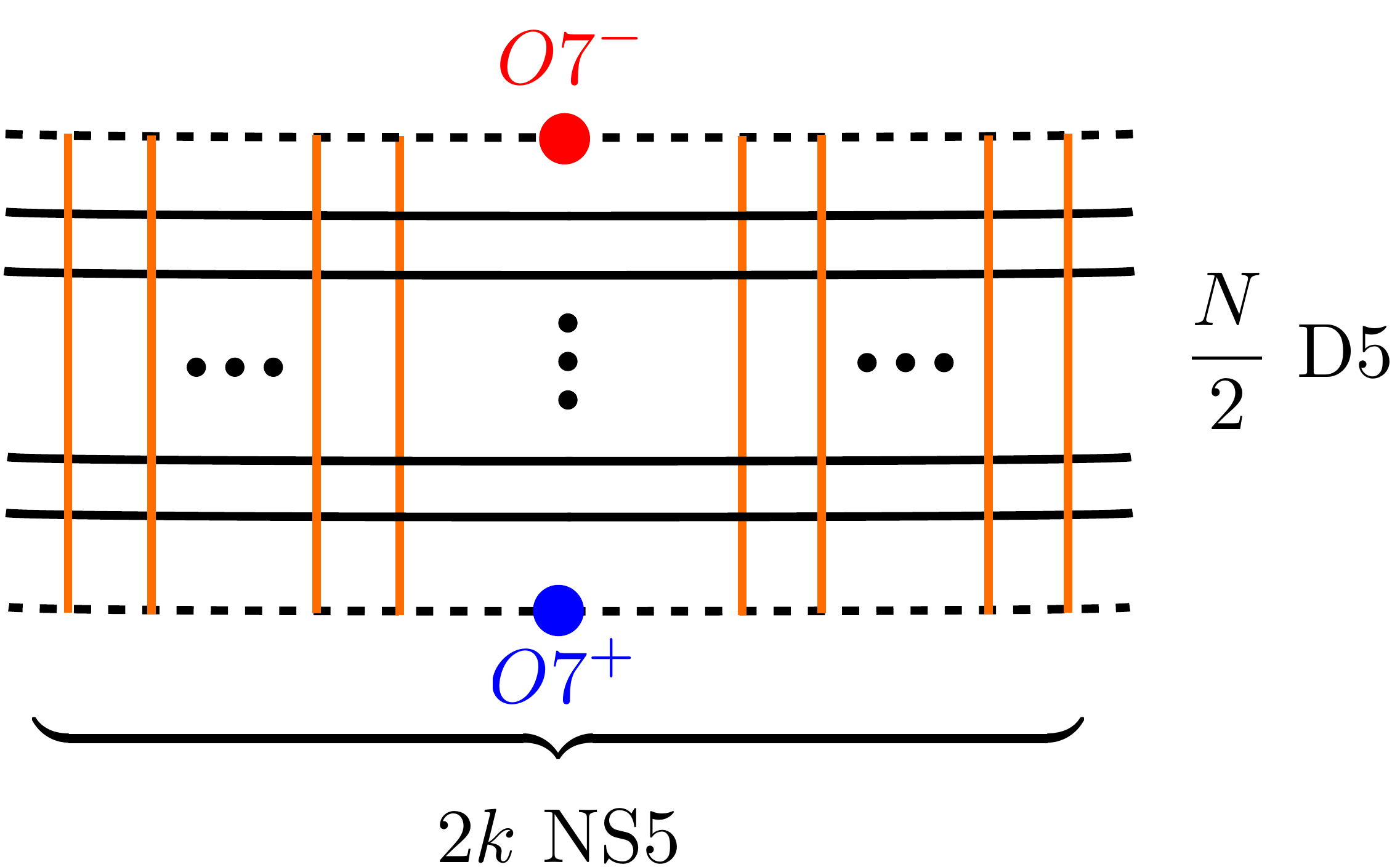}
\caption{The 5-brane web configuration after after compactifying the brane setup in Figure \ref{fig:base} on an $S^1$ with the twist and performing T-duality along the $x_5$ direction. }
\label{fig:5dSO2}
\end{figure}

It is straightforward to generalize the analysis to a 6d $SU(N)$ quiver theory given in \eqref{6dAtype}. 
We consider a 6d theory realized on the brane setup given by Figure \ref{fig:base}.
As natural generalization, we first consider $2k$ NS5-branes as well as $N$ D6-branes. Then we compactify the configuration on an $S^1$ in the $x_5$ direction with identification by a reflection and then perform T-duality along the $S^1$. The analysis is essentially the same in the case with $k=1$, and we obtain a 5-brane web given in Figure \ref{fig:5dSO2}. 
We have $\frac{N}{2}$ D5-branes, $2k$ NS5-branes and a pair of an $O7^-$-plane and an $O7^+$-plane.

\begin{figure}
\centering
\includegraphics[width=7cm]{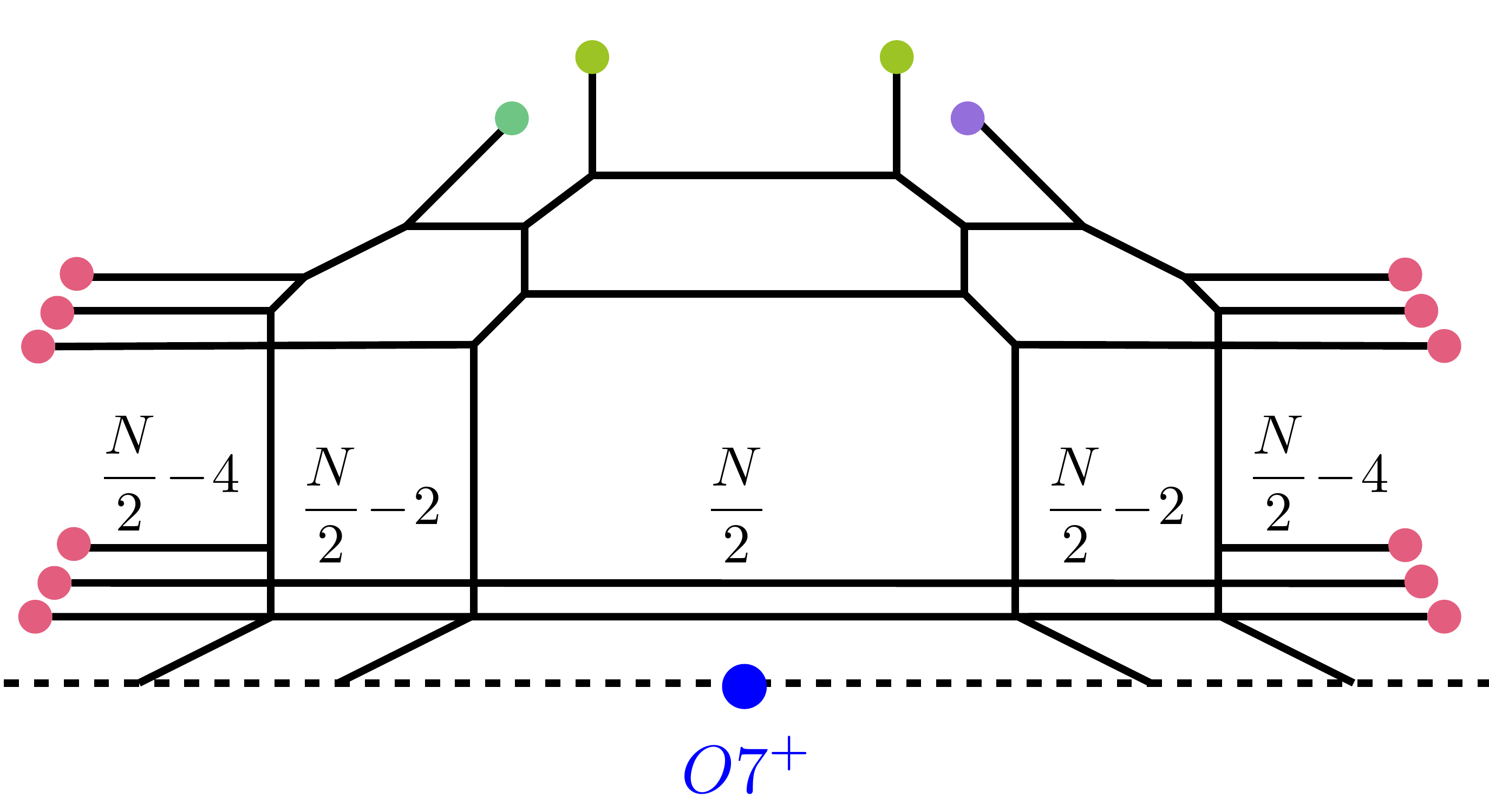}
\caption{A schematic picture of the 5-brane web configuration after splitting the $O7^-$-plane in Figure \ref{fig:5dSO2} and pulling the two 7-branes outside of the 5-brane loops. For simplicity, we draw the case of $k=2$. }
\label{fig:5dSOSU}
\end{figure}

Again at the quantum level, the $O7^-$-plane splits into two 7-branes.  This process creates 5-brane loops. Then, we try to pull out the two 7-branes outside of the 5-brane loops so that the final 5-brane web configuration yields a 5d gauge theory description. The process is essentially the same as the one done in section 4.1.2 in \cite{Hayashi:2015zka}. We will not repeat the analysis here and simply present the final 5-brane web configuration in Figure \ref{fig:5dSOSU}.
 The resulting 5d theory is 
 \begin{equation}
 5d~~SO(N+2k) - SU(N+2k-4) - SU(N+2k-8) - \cdots - SU(N-2k+4) - [N-2k+2] \label{5dSOSU}
 \end{equation} 
The number of the $SU$ gauge nodes is $k-1$. Hence, we claim that the 5d $SO$-$\prod SU$ quiver theory \eqref{5dSOSU} has the 6d UV completion given by the 6d $SU(N)$ quiver theory with the $2k-1$ gauge nodes.

\begin{figure}[t]
\begin{tabular}{cc}
\begin{minipage}{0.5\hsize}
\begin{center}
\includegraphics[width=6cm]{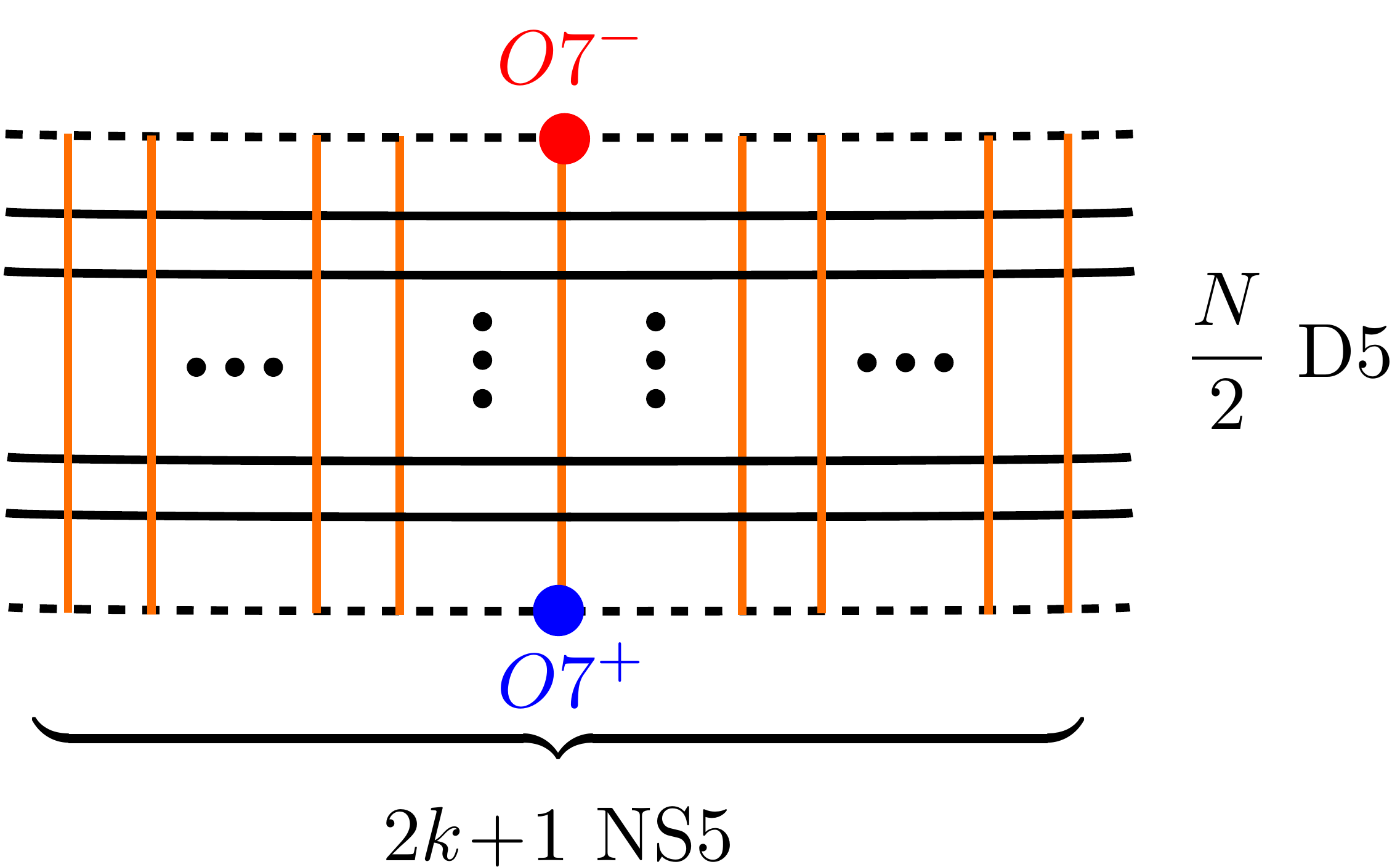}\end{center}
\end{minipage}
\begin{minipage}{0.5\hsize}
\begin{center}
\includegraphics[width=7cm]{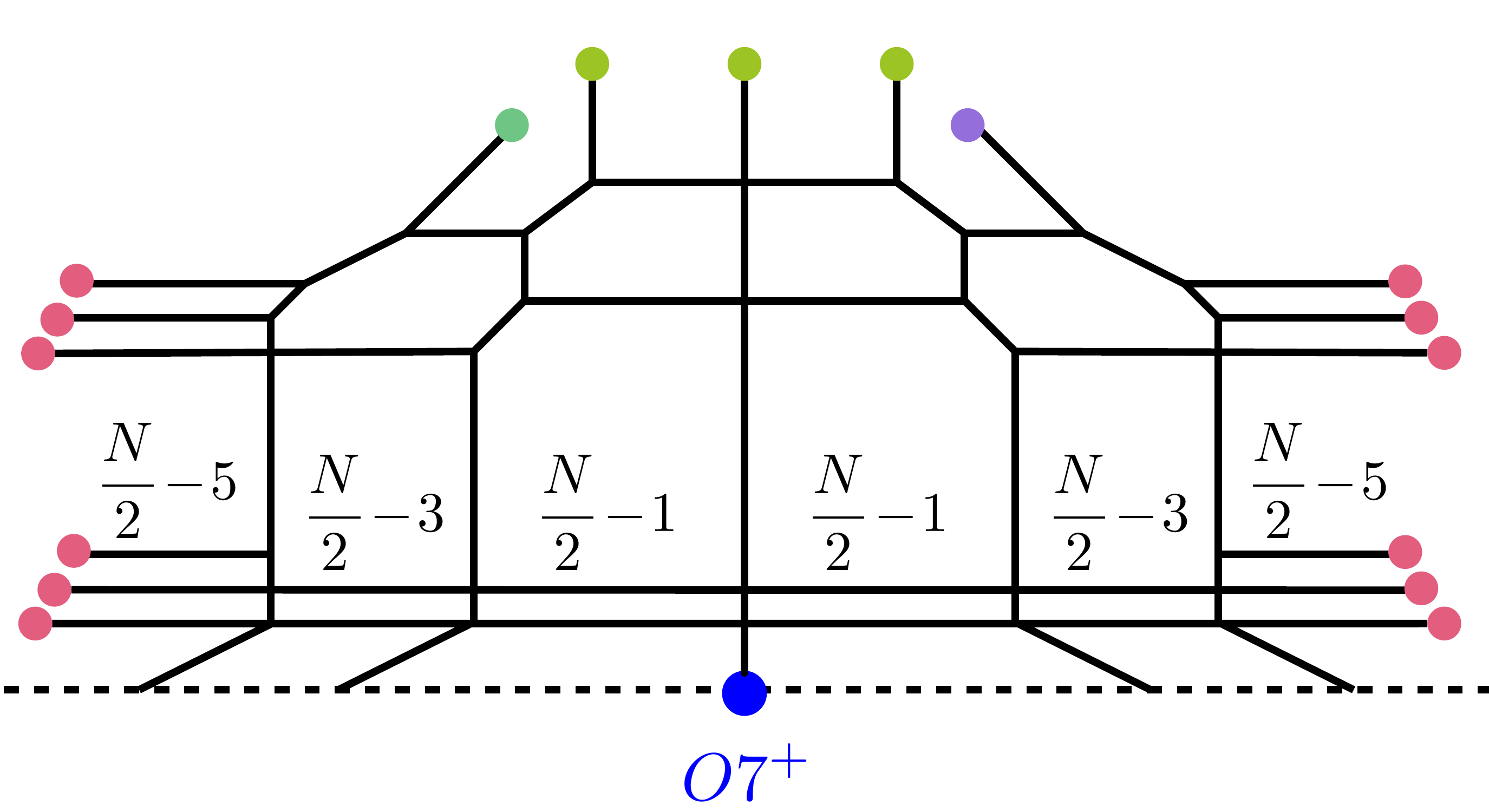}
\end{center}
\end{minipage}
\end{tabular}
\caption{The web diagram with odd number of NS5-branes.}
\label{fig:5dSUandS}
\end{figure}

The case with $2k+1$ NS5-branes is also straightforward.
This time, one of the NS5-branes connect $O7^+$ plane and $O7^-$ plane,
as in the left of Figure \ref{fig:5dSUandS}.
Analogous procedure ends up with the right of Figure \ref{fig:5dSUandS},
whose 5d theory is 
 \begin{equation}
 5d~~[S] - SU(N+2k-1) - SU(N+2k-5) - \cdots - SU(N-2k+3) - [N-2k+1] \label{5dSUS}
 \end{equation} 
The number of the $SU$ gauge nodes is $k$. In this case, the left gauge node is the $SU(N+2k-1)$ gauge group with matter in the symmetric representation which we denote by $[S]$ in \eqref{5dSUS}. Hence, we claim that the 5d $\prod SU$ quiver theory \eqref{5dSUS} with symmetric matter has the 6d UV completion given by the 6d $SU(N)$ quiver theory with the $2k$ gauge nodes. 

In particular, when $k=1$, one obtains a 5d $SU(N+1)$ gauge theory with $N-1$ fundamental hypermultiplets and also a hypermultiplet in the symmetric representation. A 5-brane web analysis implies that one can add $N-2$ fundamental matter to a 5d $SU(N+1)$ gauge theory with one symmetric matter when the UV completion is a 5d SCFT \cite{Bergman:2015dpa}. The analysis here means that when one adds one more flavor, the UV completion is the 6d $SU(N)$ quiver theory with $2$ gauge nodes.

\bigskip

\section{Conclusion and discussions}\label{sec:conclusion}
As a generalization of \cite{Hayashi:2015zka}, we studied
5d gauge theory descriptions of various 6d $\mathcal{N}=(1,0)$ SCFTs realized by
type IIA brane setup with an $ON^0$-plane with or without the presence of 
other orientifolds such as $O6$- and $O8$-planes. 

An $ON^0$-plane is an object which is S-dual to a D5-brane on top of an $O5^-$-plane. 
Brane configurations involving an $ON^0$-plane lead to various 6d quiver gauge theories of a $D$-type or an $A$-type. 5d gauge theory description whose brane setup involving such $ON^0$-plane, on the other hand, is not well studied. In this paper, 
we explored possible brane setups of $ON^0$ with/without other usual orientifolds. We gave a brane description for an $ON^0$-plane with a detailed brane web diagram in section \ref{sec:ONO5}, which allows one to see various dual descriptions through S-duality.

Among many 6d SCFTs, especially 6d SCFTs whose brane setup is realized with different types of the orientifolds, $ON^0, O6, O8$, leads to fruitful 5d gauge theory descriptions, We studied such 6d theories by first compactifying them on a circle and then taking a T-duality which yield a type IIB brane configuration with $ON^0, O5, O7$-planes. Based on the type IIB brane setup, we proposed various kinds of 5d $\mathcal{N}=1$ quiver gauge theories whose UV fixed point is an identical 6d SCFT. The main techniques that we devised to obtain a natural 5d gauge theory description is S-duality in IIB brane setup and quantum resolutions of $O7^-$-planes.

An S-duality transformation acting on the 5-brane web diagram is
often used to obtain a proper gauge theory description
and is useful to relate a 5d theory to 
different 5d gauge theories with identical UV fixed point.
In the web diagram,
this S-duality transformation is simply realized by rotating 
the $(p,q)$ 5-brane web diagram by 90 degrees.
When we include $O5$-planes as we discussed 
in section \ref{sec:ON0} and \ref{sec:ON0O8O6},
the analysis
is slightly more involved
because we do not know the field theory description
when we have an S-dual object of an $O5^{+}$-plane.
Instead of directly taking an S-duality, 
we  first recombine (as a revise procedure of splitting or as making two fractional NS5-brane unsplit) 
the fractional NS5-branes attached to the $O5$-planes
in such a way to avoid appearing $O5^{+}$-planes.
After obtaining an $ON^0$-plane as S-dual of an $O5^0$-plane,
we again split the D5-branes,
which were originally the fractional NS5-branes mentioned above.
This procedure enables us to obtain a proper 5d gauge theory description via S-duality.
As for clearer understanding of splitting D5-branes on an $ON^0$-plane,
especially in the context of the relation to its S-dual picture,
it is important to further investigate 
we need further investigation in the future.

Another 
technique is the resolution of an $O7^-$-plane.
This resolution was originally proposed by \cite{Sen:1996vd}
and applied to various brane setups to obtain non-trivial relation among 5d theories.
This resolution was also 
generalized to the case
where an $O7^-$-plane is attached to an NS5-brane \cite{Zafrir:2015rga, Hayashi:2015zka}.
In this paper, we proposed novel resolution of an $O7^-$ plane located at two intersections of an $ON^0$-plane and each $O5^0$-plane.
This is proposed by observing that the resolution of an $O7^-$-plane into
a pair of $[1,1]$ and $[1,-1]$ 7-branes
is consistent with the orientifold projection created by an $O5$-plane and an $ON^0$-plane
and by interpreting that one of them becomes the mirror image of the other.
Due to this resolution, we obtained various highly non-trivial dualities among 5d gauge theories,
one of which is the novel duality between a D-type $SU$ quiver gauge theory and
an $SO$-$Sp$ quiver gauge theory.
It would be interesting to give further evidence for these conjectured dualities
as well as for the novel resolution of the $O7^-$-plane.

Still another technique is a twisted circle compactification
which yields two different types of $O7$-planes after T-duality.
They are an $O7^-$-plane and an $O7^+$-plane maximally separated apart along the direction of the compactification radius.
Combined with the resolution of the $O7^-$-plane mentioned above,
we obtain a 5d gauge theory with an $SO$ gauge group.

Although we did not discuss in the main text, there are, of course, non-Lagrangian theories naturally arising along the procedure. For example, web diagrams leading to non-Lagrangian theory appear when one implements S-duality and the resolution of $O7^-$-plane at the same. In this paper, however, we focused on 5d web diagram giving a gauge theory description and have given proposals for 5d gauge theory description for various 6d SCFTs. 
We checked that all the proposed 5d descriptions 
have the expected number of mass parameters and Coulomb branch moduli parameters, and they all agree with those of the 6d theories.
It would be also interesting to give more evidences to support our conjectures.
One of the future work in this context will be to compute
the elliptic genus of the 6d SCFT corresponding to the original type IIA brane setup
and compare it with the 5d partition function or 5d superconformal index computed from topological string partition function.
However, the topological vertex formalism corresponding to a web diagram with an $O5$-plane and/or $ON^0$-plane is not known. The study in \cite{Hayashi:2015uka} may give a clue to this issue.


\bigskip

\acknowledgments
We thank Tohru Eguchi and Joonho Kim 
for useful discussions.
The work of H.H. is supported by the ERC Advanced Grant SPLE under contract ERC-2012-ADG-20120216-320421, by the grant FPA2012-32828 from the MINECO, and by the grant SEV-2012-0249 of the “Centro de Excelencia Severo Ochoa” Programme.
The work of K.L. is supported in part by the National Research Foundation of Korea (NRF) Grants No. 2006-0093850. 
S.K. is thankful to IPMU for hospitality during his visit. F.Y. and S.K. would like to acknowledge RIKEN where part of work is done. 
For interesting and useful discussions, we are thankful to  
various workshops we participated: 
the KIAS-YITP Joint Workshop 2015 ``Geometry in Gauge Theories and String Theory'' at KIAS, 
the 8th Taiwan String Workshop at National Tsing Hua University and the KIAS Research Station program ``Current Topics in String Theory'' in Songdo, Korea.
\bigskip

\bibliographystyle{JHEP}
\providecommand{\href}[2]{#2}\begingroup\raggedright\endgroup

\end{document}